\documentclass[aps,prx,reprint,onecolumn,10pt,superscriptaddress,nofootinbib]{revtex4-2}

\usepackage[utf8]{inputenc}
\usepackage[english]{babel}
\usepackage[T1]{fontenc}
\usepackage{amsmath}
\usepackage{amssymb}
\usepackage{mathrsfs} 
\usepackage{graphicx}
\usepackage{float}
\usepackage{tikz}
\usepackage{pgfplots}
\pgfplotsset{compat=1.18} 
\usetikzlibrary{decorations.pathmorphing, arrows.meta, positioning}

\usepackage[colorlinks=true, linkcolor=blue, citecolor=blue, urlcolor=blue]{hyperref}

\begin{document}

\title{Far-field spatial coherence driven by lossy objects: first-principles approach unifying scattering of quantum light and thermal emission}

\author{Alessandro Ciattoni}
\email{alessandro.ciattoni@spin.cnr.it}
\homepage{https://orcid.org}
\affiliation{CNR-SPIN, c/o Dip.to di Scienze Fisiche e Chimiche, Via Vetoio, 67100 Coppito (L'Aquila), Italy}

\begin{abstract}
Far-field spatial coherence dictates the interference properties of scattered light and thermal emission. Traditionally, these phenomena are treated through disjointed paradigms: classical scattering descriptions assume cold objects lacking quantum fluctuations, idealized quantum scattering schemes ignore dissipation, and semiclassical fluctuational electrodynamics relies on phenomenological noise currents, precluding the consistent treatment of incident quantum states. Here, we develop a first-principles framework based on the modified Langevin noise formalism to unify the scattering of quantum light and the intrinsic thermal emission of finite dissipative objects. We demonstrate that the outgoing far-field spatial coherence separates into an algebraic superposition of two geometry-driven mechanisms, coupled by the global unitarity of the radiation-matter dynamics. The first mechanism, elastic scattering, acts as a non-unitary spatial filter, mode-selectively attenuating and reshaping incident quantum correlations. The second mechanism, thermal emission, originates from localized material dissipation and projects the object's absorption profile into the far field, providing a quantum-vectorial derivation of the macroscopic van Cittert-Zernike theorem. Applying this framework across optical regimes, we determine operational bounds for lossy quantum photonics. Under chaotic thermal illumination, we analytically demonstrate thermal cloaking at equilibrium and show that a passive sink casts a structured thermal shadow geometrically identical to a primary emitter. Under coherent illumination, we derive a thermodynamic phase diagram bounding macroscopic phase correlations, demonstrating that subwavelength nanostructures undergo substantial coherence degradation compared to bulk objects. Finally, under spatially entangled illumination, we evaluate the geometric and thermodynamic scaling of the quantum-thermal interplay. When intrinsic thermal emission is negligible, we show that a dissipative scatterer can enhance far-field spatial coherence by selectively filtering higher-order Schmidt modes.
\end{abstract}

\maketitle

\section{Introduction}
Spatial coherence is a fundamental property of light, dictating its ability to exhibit interference \cite{Mandel1995, Glauber1963} and carrying distinct signatures of the physical processes that generated \cite{Greffet2002} or redirected it \cite{Rotter2017}. In the far field, the spatial correlation between different propagation directions is central to both classical applications, such as stellar interferometry~\cite{Michelson1921, Monnier2003} and advanced imaging~\cite{Zheng2013, Schnars1994}, and modern quantum technologies, including quantum communication~\cite{Gisin2002, Walborn2006} and continuous-variable entanglement distribution~\cite{Braunstein2005}. In the existing literature, spatial coherence in the far field of an object is typically analyzed by treating the scattering of incident partially coherent light and thermal emission as disconnected phenomena. Classical scattering theory relies on deterministic wave-propagation formalisms to describe how an obstacle redistributes the spatial correlations of incident light, neglecting the radiation spontaneously produced by the matter~\cite{Carney1998,Fischer2012,Schouten2024,Premaratne2012}. While quantum extensions of these models capture the elastic scattering of non-classical states, they typically treat the scatterers as idealized lossless dielectrics~\cite{Glauber1991, Fearn1987, Blow1990}, omitting the energy dissipation and quantum noise required to consistently describe thermal emission through Kirchhoff's law. Conversely, thermal emission is standardly modeled via semiclassical fluctuational electrodynamics \cite{Rytov1989, Joulain2005}, a framework that has provided generalizations of the van Cittert-Zernike theorem for macroscopic emitters \cite{Bertilone1997}. Nevertheless, this approach relies on phenomenological noise currents \cite{Polder1971, Eckhardt1984} and typically assumes a purely classical or chaotic background, lacking the formalism to consistently integrate the scattering of complex quantum states impinging on the very same absorptive body. As a result, a unified scheme capable of capturing all these individual phenomena and their concurrent interplay has yet to be established.

To simultaneously account for both quantum scattering and thermal emission, a rigorous framework describing the interaction of quantized fields with macroscopic dissipative matter is required. Indeed, as firmly established by classical thermodynamics through Kirchhoff's law—which strictly equates a body's capacity to emit radiation to its absorptivity—any quantum formalism omitting a rigorous description of material dissipation is inherently precluded from capturing thermal emission. The standard Langevin noise formalism (LNF) provides a quantization scheme for lossy media by introducing bosonic excitations associated with quantized dipole sources distributed throughout the material~\cite{Gruner1996, Matloob1995}; originating from dissipation, these sources vanish in the lossless limit ($\text{Im}(\varepsilon) = 0$). Yet, the LNF is formulated for unbounded absorbing media, and consequently, describing the scattering of incoming plane waves from infinity falls outside its native domain of applicability. Nevertheless, by employing specific limiting procedures ($\text{Im}(\varepsilon) \to 0^+$)~\cite{Gruner1996b, Knoll1999, Khanbekyan2003}, the LNF can be extended to treat finite-size scatterers embedded in vacuum, even though carrying out such limits becomes mathematically cumbersome in non-planar geometries. These difficulties are circumvented by the Modified Langevin Noise Formalism (MLNF), which explicitly separates the background scattering field (generated by the asymptotically incoming field, $s$-polaritons) from the medium-assisted field (generated by the quantized dipole sources, $e$- and $m$-polaritons), thereby preserving their distinct physical origins. Originally developed for one-dimensional planar structures~\cite{DiStefano2001} and later extended to non-magnetic finite objects~\cite{Drezet2017, Na2023}, the MLNF has recently been generalized to arbitrary lossy magneto-dielectric bodies~\cite{Ciattoni2024}. Crucially, the MLNF rests on a rigorous theoretical foundation, having been identified as the second-quantized form of the canonical quantization of macroscopic electromagnetism developed by Philbin~\cite{Philbin2010}: an equivalence first indirectly validated in Ref.~\cite{Ciattoni2024}, and recently proven through a direct derivation in Ref.~\cite{Ciattoni2026}. Capitalizing on its ability to accommodate quantized scattering modes, the MLNF has recently been applied to model the coupling of quantum emitters to dispersive objects~\cite{Miano2025a,Miano2025b,Miano2026}, as well as to describe the mechanical actions exerted on an absorbing object under quantum illumination~\cite{Ciattoni2026b}. Beyond these specific applications, this framework has enabled the formulation of a general quantum scattering theory for finite lossy media in vacuum~\cite{Ciattoni2025,Ciattoni2025b}, wherein Kirchhoff's law is algebraically proven to be a strict consequence of the global unitarity of the coupled radiation-matter dynamics.

Building on these theoretical advancements, in this work we develop a first-principles approach that simultaneously incorporates light scattering and thermal emission to determine the far-field spatial coherence of dissipative objects. Our framework relies on the general scattering theory established in Ref.\cite{Ciattoni2025}, which is rooted in the MLNF and connects the radiation-matter configuration in the far past (the incoming field and the object’s initial thermal state) to its far-future counterpart (the scattered radiation and the final matter distribution). Employing this formal machinery, we evaluate the outgoing far-field spatial coherence and directly compare it to its known incident counterpart. We demonstrate that this outgoing coherence strictly separates into an exact algebraic superposition of light scattering and thermal emission, both governed by the corresponding classical dyadics, thus quantitatively capturing how they are driven by the object’s electromagnetic geometry. Specifically, the scattering process acts as a non-unitary filter, mode-selectively attenuating and mapping the incident quantum correlations into the far field via the classical transmission dyadic of the object. Concurrently, thermal emission generates its own geometry-driven spatial correlations from stochastically independent internal fluctuations, effectively providing an exact quantum-vectorial derivation of the macroscopic van Cittert-Zernike theorem. This latter process projects the object's local absorption profile into the far field via the emission dyadic, in strict accordance with Kirchhoff's law. Notably, while both phenomena account for material losses, they exhibit a fundamental distinction in the transparency limit, where elastic scattering naturally survives whereas thermal emission identically vanishes. Furthermore, these two processes are intimately connected by the global unitarity of the macroscopic quantum dynamics \cite{Ciattoni2025}: thermal emission exactly compensates for the non-unitary defect of the transmission channels as strictly required by macroscopic energy conservation, an algebraic constraint that enables the full thermal contribution to be expressed solely through the classical transmission dyadic. To the best of our knowledge, this unified approach provides a generalized framework for these processes, accommodating any arbitrarily shaped inhomogeneous magnetodielectric object.
 
To demonstrate the predictive power of this framework in realistic optical scenarios where macroscopic material losses cannot be neglected, we explicitly evaluate the far-field spatial coherence across three paradigmatic illumination regimes, uncovering strict operational bounds for lossy quantum photonics:

\begin{itemize}
    \item \textbf{Chaotic thermal illumination:} We analytically solve the spatial symmetry breaking of local thermodynamics. Alongside an exact algebraic proof of thermal cloaking at global equilibrium, we formally establish that a passive cold absorber carves a structured ``thermal shadow'' into a chaotic background. This geometry-driven radiation deficit yields a spatial coherence profile structurally identical to that of an active primary emitter, explicitly linking absorption and emission phenomena at the quantum correlation level.
    
    \item \textbf{Coherent illumination:} We quantify the exact physical threshold at which macroscopic phase correlations survive stochastic material fluctuations. By demonstrating that thermal emission acts as a non-unitary decoherence channel, we derive a universal thermodynamic phase diagram bounding spatial coherence. This reveals that while the decoherence of macroscopic objects is governed by the pure thermal energy scale, subwavelength nanostructures act as highly efficient thermal antennas that severely suppress the coherence survival threshold.
    
    \item \textbf{Spatially entangled illumination:} Evaluating the geometric and thermodynamic scaling of the quantum-thermal interplay, we establish the operational regimes governing the competition between non-local correlations and local dissipation. Furthermore, we demonstrate that, under conditions where intrinsic thermal emission is negligible, a dissipative scatterer can enact geometric purification of the incident field via selective spatial mode filtering, enhancing macroscopic spatial coherence at the strict expense of the transmitted biphoton flux.
\end{itemize}

Ultimately, by providing a single theoretical framework that rigorously unifies the quantum optical scattering of generally non-pure states with intrinsic thermal emission, our results capture their previously inaccessible interplay, establishing a generalized, first-principles foundation for the design of lossy quantum optical components, robust structured-light networks, and realistic thermal metamaterials.

\section{Theoretical background}
To investigate the far-field spatial coherence, we consider an arbitrary dispersive and absorbing magneto-dielectric object occupying a finite volume $V$ embedded in vacuum, characterized by a complex dielectric permittivity $\varepsilon_\omega({\bf r})$ and magnetic permeability $\mu_\omega({\bf r})$. Our analysis relies directly on the Modified Langevin Noise Formalism (MLNF) and the associated quantum scattering approach \cite{Ciattoni2024, Ciattoni2026, Ciattoni2025}. By providing a canonical description of the interaction between quantized fields and macroscopic matter, this framework allows us to simultaneously describe both the scattering of incoming quantum light and the intrinsic thermal emission of the dissipative object. A detailed mathematical summary is provided in Appendix A; here we outline the key relations governing the far-field dynamics.

Within the MLNF, the coupled radiation-matter system is described by continuous sets of bosonic excitations. The incoming radiation from spatial infinity is described by the scattering ($s$) polariton operators ${\bf{\hat g}}_{\omega s} \left( {\bf{n}} \right)$, representing the external vacuum modes propagating along the direction ${\bf{n}}$. The object's internal dissipation is modeled by localized electric ($e$) and magnetic ($m$) polariton operators, ${\bf{\hat f}}_{\omega e } \left( {\bf{r}} \right)$ and ${\bf{\hat f}}_{\omega m } \left( {\bf{r}} \right)$. These act as quantized sources confined to the object's volume $V$ and vanish in the lossless limit. In the MLNF, the positive-frequency electric field operator ${\bf{\hat E}}_\omega  \left( {\bf{r}} \right)$ is expressed as the sum of two distinct contributions:
\begin{equation} \label{E_om_Main}
{\bf{\hat E}}_\omega  \left( {\bf{r}} \right) = \int {do_{\bf{n}} } {\cal F}_{\omega s} \left( {\left. {\bf{r}} \right|{\bf{n}}} \right) \cdot {\bf{\hat g}}_{\omega s} \left( {\bf{n}} \right) + \sum\limits_{\nu=e,m}  {\int_V {d^3 {\bf{r}}'\,} {\cal G}_{\omega \nu } \left( {\left. {\bf{r}} \right|{\bf{r}}'} \right) \cdot {\bf{\hat f}}_{\omega \nu } \left( {{\bf{r}}'} \right)}. 
\end{equation}
The first term represents the \textit{scattering field}, weighted by the scattering kernel ${\cal F}_{\omega s}$---a quantity constructed from the classical modal dyadic that describes the scattering of the external incoming vacuum modes by the object. The second term is the \textit{medium-assisted field}, weighted by the electric and magnetic kernels ${\cal G}_{\omega e}$ and ${\cal G}_{\omega m}$---quantities constructed from the classical dyadic Green's function that are proportional to the imaginary parts of the object's permittivity and permeability, respectively, evaluated at the source point ${\bf r}'$, describing the radiation generated by the quantized material fluctuations. The independence of the polaritonic excitations is reflected in the system's Hamiltonian, which decomposes into distinct scattering and material components, $\hat{H} = \hat{H}_s + \hat{H}_{em}$:
\begin{align} \label{Hs_Hem_Main}
\hat{H}_s &= \int {d\omega } \;\hbar \omega \int {do_{\bf{n}} } {\bf{\hat g}}_{\omega s}^\dag  \left( {\bf{n}} \right) \cdot {\bf{\hat g}}_{\omega s} \left( {\bf{n}} \right), &
\hat{H}_{em} &= \int {d\omega } \;\hbar \omega \sum\limits_{\nu=e,m} \int_V {d^3 {\bf{r}} \,}  {{\bf{\hat f}}_{\omega \nu }^\dag  \left( {\bf{r}} \right) \cdot {\bf{\hat f}}_{\omega \nu } \left( {\bf{r}} \right)}.
\end{align}

Scattering observables are evaluated from the asymptotic behavior of the radiated field. The positive-frequency time-dependent field ${\bf{\hat E}}^{(+)}({\bf{r}},t) = \int d\omega \, e^{-i\omega t} {\bf{\hat E}}_\omega({\bf{r}})$ takes the following far-field ($r \to +\infty$) forms in the far-past ($t \to -\infty$) and far-future ($t \to +\infty$) limits:
\begin{equation} \label{E_asymptotic_Main}
{\bf{\hat E}}^{(+)}(r{\bf{n}},t) \mathop \approx \limits_{r \to +\infty} \frac{1}{ir} \int d\omega \sqrt{\frac{\hbar k_\omega}{4\pi \varepsilon_0}} 
\begin{cases} 
- e^{-ik_\omega(r + ct)} {\bf{\hat g}}_{\omega s}(-{\bf{n}}), & t \to -\infty, \\
\phantom{-} e^{ik_\omega(r - ct)} {\bf{\hat G}}_{\omega s}({\bf{n}}), & t \to +\infty,
\end{cases}
\end{equation}
where $k_\omega = \omega/c$. The far-past behavior consists of converging spherical waves associated with the ingoing operators ${\bf{\hat g}}_{\omega s}$, whereas the far-future behavior consists of diverging spherical waves associated with the outgoing scattering polariton operators ${\bf{\hat G}}_{\omega s}$. As part of the input-output relation detailed in Appendix~A, $\hat{\mathbf{G}}_{\omega s}$ is expressed as a linear superposition of the fundamental bosonic operators $\hat{\mathbf{g}}_{\omega s}$, $\hat{\mathbf{f}}_{\omega e}$, and $\hat{\mathbf{f}}_{\omega m}$, thereby accounting for both the transmitted incoming field and the emitted thermal fluctuations:
\begin{equation} \label{G_operator_Main}
{\bf{\hat G}}_{\omega s} \left( {\bf{n}} \right) = \int do_{\bf{n}'} {\cal T}_{\omega ss} \left( {\left. {\bf{n}} \right|{\bf{n}}'} \right) \cdot {\bf{\hat g}}_{\omega s} \left( {{\bf{n}}'} \right) + \sum\limits_{\nu = e,m} \int_V d^3 {\bf{r}}\, {{\cal E}_{\omega s\nu } \left( {\left. {\bf{n}} \right|{\bf{r}}} \right) \cdot {\bf{\hat f}}_{\omega \nu } \left( {\bf{r}} \right)}.
\end{equation}
Here, ${\cal T}_{\omega ss}$ is the classical transmission dyadic, describing the geometric redistribution of the incident field, which is related to the classical scattering dyadic ${\cal S}_\omega$ by the relation \cite{Kristensson2016}:
\begin{equation} \label{T_dyadic}
{\cal T}_{\omega ss} \left( {\left. {\bf{n}} \right|{\bf{n}}'} \right) = \delta \left( {o_{\bf{n}}  - o_{{\bf{n}}'} } \right){\cal I}_{\bf{n}}  + \frac{{ik_\omega  }}{{2\pi }}{\cal S}_\omega  \left( {\left. {\bf{n}} \right|{\bf{n}}'} \right).
\end{equation}
The classical emission dyadics $\mathcal{E}_{\omega s\nu }$ describe the propagation of the local internal quantum fluctuations into the far field. Crucially, as established within the quantum scattering formalism, these emission dyadics are linked to the classical absorption dyadics $\mathcal{A}_{\omega \nu s}$ through the microscopic symmetry relation
\begin{equation} \label{Kirchhoff_Symmetry}
\mathcal{A}_{\omega \nu s}(\mathbf{r}|\mathbf{n}) = \mathcal{E}_{\omega s\nu}^T(-\mathbf{n}|\mathbf{r}).
\end{equation}
This symmetry strictly equates the object's emission and absorption capacities, thereby enforcing Kirchhoff's law. Notably, within this quantum optical scattering framework, Kirchhoff's law is not introduced \textit{a priori} as a phenomenological thermodynamic assumption, but rather emerges from the foundational input-output relations as a strict consequence of the global unitarity of the coupled radiation-matter quantum dynamics. Physically, these emission and absorption dyadics are proportional to the imaginary parts of the object's permittivity and permeability, respectively, anchoring the macroscopic thermal radiation to the localized internal electromagnetic dissipation.

Furthermore, macroscopic energy conservation---manifesting mathematically as the unitarity of the input-output relations---requires the transmission and emission dyadics to satisfy the exact integral identity:
\begin{equation} \label{Unitarity_Relation_Main}
\int {do_{\bf{m}} } {\cal T}_{\omega ss} \left( {\left. {\bf{n}} \right|{\bf{m}}} \right) \cdot {\cal T}_{\omega ss}^{T*} \left( {\left. {{\bf{n}}'} \right|{\bf{m}}} \right) + \sum\limits_\nu  \int_V {d^3 {\bf{r}} \,}  {{\cal E}_{\omega s\nu } \left( {\left. {\bf{n}} \right|{\bf{r}}} \right) \cdot {\cal E}_{\omega s\nu }^{T*} \left( {\left. {{\bf{n}}'} \right|{\bf{r}}} \right)}  = \delta \left( {o_{\bf{n}}  - o_{{\bf{n}}'} } \right){\cal I}_{\bf{n}}.
\end{equation}
Equation~(\ref{Unitarity_Relation_Main}) demonstrates that, for a dissipative object, the transmission operator $\mathsf{T}_{\omega ss}$ (see Eq.(\ref{Toss})) is not unitary and that its unitarity defect is exactly compensated by the emission channels.

\section{Spatial coherence in the far field of a lossy object}

Building on the quantum scattering framework, we now evaluate the far-field spatial coherence of a dissipative object. We first introduce the far-field correlation dyadic and the generalized degree of coherence. We then show that the outgoing spatial coherence arises from the interplay of two distinct mechanisms: the elastic scattering of the incident quantum field and the object's intrinsic thermal emission. Finally, by exploiting the unitarity of the global radiation-matter dynamics, we demonstrate that the full far-field coherence can be expressed relying exclusively on the classical transmission dyadics.

\paragraph{Far-field correlation dyadic and degree of coherence.} In a general far-field measurement, radiation is collected by detectors located at a large distance from the object volume $V$. For simplicity, we assume that all detectors have identical acceptance; thus, any given detector subtends a small but finite solid angle $\Delta \Omega_{\bf n}$ centered around its observation direction ${\bf n}$, whose scalar measure is equal to $\Delta \Omega$. This finite acceptance effectively discretizes the continuous set of observation directions, implying that two directions ${\bf n}$ and ${\bf n}'$ are considered to coincide if and only if their corresponding detection solid angles $\Delta \Omega_{\bf n}$ and $\Delta \Omega_{{\bf n}'}$ coincide. Accordingly, the spatial correlation between two observation directions ${\bf n}$ and ${\bf n}'$ at time $t$ is quantified by the far-field correlation dyadic ${\cal C}\left( {{\bf{n}},{\bf{n}}';r,t} \right)$, which is extracted via the asymptotic limit
\begin{equation} \label{C_dyad_def}
\int_{\Delta \Omega_{\bf{n}} } {do_{\bf{m}} } \int_{\Delta \Omega_{{\bf{n}}'} } {do_{{\bf{m}'}} } \langle {{\bf{\hat E}}^{(+) \dagger} (r{\bf{m}},t) {\bf{\hat E}}^{(+)}(r{\bf{m}}',t)} \rangle \mathop \approx \limits_{r \to +\infty} \frac{\hbar }{{4\pi \varepsilon _0 cr^2 }}{\cal C}\left( {{\bf{n}},{\bf{n}}';r,t} \right),
\end{equation}
as the leading-order term in the $r \to +\infty$ asymptotic expansion of the normally ordered correlation dyadic integrated over the detector apertures, where the dimensional prefactor $\hbar / (4\pi \varepsilon_0 c r^2)$ has been factored out to simplify the subsequent expressions. The quantum expectation values are evaluated as $\langle \hat{O} \rangle = \text{Tr}(\hat{\rho} \hat{O})$, where $\text{Tr}$ denotes the operator trace over the entire MLNF Fock space and, in the Heisenberg picture we employ, $\hat{\rho}$ is the time-independent ``initial'' density operator representing the global radiation-matter state prepared in the far-past. While the far-field correlation dyadic in Eq.(\ref{C_dyad_def}) is explicitly defined at any time $t$, focusing on its large-time behaviors, namely ${\cal C}_{in}({\bf n},{\bf n}';r,t)$ in the far-past ($t \to -\infty$) and ${\cal C}_{out}({\bf n},{\bf n}';r,t)$ in the far-future ($t \to +\infty$), simplifies the formalism and enables their comparison to physically isolate the quantum signature left by the material on the far-field spatial coherence. Substituting the large-time expressions of the far field of ${\bf{\hat E}}^{(+)}$ from Eq.(\ref{E_asymptotic_Main}) into Eq.(\ref{C_dyad_def}) directly yields
\begin{align} \label{C_dyads}
 {\cal C}_{in} \left( {{\bf{n}},{\bf{n}}';r,t} \right) &= \int {d\omega } \int {d\omega '} \sqrt {\omega \omega '} e^{i\left( {\omega  - \omega '} \right)\left( {t + \frac{r}{c}} \right)} \int_{\Delta \Omega _{-{\bf{n}}} } {do_{\bf{m}} } \int_{\Delta \Omega _{-{\bf{n}}'} } {do_{{\bf{m}}'} } \langle {{\bf{\hat g}}_{\omega s}^\dag  \left( {\bf{m}} \right){\bf{\hat g}}_{\omega 's} \left( {\bf{m}}' \right)} \rangle , \nonumber \\ 
 {\cal C}_{out} \left( {{\bf{n}},{\bf{n}}';r,t} \right) &= \int {d\omega } \int {d\omega '} \sqrt {\omega \omega '} e^{i\left( {\omega  - \omega '} \right)\left( {t - \frac{r}{c}} \right)} \int_{\Delta \Omega _{\bf{n}} } {do_{\bf{m}} } \int_{\Delta \Omega _{{\bf{n}}'} } {do_{{\bf{m}}'} } \langle {{\bf{\hat G}}_{\omega s}^\dag  \left( {\bf{m}} \right){\bf{\hat G}}_{\omega 's} \left( {{\bf{m}}'} \right)} \rangle .
\end{align}
In the far-past, the far-field correlation dyadic ${\cal C}_{in}$ depends on the ingoing spectral correlation dyadic $\langle {\bf{\hat g}}_{\omega s}^\dag({\bf m}) {\bf{\hat g}}_{\omega 's}({\bf m}') \rangle$, with the negative signs in the integration domains $\Delta \Omega_{-{\bf n}}$ and $\Delta \Omega_{-{\bf n}'}$ geometrically indicating radiation propagating inward from spatial infinity. Physically, ${\cal C}_{in}$ contains no contribution from the material itself since the field has not yet interacted with the scatterer; consequently, the spatial coherence is determined by the specific quantum illumination prepared in $\hat{\rho}_s$. Conversely in the far-future, the far-field correlation dyadic ${\cal C}_{out}$ is governed by the outgoing spectral correlation dyadic $\langle {\bf{\hat G}}_{\omega s}^\dag({\bf m}) {\bf{\hat G}}_{\omega 's}({\bf m}') \rangle$ which, via the outgoing scattering polariton operators defined in Eq.(\ref{G_operator_Main}), encodes the macroscopic presence of the object through the classical dyadics ${\cal T}_{\omega ss}$ and ${\cal E}_{\omega s \nu}$. Furthermore, its dependence on the unit vectors ${\bf m}$ and ${\bf m}'$ indicates that the radiation has interacted with the scatterer and propagates outward, thereby carrying the spectral and spatial signature of the material. 

To quantify the spatial correlation of the quantized vectorial field evaluated along two arbitrary far-field directions, ${\bf{n}}$ and ${\bf{n}}'$, it is convenient to introduce the generalized electromagnetic degree of coherence $\mu^2$. Extending the classical formulation of Ref.~\cite{Tervo2003} to the quantum domain, this metric is defined as the normalized squared Frobenius norm of the far-field correlation dyadic ${\cal C}$, namely
\begin{equation} \label{mu2_definition}
\mu ^2 \left( {{\bf{n}},{\bf{n}}';r,t} \right) = \frac{{{\rm tr}\left[ {{\cal C}\left( {{\bf{n}},{\bf{n}}';r,t} \right) \cdot {\cal C}^{T*} \left( {{\bf{n}},{\bf{n}}';r,t} \right)} \right]}}{{{\rm tr}\left[ {{\cal C}\left( {{\bf{n}},{\bf{n}};r,t} \right)} \right]{\rm tr}\left[ {{\cal C}\left( {{\bf{n}}',{\bf{n}}';r,t} \right)} \right]}},
\end{equation}
where ${\rm tr}$ denotes the dyadic trace. Owing to the vectorial nature of the field, this quantity is physically insightful even when evaluated along a single direction (${\bf{n}} = {\bf{n}}'$), where it yields the local degree of partial polarization. This scalar metric allows for a direct comparison between the incident and outgoing radiation by substituting ${\cal C}_{in}$ and ${\cal C}_{out}$, respectively, into Eq.(\ref{mu2_definition}).

\paragraph{Quantum scattering and thermal emission interplay.} In this work, our analysis focuses on the outgoing spatial coherence in the far field, produced by an arbitrary initial quantum illumination impinging on an object in thermal equilibrium at temperature $T_{em}$. Assuming that the incident radiation is statistically independent of the object's intrinsic material excitations, the global state factorizes as $\hat{\rho} = \hat{\rho}_s \hat{\rho}_{em}$, where $\hat{\rho}_s$ operates within the scattering sector of the Fock space and specifies the quantum-statistical state of the ingoing $s$-polaritons, whereas $\hat{\rho}_{em} = e^{-\hat{H}_{em} / k_B T_{em}} / {\rm Tr} ( e^{-\hat{H}_{em} / k_B T_{em}} )$ represents the canonical distribution over the material sector of the Fock space, governed by the Hamiltonian $\hat{H}_{em}$ (see Eq.~(\ref{Hs_Hem_Main})). To evaluate the outgoing spectral correlation dyadic $\langle {\bf{\hat G}}_{\omega s}^\dag({\bf m}) {\bf{\hat G}}_{\omega' s}({\bf m}') \rangle$, we note that this state preparation implies that the expectation values of the material polariton operators vanish, $\langle {\bf{\hat f}}_{\omega \nu} ({\bf r}) \rangle = 0$, whereas their two-point correlations follow directly from thermal equilibrium:
\begin{equation} \label{Thermal_Averages}
\langle {\bf{\hat f}}_{\omega \nu}^\dag({\bf r}) {\bf{\hat f}}_{\omega' \nu'}({\bf r}') \rangle = n_\omega(T_{em}) \delta(\omega-\omega') \delta_{\nu \nu'} \delta({\bf r}-{\bf r}') {\cal I},
\end{equation}
where $n_\omega(T_{em}) = (e^{\hbar \omega / k_B T_{em}} - 1)^{-1}$ is the standard Bose-Einstein distribution factor. Furthermore, owing to the condition $\langle {\bf{\hat f}}_{\omega \nu} ({\bf r}) \rangle = 0$ alongside the statistical independence of the initial states, all cross-correlation terms vanish (e.g., $\langle {\bf{\hat g}}_{\omega s}^\dag({\bf p}) {\bf{\hat f}}_{\omega' \nu'}({\bf r}') \rangle = 0$). As a result, we directly obtain the explicit decomposition of the outgoing spectral correlation dyadic into distinct scattering and thermal emission contributions:
\begin{align} \label{Out_Spec_Corr}
\langle {\bf{\hat G}}_{\omega s}^\dag({\bf m}) {\bf{\hat G}}_{\omega' s}({\bf m}') \rangle &= \int do_{\bf p} \int do_{{\bf p}'} {\cal T}_{\omega ss}^*({\bf m}|{\bf p}) \cdot \langle {\bf{\hat g}}_{\omega s}^\dag({\bf p}) {\bf{\hat g}}_{\omega' s}({\bf p}') \rangle \cdot {\cal T}_{\omega' ss}^T({\bf m}'|{\bf p}') \nonumber \\
&\quad + \delta(\omega-\omega') n_\omega(T_{em}) \sum_\nu \int_V d^3 {\bf r} \, {\cal E}_{\omega s \nu}^*({\bf m}|{\bf r}) \cdot {\cal E}_{\omega s \nu}^T({\bf m}'|{\bf r}).
\end{align}

The first term on the right-hand side of Eq.~(\ref{Out_Spec_Corr}) represents the scattering contribution to the outgoing correlations. Physically, this term dictates how a structured spatial coherence is imprinted onto the far field from the quantum fluctuations $\langle {\bf{\hat g}}_{\omega s}^\dag({\bf p}) {\bf{\hat g}}_{\omega' s}({\bf p}') \rangle$ of the incident illumination, a process driven by the object's electromagnetic geometry encoded within the classical transmission dyadics ${\cal T}_{\omega ss}^*({\bf m}|{\bf p})$ and ${\cal T}_{\omega' ss}^T({\bf m}'|{\bf p}')$. The second term on the right-hand side of Eq.~(\ref{Out_Spec_Corr}) represents the thermal emission contribution. Explicitly, it is scaled by the frequency Dirac delta $\delta(\omega-\omega')$ as a consequence of the temporal stationarity of the emission process, and weighted by the Bose-Einstein distribution $n_\omega(T_{em})$ governing the thermal population of the modes. Notably, this term generates structured spatial correlations in the far field despite the underlying internal thermal sources being spatially uncorrelated, as established by Eq.~(\ref{Thermal_Averages}). Physically, this emergence of macroscopic coherence is mediated by the classical emission dyadics, which map the localized material fluctuations onto the asymptotic observation directions. Because ${\cal E}_{\omega s e}$ and ${\cal E}_{\omega s m}$ are proportional to ${\rm Im}[\varepsilon _\omega ({\bf r})]$ and ${\rm Im}[\mu _\omega ({\bf r})]$ (see Appendix A), the thermal emission contribution propagates the object's local absorption profile onto the outgoing transverse spatial correlations. Inserting Eq.~(\ref{Out_Spec_Corr}) into the second of Eqs.~(\ref{C_dyads}) yields the far-future correlation dyadic 
\begin{align} \label{General_C_out}
{\cal C}_{out} \left( {{\bf{n}},{\bf{n}}';r,t} \right) &= \int d\omega \int d\omega ' \sqrt {\omega \omega '} e^{i\left( {\omega  - \omega '} \right)\left( {t - \frac{r}{c}} \right)} \int_{\Delta \Omega _{\bf{n}} } do_{\bf{m}} \int_{\Delta \Omega _{{\bf{n}}'} } do_{{\bf{m}}'} \int do_{\bf{p}}  \int do_{{\bf{p}}'} \nonumber \\ 
&\quad \times {\cal T}_{\omega ss}^* \left( {\left. {\bf{m}} \right|{\bf{p}}} \right) \cdot \left\langle {{\bf{\hat g}}_{\omega s}^\dag  \left( {\bf{p}} \right){\bf{\hat g}}_{\omega 's} \left( {{\bf{p}}'} \right)} \right\rangle \cdot {\cal T}_{\omega 'ss}^T \left( {\left. {{\bf{m}}'} \right|{\bf{p}}'} \right) \nonumber \\
&\quad + \int d\omega \, \omega n_\omega  \left( {T_{em} } \right){\cal V}_\omega  \left( {{\bf{n}},{\bf{n}}'} \right),
\end{align}
where the structured dyadic ${\cal V}_\omega  \left( {{\bf{n}},{\bf{n}}'} \right)$ is defined as
\begin{equation} \label{DyadV}
{\cal V}_\omega  \left( {{\bf{n}},{\bf{n}}'} \right) = \int_{\Delta \Omega _{\bf{n}} } do_{\bf{m}} \int_{\Delta \Omega _{{\bf{n}}'} } do_{{\bf{m}}'} \int d^3 {\bf{r}} \sum\limits_\nu  {{\cal E}_{\omega s\nu }^* \left( {\left. {\bf{m}} \right|{\bf{r}}} \right) \cdot {\cal E}_{\omega s\nu }^T \left( {\left. {{\bf{m}}'} \right|{\bf{r}}} \right)}.
\end{equation}
Equation~(\ref{General_C_out}) constitutes the central theoretical result of this framework. Unlike standard semiclassical approaches that append phenomenological noise currents to classical scattering, or idealized quantum models that neglect dissipation, Eq.~(\ref{General_C_out}) provides a first-principles unification. It demonstrates that the outgoing macroscopic spatial coherence exactly separates into two distinct contributions: the elastic scattering of the incident quantum field and the geometry-driven intrinsic thermal emission arising from localized dissipation. The first term, representing the scattered illumination, generally retains an explicit dependence on the radial distance $r$ and observation time $t$ due to the arbitrary spectral shape of the incident field. Conversely, the second term, governed by ${\cal V}_\omega \left( {{\bf{n}},{\bf{n}}'} \right)$, isolates the intrinsic thermal emission and is independent of both $r$ and $t$. This space-time independence is a direct consequence of the temporal stationarity of thermal equilibrium, which enforces the frequency Dirac delta $\delta(\omega-\omega')$. Substituting this total correlation dyadic into Eq.~(\ref{mu2_definition}) yields the generalized degree of coherence $\mu^2_{out}$. While its explicit algebraic expression is omitted for brevity, this metric formally captures how the object's electromagnetic geometry shapes the outgoing spatial correlations purely through classical dyadics.

\paragraph{Pure scattering regime.} The thermal emission contribution to the far-future correlation dyadic in Eq.~(\ref{General_C_out}) vanishes in two specific macroscopic limits: when the object is ideally transparent (${{\rm Im} [ \varepsilon _\omega ] = {\rm Im} [ \mu _\omega ] = 0}$) due to the exact vanishing of the structured emission dyadic ${\cal V}_\omega$, and when the object is a cold absorber ($T_{em} = 0$) due to the strict suppression of the thermal population factor. In both scenarios, the outgoing correlation dyadic ${\cal C}_{out}$ reduces purely to the first integral of Eq.~(\ref{General_C_out}), meaning the far-field spatial coherence is governed solely by the elastic scattering of the initial quantum illumination. Although both regimes isolate this pure scattering contribution, their underlying physical mechanisms shaping the far-field correlations are distinct. In the transparent case, the lack of absorption renders the transmission operator $\mathsf{T}_{\omega ss}$ unitary, as dictated by Eq.~(\ref{Unitarity_Relation_Main}) when the emission channels vanish. Consequently, from Eq.(\ref{Out_Spec_Corr}), the scattering process conserves the total spectral optical power formally expressed by the macroscopic angular trace 
\begin{equation} \label{Transp_Conserv}
\int do_{\bf m} \, {\rm tr} \left[ \langle {\bf{\hat G}}_{\omega s}^\dag({\bf m}) {\bf{\hat G}}_{\omega s}({\bf m}) \rangle \right] = \int do_{\bf p} \, {\rm tr} \left[ \langle {\bf{\hat g}}_{\omega s}^\dag({\bf p}) {\bf{\hat g}}_{\omega s}({\bf p}) \rangle \right]
\end{equation}
and the resulting far-field correlation dyadic is governed solely by the lossless angular redistribution of the incident field. Conversely, for a cold dissipative scatterer, irreversible photon absorption renders the transmission operator $\mathsf{T}_{\omega ss}$ non-unitary, breaking the conservation of spectral optical power. The object acts as a passive macroscopic sink. Consequently, the outgoing spatial correlations are governed not merely by geometric routing, but by the mode-dependent differential attenuation applied by the non-unitary transmission channels to the incoming quantum fluctuations. This demonstrates that material losses actively reshape spatial coherence through deterministic filtering, altering the correlation profile even without intrinsic thermal emission. Its ultimate impact on the normalized coherence degree $\mu^2_{out}$ depends strictly on the overlap between the object's attenuation profile and the statistical properties of the incident quantum state.

\paragraph{Pure thermal emission regime.} In the absence of external quantum illumination, the incident field resides in the vacuum state $|0_s\rangle$. Because the normally ordered expectation value for the vacuum state vanishes ($\langle {\bf{\hat g}}_{\omega s}^\dag({\bf p}) {\bf{\hat g}}_{\omega' s}({\bf p}') \rangle = 0$), the elastic scattering contribution in Eq.~(\ref{General_C_out}) is suppressed. Consequently, the outgoing correlation dyadic reduces purely to the stationary thermal emission term:
\begin{equation} \label{Emission_C_out}
{\cal C}_{out}({\bf n}, {\bf n}') = \int d\omega \, \omega n_\omega(T_{em}) {\cal V}_\omega({\bf n}, {\bf n}').
\end{equation}
Equation~(\ref{Emission_C_out}) explicitly isolates how the dyadic ${\cal V}_\omega({\bf n}, {\bf n}')$ generates macroscopic spatial correlations from stochastically uncorrelated internal thermal fluctuations. While this emergence of far-field coherence from an incoherent source is phenomenologically described by the classical van Cittert-Zernike theorem \cite{Mandel1995} and its three-dimensional generalizations \cite{Carter1981}, our formulation provides its exact quantum-vectorial derivation. It establishes the spatial coherence profile of an arbitrary macroscopic emitter directly from the first principles of canonical quantization, entirely bypassing the need for phenomenological noise currents. Furthermore, evaluating the generalized degree of spatial coherence $\mu^2_{out}$ via Eq.~(\ref{mu2_definition}) establishes a formal upper bound. As detailed in Appendix B, the spatial delta-correlation of the internal thermal fluctuations [Eq.~(\ref{Thermal_Averages})] reduces the outgoing correlation dyadic to a single volume integral and a single polaritonic summation. This specific algebraic structure allows its Frobenius norm to be expressed in terms of a functional inner product, where the Cauchy-Schwarz inequality demonstrates that the limit $\mu^2_{out} = 1$ would require a strict linear dependence between the fields radiated into distinct observation directions. However, because these fields are integrated over strictly disjoint solid angles and accumulate distinct propagation phases, such linear dependence is precluded. Consequently, the spatial coherence remains strictly bounded below unity ($\mu^2_{out} < 1$), providing a first-principles derivation of the standard thermodynamic result that intrinsic thermal emission from localized dissipation inherently yields a partially coherent field.

\section{Thermal illumination}
\paragraph{Core setup.}  Serving as a primary application of our general framework, we explore the physical interplay between elastic scattering and intrinsic emission under thermodynamic non-equilibrium. Specifically, we consider a dissipative object illuminated by an isotropic blackbody bath at a temperature $T_s$, generally distinct from its internal temperature $T_{em}$. In this regime, the density operator of the ingoing $s$-polaritons is $\hat{\rho}_s = e^{-\hat{H}_s / k_B T_s} / {\rm Tr} ( e^{-\hat{H}_s / k_B T_s} )$, representing the canonical distribution over the scattering sector of the Fock space governed by the Hamiltonian $\hat{H}_s$ (see Eq.(\ref{Hs_Hem_Main})). Accordingly, the ingoing spectral correlation dyadic assumes the diagonal form
\begin{equation} \label{Ingoing_Thermal}
\langle {\bf{\hat g}}_{\omega s}^\dag({\bf m}) {\bf{\hat g}}_{\omega' s}({\bf m}') \rangle = \delta(\omega-\omega') \delta(o_{\bf m} - o_{{\bf m}'}) n_\omega(T_s) {\cal I}_{\bf m},
\end{equation}
where $n_\omega(T_s) = (e^{\hbar \omega / k_B T_s} - 1)^{-1}$ is the Bose-Einstein distribution. Equation (\ref{Ingoing_Thermal}) captures the second-order coherence properties of the ingoing thermal radiation, where the frequency and angular Dirac deltas, $\delta(\omega-\omega')$ and $\delta(o_{\bf m} - o_{{\bf m}'})$, reflect its stationary nature and spatial incoherence, while the transverse identity tensor ${\cal I}_{\bf m}$ signifies that the field is unpolarized. By substituting the ingoing spectral correlation dyadic of Eq.(\ref{Ingoing_Thermal}) into the first of Eqs.(\ref{C_dyads}), and exploiting both the aforementioned direction discretization and the smallness of the detection solid angles to evaluate the angular integrals as 
\begin{equation} \label{Integral}
\int_{\Delta \Omega _{\bf{n}} } {do_{\bf{m}} } \int_{\Delta \Omega _{{\bf{n}}'} } {do_{{\bf{m}}'} } \delta \left( {o_{\bf{m}}  - o_{{\bf{m}}'} } \right){\cal I}_{\bf{m}}  \simeq  \Delta \Omega  \delta _{{\bf{n}},{\bf{n}}'} {\cal I}_{\bf{n}}
\end{equation}
(where $\delta_{{\bf n},{\bf n}'}$ is the Kronecker delta), the ingoing correlation dyadic evaluates to
\begin{equation} \label{therm_C_in}
{\cal C}_{in} \left( {{\bf{n}},{\bf{n}}'} \right) = \left[\int {d\omega } \omega n_\omega \left( {T_s } \right) \right]\Delta \Omega \delta _{{\bf{n}},{\bf{n}}'} {\cal I}_{\bf{n}}.
\end{equation}
Besides, by substituting $  {\cal C}_{in}$ into Eq.(\ref{mu2_definition}), a straightforward calculation yields
\begin{equation}
\mu_{in}^2({\bf n}, {\bf n}') = \frac{1}{2} \delta_{{\bf n},{\bf n}'}.
\end{equation}
As expected for a stationary process, the resulting coherence degree is time-independent. Furthermore, it exhibits no explicit dependence on the source temperature, as the thermal intensity factors intrinsically cancel out in the normalized definition of $\mu^2$. Physically, the coherence degree vanishes for ${\bf n} \neq {\bf n}'$, confirming that distinct observation directions are uncorrelated, whereas for coinciding directions (${\bf n} = {\bf n}'$), the value of $1/2$ is a direct consequence of the unpolarized nature of the thermal field \cite{Tervo2003}. To evaluate the total outgoing correlation dyadic under thermal illumination, we directly substitute the ingoing thermal field correlations of Eq.~(\ref{Ingoing_Thermal}) into the general expression derived in Eq.~(\ref{General_C_out}). Owing to the stationarity of the incident thermal state, the temporal phase factor vanishes, and the scattering contribution reduces to an integral governed by the product of the transmission dyadics, specifically $\int do_{\bf p} \, {\cal T}_{\omega ss}^* \left( {\left. {\bf{m}} \right|{\bf{p}}} \right) \cdot {\cal T}_{\omega ss}^T \left( {\left. {{\bf{m}}'} \right|{\bf{p}}} \right)$. By leveraging the global unitarity relation established in Eq.~(\ref{Unitarity_Relation_Main}), this transmission product can be expressed in terms of the emission dyadics. Evaluating the remaining integrations over the finite detection solid angles already present in the first term of Eq.~(\ref{General_C_out}), and combining the result with the intrinsic thermal emission term, the far-field outgoing correlation dyadic simplifies to
\begin{equation} \label{mu2_C_out_therm}
{\cal C}_{out} \left( {{\bf{n}},{\bf{n}}'} \right) = \int d\omega \, \omega \left\{ n_\omega(T_s) \Delta \Omega \delta_{{\bf{n}},{\bf{n}}'} {\cal I}_{\bf{n}}  + \left[ n_\omega(T_{em}) - n_\omega(T_s) \right] {\cal V}_\omega \left( {{\bf{n}},{\bf{n}}'} \right) \right\},
\end{equation}
where ${\cal V}_\omega \left( {{\bf{n}},{\bf{n}}'} \right)$ is the structured emission dyadic previously defined in Eq.~(\ref{DyadV}). As formalized in Eq.~(\ref{mu2_C_out_therm}), the dyadic ${\cal C}_{out} \left( {{\bf{n}},{\bf{n}}'} \right)$ naturally splits into an incoherent isotropic thermal background, spectrally weighted by $n_\omega(T_s)$ and governed by the diagonal dyadic $\Delta \Omega \delta _{{\bf{n}},{\bf{n}}'} {\cal I}_{\bf{n}}$, and a structured term, spectrally weighted by the thermal contrast $\left[ n_\omega(T_{em}) - n_\omega(T_s) \right]$ and governed by the dyadic ${\cal V}_\omega \left( {{\bf{n}},{\bf{n}}'} \right)$. Crucially, while the diagonal background remains completely uncorrelated for distinct observation directions (${\bf n} \neq {\bf n}'$), this structured dyadic ${\cal V}_\omega$ acts as the primary mechanism generating spatial coherence in the far field. 

To physically quantify this effect, we shift our analysis to the generalized degree of spatial coherence $\mu_{out}^2({\bf n}, {\bf n}')$ obtained by substituting Eq.~(\ref{mu2_C_out_therm}) into Eq.~(\ref{mu2_definition}), whose explicit general formula for an arbitrary object is mathematically cumbersome and therefore omitted here. In the numerator ${\rm Tr} \left[ {{\cal C}_{out} \left( {{\bf{n}},{\bf{n}}'} \right) \cdot {\cal C}_{out}^{T*}\left( {{\bf{n}},{\bf{n}}'} \right)} \right]$, which evaluates the cross-correlation between distinct directions (${\bf n} \neq {\bf n}'$), the isotropic background vanishes ($\delta_{{\bf n},{\bf n}'} = 0$), thereby leaving the emergent spatial coherence to be determined solely by the spatially structured profile. Conversely, the denominator of $\mu_{out}^2$ normalizes this correlation using the product of the total local intensities ${\rm Tr} \left[ {{\cal C}_{out} \left( {{\bf{n}},{\bf{n}}} \right)} \right]$ and ${\rm Tr} \left[ {{\cal C}_{out} \left( {{\bf{n}}',{\bf{n}}'} \right)} \right]$, where the non-vanishing Kronecker delta ($\delta_{{\bf n},{\bf n}} = 1$) forces these local evaluations to incorporate the uncorrelated isotropic thermal background proportional to $\int d\omega \, \omega n_\omega \left( {T_s} \right)\Delta \Omega {\cal I}_{\bf{n}}$. Consequently, by increasing the total detected intensity at the denominator without contributing to the structured cross-correlation at the numerator, this chaotic background acts as a strong diluting noise that suppresses the observable magnitude of $\mu_{out}^2$. This demonstrates that the measurable far-field spatial coherence is governed by a competition between the spatial correlations generated by the thermal contrast and the incoherent isotropic background resulting from the incident thermal illumination.

To contextualize the unifying capability of our framework, we examine four fundamental thermodynamic limits. This analysis not only recovers established macroscopic principles from quantum first principles, but also unveils fundamental physical symmetries—such as the exact equivalence between the structured coherence of an active thermal emitter and that of a passive thermal shadow. Crucially, it demonstrates how a single cohesive formalism captures the simultaneous interplay between elastic light scattering and thermal emission across all these diverse regimes. We finally illustrate these general results through the model of a subwavelength spherical scatterer, which admits a fully analytical treatment of the spatial coherence degree.

\begin{figure}[!t]
\centering
\begin{tikzpicture}[
    % Stili generali
    photon/.style={decorate, decoration={snake, post length=1.5mm, amplitude=1mm, segment length=3mm}, ->, >=Stealth, red!80, thick},
    wavefront/.style={cyan!80!blue, thick},
    correlation/.style={<->, >=Stealth, dashed, thick, purple},
    vector/.style={->, >=Stealth, thick, black}
]

% Definiamo i limiti del riquadro: più spazio in basso per contenere le onde
\def\fw{3.6}      % Mezza-larghezza 
\def\fhtop{2.5}   % Tetto (leggermente alzato)
\def\fhbot{-3.0}  % Pavimento (notevolmente abbassato per non tagliare le onde)

% ================= PANEL (a): Ideally Transparent =================
\begin{scope}[shift={(0,0)}]
    \draw[thin, black] (-\fw, \fhbot) rectangle (\fw, \fhtop);
    
    \node[anchor=north west, inner sep=6pt] at (-\fw, \fhtop) {\textbf{(a)}};
    \node[anchor=north, inner sep=6pt] at (0, \fhtop) {Transparent (${\rm Im}(\varepsilon)=0$)};
    
    % SCOPE INTERNO: Abbasso tutto di 0.6cm per proteggere i titoli
    \begin{scope}[yshift=-0.6cm]
        % Profilo asimmetrico arbitrario al posto del cerchio
        \fill[blue!10, draw=blue!50, thick] plot [smooth cycle, tension=0.7] coordinates { (0:0.8) (50:0.6) (100:0.85) (160:0.7) (210:0.8) (260:0.6) (310:0.75) };
        
        % Illuminazione Termica Isotropa (accorciata a r=2.3 per stare nei bordi)
        \foreach \ang in {120, 160, 200, 240, 280, 320} {
            \draw[photon] (\ang:2.3) -- (\ang:0.9);
        }
        
        % Vettori (accorciati a r=2.3)
        \draw[vector] (0,0) -- (55:2.3) node[above right, inner sep=2pt] {${\bf n}$};
        \draw[vector] (0,0) -- (10:2.3) node[right, inner sep=3pt] {${\bf n}'$};
        
        \node[purple] at (32.5:1.9) {$\mu^2 = 0$};
    \end{scope}
\end{scope}

% ================= PANEL (b): Global Equilibrium =================
\begin{scope}[shift={(7.5,0)}] 
    \draw[thin, black] (-\fw, \fhbot) rectangle (\fw, \fhtop);
    
    \node[anchor=north west, inner sep=6pt] at (-\fw, \fhtop) {\textbf{(b)}};
    \node[anchor=north, inner sep=6pt] at (0, \fhtop) {Equilibrium ($T_s = T_{em}$)};
    
    \begin{scope}[yshift=-0.6cm]
        % Profilo asimmetrico arbitrario
        \fill[red!30, draw=red!80, thick] plot [smooth cycle, tension=0.7] coordinates { (0:0.8) (50:0.6) (100:0.85) (160:0.7) (210:0.8) (260:0.6) (310:0.75) };
        
        \foreach \ang in {120, 160, 200, 240, 280, 320} {
            \draw[photon] (\ang:2.3) -- (\ang:0.9);
        }
        
        \draw[vector] (0,0) -- (55:2.3) node[above right, inner sep=2pt] {${\bf n}$};
        \draw[vector] (0,0) -- (10:2.3) node[right, inner sep=3pt] {${\bf n}'$};
        
        \node[purple] at (32.5:1.9) {$\mu^2 = 0$};
    \end{scope}
\end{scope}

% ================= PANEL (c): Cold Vacuum (Primary Emitter) =================
% Spaziatura verticale aumentata (da -5.3 a -5.8) per via dei box più alti
\begin{scope}[shift={(0,-5.8)}] 
    \draw[thin, black] (-\fw, \fhbot) rectangle (\fw, \fhtop);
    
    \node[anchor=north west, inner sep=6pt] at (-\fw, \fhtop) {\textbf{(c)}};
    \node[anchor=north, inner sep=6pt] at (0, \fhtop) {Cold Vacuum ($T_s = 0$)};
    
    \begin{scope}[yshift=-0.6cm]
        % Profilo asimmetrico arbitrario
        \fill[orange!70!yellow, draw=red, thick] plot [smooth cycle, tension=0.7] coordinates { (0:0.8) (50:0.6) (100:0.85) (160:0.7) (210:0.8) (260:0.6) (310:0.75) };
        
        % Fronti d'onda
        \draw[wavefront] (0,0) +(10:1.2) arc (10:55:1.2);
        \draw[wavefront] (0,0) +(10:1.6) arc (10:55:1.6);
        \draw[wavefront] (0,0) +(10:1.9) arc (10:55:1.9);
        
        \draw[vector] (0,0) -- (55:2.3) node[above right, inner sep=2pt] {${\bf n}$};
        \draw[vector] (0,0) -- (10:2.3) node[right, inner sep=3pt] {${\bf n}'$};
        
        \draw[correlation] (10:2.1) arc (10:55:2.1) node[midway, right=2mm, purple] {$\mu^2 \neq 0$};
    \end{scope}
\end{scope}

% ================= PANEL (d): Cold Scatterer (Passive Sink) =================
\begin{scope}[shift={(7.5,-5.8)}]
    \draw[thin, black] (-\fw, \fhbot) rectangle (\fw, \fhtop);
    
    \node[anchor=north west, inner sep=6pt] at (-\fw, \fhtop) {\textbf{(d)}};
    \node[anchor=north, inner sep=6pt] at (0, \fhtop) {Cold Scatterer ($T_{em} = 0$)};
    
    \begin{scope}[yshift=-0.6cm]
        % Profilo asimmetrico arbitrario
        \fill[black!80, draw=black, thick] plot [smooth cycle, tension=0.7] coordinates { (0:0.8) (50:0.6) (100:0.85) (160:0.7) (210:0.8) (260:0.6) (310:0.75) };
        
        \foreach \ang in {120, 160, 200, 240, 280, 320} {
            \draw[photon] (\ang:2.3) -- (\ang:0.9);
        }
        
        \draw[wavefront] (0,0) +(10:1.2) arc (10:55:1.2);
        \draw[wavefront] (0,0) +(10:1.6) arc (10:55:1.6);
        \draw[wavefront] (0,0) +(10:1.9) arc (10:55:1.9);
        
        \draw[vector] (0,0) -- (55:2.3) node[above right, inner sep=2pt] {${\bf n}$};
        \draw[vector] (0,0) -- (10:2.3) node[right, inner sep=3pt] {${\bf n}'$};
        
        \draw[correlation] (10:2.1) arc (10:55:2.1) node[midway, right=2mm, purple] {$\mu^2 \neq 0$};
    \end{scope}
\end{scope}

\end{tikzpicture}
\caption{Schematic illustration of far-field spatial coherence generation under fundamental thermodynamic limits. Wavy lines represent incoherent thermal illumination converging from the surrounding isotropic bath, while linked phase-fronts denote emergent transverse spatial correlations between distinct far-field observation directions (${\bf n}$ and ${\bf n}'$). \textbf{(a)} Ideally transparent limit: A lossless scatterer merely redirects incoming thermal photons, preserving the completely uncorrelated nature of the radiation ($\mu^2 = 0$ for distinct directions). \textbf{(b)} Thermal cloaking: At global thermodynamic equilibrium ($T_s = T_{em}$), deterministic scattering and intrinsic emission perfectly balance, rendering the object indistinguishable from the uncorrelated background. \textbf{(c)} Cold vacuum limit ($T_s=0$): In an empty background, the localized thermal fluctuations of the hot dissipative object generate an active outward heat flux, imprinting structured spatial correlations in the far field (van Cittert-Zernike theorem). \textbf{(d)} Cold scatterer limit ($T_{em}=0$): A totally passive absorbing sink placed in a hot bath drives an inward heat flux.By irreversibly filtering incident photons, it acts as an effective negative emitter, casting a "thermal shadow" into the background and generating a spatial coherence profile structurally identical to that of an active primary emitter.}
\label{fig:thermodynamic_limits}
\end{figure}
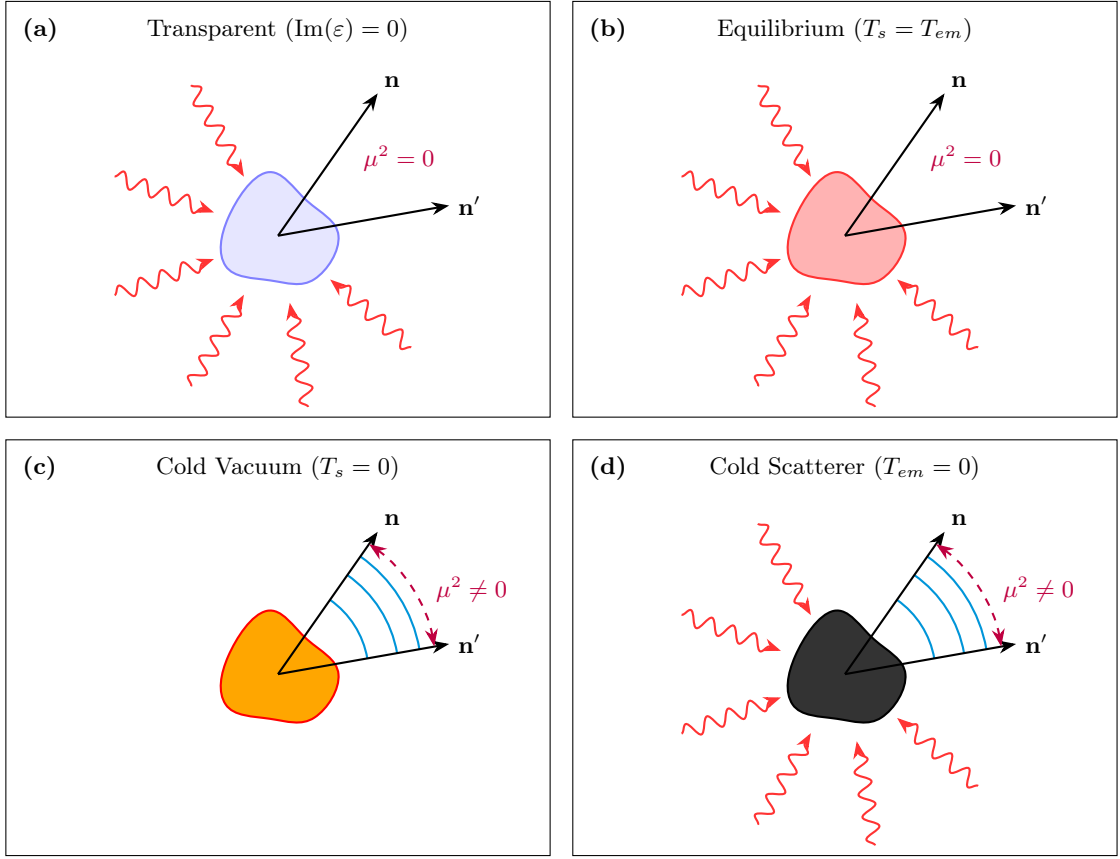

\paragraph{Ideally transparent limit ($\text{Im}(\varepsilon_\omega) = \text{Im}(\mu_\omega) = 0$).} 
As established in our general analysis of the pure scattering regime, the absence of material losses suppresses thermal emission and forces the structured dyadic to vanish (${\cal V}_\omega \left( {{\bf{n}},{\bf{n}}'} \right) = 0$). Substituting this directly into Eq.~(\ref{mu2_C_out_therm}), the outgoing correlation dyadic simplifies entirely to the incident background term, matching its ingoing counterpart ${\cal C}_{in}\left( {{\bf{n}},{\bf{n}}'} \right)$ from Eq.~(\ref{therm_C_in}). Consequently, the outgoing degree of spatial coherence remains purely diagonal, $\mu_{out}^2({\bf n}, {\bf n}') = \frac{1}{2}\delta_{{\bf n},{\bf n}'}$ (see Fig.~\ref{fig:thermodynamic_limits}(a)). While it is a known macroscopic expectation that the purely elastic scattering of a chaotic bath preserves its uncorrelated nature, our exact formulation analytically proves it directly from the first principles of the MLNF. It establishes that a lossless object cannot generate transverse spatial coherence, confirming that the emergence of spatial correlations requires a symmetry breaking induced by material absorption.

\paragraph{Global thermodynamic equilibrium ($T_s = T_{em} \equiv T$).} 
In this case, the vanishing of the thermal contrast in Eq.~(\ref{mu2_C_out_therm}) exactly cancels the structured term governed by ${\cal V}_\omega$. The outgoing correlation dyadic reduces to $  {\cal C}_{out} \left( {{\bf{n}},{\bf{n}}'} \right)  = \left[ \int d\omega \, \omega n_\omega  \left( T \right) \right] \Delta \Omega \delta _{{\bf{n}},{\bf{n}}'} {\cal I}_{\bf{n}}$, which identically matches the ingoing correlation dyadic. Consequently, the far-field spatial coherence retains its uncorrelated profile, $\mu_{out}^2({\bf n}, {\bf n}') = \frac{1}{2} \delta_{{\bf n},{\bf n}'}$ (see Fig.~\ref{fig:thermodynamic_limits}(b)). Physically, the deterministic elastic scattering of the incident chaotic bath and the intrinsic thermal emission perfectly balance, reconstructing the uniform spatial profile of the background and rendering the object fully undetectable via spatial coherence measurements. While this thermal cloaking is a known phenomenological requirement of macroscopic thermodynamics \cite{Planck1914}, our formulation provides a rigorous algebraic proof directly from the canonical MLNF. It demonstrates that this balance holds exactly for any arbitrary lossy magnetodielectric object, guaranteed by the unitarity of the underlying quantum dynamics.

\paragraph{Cold vacuum limit ($T_s = 0$).} In this regime, the incident thermal bath is devoid of photons ($n_\omega(T_s) = 0$), implying that the ingoing scattering polaritons reside in the vacuum state $|0_s\rangle$ (see Fig.~1). Consequently, this configuration physically coincides with the pure thermal emission scenario detailed in the previous section. Evaluating Eq.~(\ref{mu2_C_out_therm}) under this limit collapses the outgoing correlation dyadic to the intrinsic emission term, ${\cal C}_{out}({\bf n}, {\bf n}') = \int d\omega \, \omega n_\omega(T_{em}) {\cal V}_\omega({\bf n}, {\bf n}')$. The body acts purely as a primary thermal emitter into free space, generating the same bounded macroscopic spatial coherence profile ($\mu^2_{out} < 1$) established previously (see Fig.~\ref{fig:thermodynamic_limits}(c)).

\paragraph{Cold scatterer limit ($T_{em} = 0$).} 
In the absence of intrinsic thermal emission, this scenario constitutes a pure scattering regime. The thermal contrast reduces to $-n_\omega(T_s)$, yielding the correlation dyadic $ {\cal C}_{out} \left( {{\bf{n}},{\bf{n}}'} \right) = \int d\omega \, \omega n_\omega \left( {T_s } \right) \left[ \Delta \Omega \delta _{{\bf{n}},{\bf{n}}'} {\cal I}_{\bf{n}} - {\cal V}_\omega \left( {{\bf{n}},{\bf{n}}'} \right) \right]$. This expression reveals the emergence of macroscopic spatial coherence exclusively governed by the negative structured term $-{\cal V}_\omega \left( {{\bf{n}},{\bf{n}}'} \right)$. The geometric profile of this spatial correlation is dictated by the exact same dyadic responsible for intrinsic thermal emission. This exact structural symmetry emerges strictly from the operator-level enforcement of Kirchhoff's law within the MLNF. Physically, the irreversible absorption of incident chaotic photons breaks the spatial symmetry of the thermal bath. The object effectively acts as a localized passive sink, casting a spatially structured radiation deficit—a macroscopic "thermal shadow"—into the far-field uncorrelated background. While phenomenological scattering variants of the van Cittert-Zernike theorem have been previously explored classically \cite{Bykovskii1985,Zarubin1993}, our exact framework formally establishes such phenomena directly from quantum first principles. It demonstrates that a passive cold sink filtering a chaotic bath generates a transverse spatial coherence profile structurally identical to that of an active primary emitter in vacuum, strictly linking absorption and emission phenomena at the quantum correlation level (see Fig.~\ref{fig:thermodynamic_limits}(d)).

\paragraph{Subwavelength spherical scatterer.}
To physically illustrate the general framework under thermal illumination, we explicitly evaluate the far-field spatial coherence for a homogeneous, isotropic, and nonmagnetic sphere of radius $a$ in the limit $a \ll \lambda$ (Rayleigh regime). As discussed in Appendix C, this quasi-static approximation is justified provided the particle size remains much smaller than the shortest characteristic thermal wavelength of the system, a condition dictated by Wien's displacement law as $a \ll b / [2\pi \max(T_s, T_{em})]$, where $b$ is Wien's displacement constant. Furthermore, by utilizing the exact dressed polarizability and enforcing the macroscopic optical theorem, the generalized degree of spatial coherence evaluates to
\begin{equation} \label{mu2_Rayleigh}
\mu_{out}^2 \left( {{\bf{n}},{\bf{n}}'} \right) = \frac{1}{2}\delta _{{\bf{n}},{\bf{n}}'} + \left( {\frac{\Gamma }{{1 + \Gamma }}} \right)^2 \frac{{1 + \left( {{\bf{n}} \cdot {\bf{n}}'} \right)^2 }}{4}\left( {1 - \delta _{{\bf{n}},{\bf{n}}'} } \right),
\end{equation}
where the dimensionless parameter $\Gamma$, which acts as an effective thermodynamic signal-to-noise ratio driving the magnitude of the far-field spatial correlations, is defined as
\begin{equation} \label{Gamma_Rayleigh}
\Gamma  = \Delta \Omega \frac{{\int d\omega \omega ^3 \left[ {n_\omega  \left( {T_{em} } \right) - n_\omega  \left( {T_s } \right)} \right] \sigma _\omega ^{\left( {abs} \right)} }}{{\left( {2\pi c} \right)^2 \int d\omega \omega n_\omega  \left( {T_s } \right)}},
\end{equation}
with $\sigma _\omega ^{\left( {abs} \right)}$ being the absorption cross-section of the sphere in the Rayleigh regime. 

Equations~(\ref{mu2_Rayleigh}) and (\ref{Gamma_Rayleigh}) analytically encapsulate the four fundamental limiting scenarios discussed previously. First, in the ideally transparent limit ($\text{Im}(\varepsilon_\omega) = \text{Im}(\mu_\omega) = 0$), the optical theorem enforces $\sigma _\omega ^{\left( {abs} \right)} = 0$ (see Appendix~C). This yields $\Gamma=0$, immediately reducing the spatial coherence to the purely diagonal background $\mu_{out}^2 = \frac{1}{2}\delta _{{\bf{n}},{\bf{n}}'}$. Second, at global thermodynamic equilibrium ($T_s = T_{em}$), the spectral contrast vanishes, again yielding $\Gamma = 0$ and recovering the diagonal thermal cloaking signature $\mu_{out}^2 = \frac{1}{2} \delta_{{\bf n},{\bf n}'}$. Third, in the cold vacuum limit ($T_s = 0$), the absence of incident background intensity ($n_\omega(T_s)=0$) drives $|\Gamma| \to \infty$, saturating the modulating pre-factor to unity ($\left[\Gamma/(1+\Gamma)\right]^2 \to 1$). Consequently, for distinct observation directions (${\bf n} \neq {\bf n}'$), $\mu_{out}^2$ collapses exactly onto the transverse dipole profile $[1 + ({\bf n} \cdot {\bf n}')^2]/4$. In this regime, the explicit cancellation of both $T_{em}$ and $\Delta \Omega$ algebraically isolates the source's intrinsic geometric correlation without any environmental dilution. Fourth, in the cold scatterer limit ($T_{em} = 0$), the strictly negative thermal contrast yields $\Gamma < 0$. Since Eq.~(\ref{mu2_Rayleigh}) scales with the square $[\Gamma/(1+\Gamma)]^2$, this negative sign is absorbed. This mathematically proves that a net inward energy flux ($\Gamma < 0$) reconstructs the transverse dipole profile of an active primary emitter, although dynamically modulated in magnitude by the overall pre-factor.
 
Whenever spatial correlations emerge ($\Gamma \neq 0$), their structure is uniquely dictated by the far-field isotropic dipole profile $[1 + ({\bf n} \cdot {\bf n}')^2]/4$. Crucially, this exact analytical limit reveals a fundamental limitation of purely scalar models for incoherent subwavelength sources, which standardly predict a uniform spatial coherence across all observation directions \cite{Mandel1995, Carter1981}. By fully accounting for the vectorial nature of the field, our treatment uncovers an irreducible geometric signature. The transverse character of electromagnetic fields inherently breaks this scalar uniformity, establishing a non-zero correlation baseline of $1/4$ even for strictly orthogonal observation directions (${\bf n} \cdot {\bf n}' = 0$). Because evaluating the generalized degree of coherence for unpolarized light traces over the local transverse plane of each detector, strictly orthogonal observation directions still possess detection planes that geometrically intersect along a common polarization axis. Consequently, source fluctuations along this shared axis contribute simultaneously to both detectors, generating an intrinsic spatial correlation that purely scalar formalisms---which inherently lack directional filtering---structurally overlook. 

Furthermore, because the parameter $\Gamma$ embeds both the particle's scattering efficiency and its thermodynamic contrast with the bath, the pre-factor $[\Gamma/(1+\Gamma)]^2$ continuously bridges the formalisms of thermal emission and elastic scattering. For weak thermodynamic imbalances ($|\Gamma| \ll 1$), the structured correlations scale quadratically ($\approx \Gamma^2$), demonstrating strong dilution by the dominant incoherent noise of the isotropic bath. Conversely, under strong thermodynamic driving ($|\Gamma| \gg 1$), the pre-factor asymptotically saturates to unity. In this limit, the thermal noise background is overcome, allowing the measurable far-field spatial coherence to attain its maximum theoretical visibility and isolating the intrinsic geometric signature of the subwavelength source.

\section{Coherent illumination}
\paragraph{Core setup.} Transitioning from a completely spatially incoherent thermal bath to the regime of perfect incident macroscopic coherence, we examine the lossy object driven by a deterministic coherent field, such as laser light. This scenario constitutes the foundational setup for virtually any modern optical experiment, providing the theoretical platform to assess how intrinsic material dissipation degrades pure phase correlations. In this regime, the scattering sector density operator is $\hat{\rho}_s = \left| \boldsymbol{\alpha} \right\rangle \left\langle \boldsymbol{\alpha} \right|$, where 
\begin{equation}
\left| \boldsymbol{\alpha} \right\rangle  = \exp \left\{ \int d\omega \int do_{\bf{n}} \left[ {\boldsymbol{\alpha}_\omega  \left( {\bf{n}} \right) \cdot {\bf{\hat g}}_{\omega s}^\dag  \left( {\bf{n}} \right) - \boldsymbol{\alpha}_\omega ^* \left( {\bf{n}} \right) \cdot {\bf{\hat g}}_{\omega s} \left( {\bf{n}} \right)} \right] \right\} \left| 0_s \right\rangle
\end{equation}
is a multimode coherent state with the vector spectral amplitude $\boldsymbol{\alpha}_\omega \left( {\bf{n}} \right)$ specifying the spatio-temporal profile of the incident wavepacket, which satisfies the transversality condition $\boldsymbol{\alpha}_\omega \left( {\bf{n}} \right) \cdot {\bf{n}} = 0$. Because the coherent state is an eigenstate of the annihilation operator, i.e., ${\bf{\hat g}}_{\omega s} \left( {\bf{n}} \right)\left| \boldsymbol{\alpha} \right\rangle  = \boldsymbol{\alpha}_\omega  \left( {\bf{n}} \right)\left| \boldsymbol{\alpha} \right\rangle$, the ingoing spectral correlation dyadic assumes the factorized form:
\begin{equation} \label{COH_ing_spe_cor_dya}
\langle {\bf{\hat g}}_{\omega s}^\dag({\bf m}) {\bf{\hat g}}_{\omega' s}({\bf m}') \rangle = \boldsymbol{\alpha}_\omega^*({\bf m}) \boldsymbol{\alpha}_{\omega'}({\bf m}').
\end{equation}
Notably, the absence of the frequency and angular Dirac deltas characteristic of thermal fields formally reflects the deterministic, phase-locked nature of coherent illumination. As a direct consequence of this exact factorization, a calculation proves that $\mu_{in}^2({\bf n}, {\bf n}'; r, t) = 1$ uniformly. This establishes a condition of maximal spatial correlation, providing a physical counterpart to the previously discussed thermal illumination scenario, where the incident field is spatially uncorrelated. 

\begin{figure}[!t]
\centering

\begin{tikzpicture}[
    laser/.style={->, >=Stealth, red, thick},
    wavefront/.style={red!80, thick},
    thermal/.style={decorate, decoration={snake, post length=1.5mm, amplitude=0.8mm, segment length=2.5mm}, ->, >=Stealth, orange, thick},
    scattered/.style={blue!80, thick, dashed},
    vector/.style={->, >=Stealth, thick, black}
]

% ========================================================
% PANNELLO (a): SCHEMA FISICO
% ========================================================
% Riquadro A (Da X=0 a X=15.2, Y=7 a Y=11.5)
\draw[thin, black] (0, 7) rectangle (15.0, 11.5);
\node[anchor=north west, inner sep=6pt] at (0, 11.5) {\textbf{(a)} Coherent and Thermal Fields Interplay};

% Laser incidente (spostato leggermente indietro per far spazio)
\foreach \x in {3.2, 3.7, 4.2} {
    \draw[wavefront] (\x, 8.25) -- (\x, 10.25);
}
\draw[laser] (4.6, 9.25) -- (6.3, 9.25) node[midway, above, black] {$I_{inc}$};

% Scatteratore IRREGOLARE (Centro focale in 7.8, 9.25)
% Usiamo uno smooth cycle per creare una "patata" asimmetrica
\fill[ball color=gray!40, draw=black, thick] 
    plot [smooth cycle, tension=0.7] coordinates { 
        (8.5, 9.25)   % Punta destra
        (8.2, 10.0)   % Alto destra
        (7.5, 10.2)   % Alto centro
        (6.9, 9.7)    % Alto sinistra
        (6.8, 9.0)    % Basso sinistra
        (7.4, 8.5)    % Basso centro
        (8.1, 8.7)    % Basso destra
    };
% Etichetta T_em spostata a sinistra dentro l'oggetto
\node at (7.2, 9.3) {$T_{em}$};

% Emissione Termica (origine dal centro focale 7.8, 9.25)
\foreach \ang in {45, 90, 135, 225, 270, 315} {
    \draw[thermal] (7.8, 9.25) +(\ang:0.7) -- +(\ang:1.6);
}

% Scattering Deterministico Coerente (origine dal centro focale)
\draw[scattered] (7.8, 9.25) +(30:1.1) arc (30:70:1.1);
\draw[scattered] (7.8, 9.25) +(30:1.5) arc (30:70:1.5);
\draw[scattered] (7.8, 9.25) +(30:1.9) arc (30:70:1.9);

\draw[scattered] (7.8, 9.25) +(-30:1.1) arc (-30:-70:1.1);
\draw[scattered] (7.8, 9.25) +(-30:1.5) arc (-30:-70:1.5);
\draw[scattered] (7.8, 9.25) +(-30:1.9) arc (-30:-70:1.9);

% Vettori (origine dal centro focale)
\draw[vector] (7.8, 9.25) -- +(45:2.5) node[above right] {${\bf n}$};
\draw[vector] (7.8, 9.25) -- +(-45:2.5) node[below right] {${\bf n}'$};

% Equazione
\node[right, align=center] at (10.1, 9.25) {
    $\displaystyle \eta \sim \frac{\color{blue}{\mathrm{Coherent\ Signal}}}{\color{orange}{\mathrm{Thermal\ Noise}}}
$ 
};

% ========================================================
% PANNELLO (b): DIAGRAMMA DI FASE
% ========================================================
% Riquadro B (Da X=0 a X=7.2, Y=0 a Y=6.5)
\draw[thin, black] (0, 0) rectangle (7.2, 6.5);
\node[anchor=north west, inner sep=6pt] at (0, 6.5) {\textbf{(b)} Coherence Degradation Threshold};

\begin{axis}[
    at={(4.2cm, 3.3cm)},
    anchor=center,
    width=6.6cm, height=5.6cm,
    xmode=log, ymode=log,
    xmin=1e-3, xmax=1e1,
    ymin=1e-5, ymax=1e1,
    xlabel={$a/\bar{\lambda}$},
    ylabel={$T_{th}/T_{macro}$},
    axis background/.style={fill=white},
    grid=both, grid style={line width=.1pt, draw=gray!20}
]
    % Curva esatta con la statistica di Bose-Einstein
    \addplot[domain=1e-3:1e1, samples=100, thick, black] {1e-3 / ln(1 + 1e-3 * (1 + 1/(x^3)))};
    
    % Testi termici/scattering
    \node[text=red!70!black, anchor=center, align=center] at (axis cs: 6e-3, 1.5e0) {\textbf{Thermal} \\ \textbf{Dominated}};
    \node[text=red!70!black, anchor=center] at (axis cs: 6e-3, 1.5e-1) {$\eta \lesssim 1$};
    
    \node[text=blue!70!black, anchor=center, align=center] at (axis cs: 1.5e0, 5e-4) {\textbf{Scattering} \\ \textbf{Dominated}};
    \node[text=blue!70!black, anchor=center] at (axis cs: 1.5e0, 5e-5) {$\eta \gg 1$};
    
    % Linea divisoria e testi spostati verso l'alto
    \draw[dashed, gray] (axis cs: 0.1, 1e-5) -- (axis cs: 0.1, 1e1);
    \node[text=gray, anchor=north, rotate=90] at (axis cs: 0.04, 1e-1) {\small Rayleigh};
    \node[text=gray, anchor=south, rotate=90] at (axis cs: 0.25, 1e-1) {\small Macro};
\end{axis}

% ========================================================
% PANNELLO (c): LIMITI ANALITICI
% ========================================================
% Riquadro C (Da X=7.8 a X=15.2, Y=0 a Y=6.5)
\draw[thin, black] (7.8, 0) rectangle (15, 6.5);
\node[anchor=north west, inner sep=6pt] at (7.8, 6.5) {\textbf{(c)} Rayleigh Regime ($a \ll \bar{\lambda}$)};

\begin{axis}[
    at={(11.7cm, 3.3cm)}, 
    anchor=center,
    width=6.6cm, height=5.6cm,
    xmode=log, ymode=linear,
    xmin=1e-2, xmax=1e2,
    ymin=0, ymax=1.1,
    xlabel={$|{\bf J}|^2$},
    ylabel={$\mu_{out}^2$},
    legend style={
        at={(1.1,0.05)}, anchor=south east, 
        draw=gray!50, fill=white, fill opacity=0.9, text opacity=1, 
        font=\small, row sep=10pt, 
        nodes={anchor=center, inner ysep=4pt, inner xsep=4pt}, 
        legend cell align=center
    },
    grid=major, grid style={line width=.1pt, draw=gray!20}
]
    % Curva 1: Collineare
    \addplot[domain=1e-2:1e2, samples=100, thick, purple] {(x^2 + x + 0.5) / (x + 1)^2};
    \addlegendentry{\begin{tabular}[c]{@{}c@{}}Collinear \\ (${\bf n} = {\bf n}'$)\end{tabular}}
    
    % Curva 2: Ortogonale
    \addplot[domain=1e-2:1e2, samples=100, thick, teal] {(x^2 + 0.25) / (x + 1)^2};
    \addlegendentry{\begin{tabular}[c]{@{}c@{}}Orthogonal \\ (${\bf n} \cdot {\bf n}' = 0$)\end{tabular}}
    
    % Asintoti
    \addplot[domain=1e-2:1e2, dotted, thick, black] {1};
    \addplot[domain=1e-2:1e2, dashed, teal!60] {0.25};
    \addplot[domain=1e-2:1e2, dashed, purple!60] {0.5};
    
    % Etichette asintoti
    \node[anchor=west, text=black]  at (axis cs: 1.1e-2, 1.05) {$1$};
    \node[anchor=west, text=purple] at (axis cs: 1.1e-2, 0.55) {$1/2$};
    \node[anchor=west, text=teal]   at (axis cs: 1.1e-2, 0.30) {$1/4$};
\end{axis}

\end{tikzpicture}

\caption{\textbf{Degradation of far-field spatial coherence under coherent illumination.} \textbf{(a)} Schematic representation of the physical interplay: a deterministic incident laser field (red plane waves) is elastically scattered by an arbitrary dissipative object (blue dashed spherical phase fronts), simultaneously competing with the stochastic intrinsic thermal emission (orange wavy arrows). \textbf{(b)} Thermodynamic phase diagram illustrating the normalized temperature threshold $T_{th}/T_{macro}$ (from Eq.~(\ref{T_th})) at which spatial coherence breaks down, as a function of the size parameter $a/\bar{\lambda}$ for a fixed incident illumination (constant $I_{inc}$ and $\bar{\omega}$). The normalizing factor $T_{macro}$ represents the asymptotic plateau temperature reached in the macroscopic limit ($a \gg \bar{\lambda}$), where the absorption-to-scattering ratio converges to a size-independent bulk constant. Conversely, in the Rayleigh regime ($a \ll \bar{\lambda}$), the sharp decrease in scattering efficiency causes $T_{th}$ to decrease rapidly, rendering the spatial coherence highly vulnerable to intrinsic thermal noise. \textbf{(c)} Exact analytical spatial coherence degree $\mu_{out}^2$ of a subwavelength sphere as a function of the operational signal-to-noise ratio $|{\bf J}|^2$. In the scattering-dominated limit ($|{\bf J}| \to \infty$), macroscopic coherence is preserved ($\mu^2 \to 1$). In the thermal-dominated limit ($|{\bf J}| \to 0$), the coherence collapses onto the intrinsic van Cittert-Zernike geometric limits: $1/2$ for collinear observation directions (${\bf n} = {\bf n}'$) and $1/4$ for orthogonal directions (${\bf n} \cdot {\bf n}' = 0$).}
\label{fig:coherent_interplay}
\end{figure}
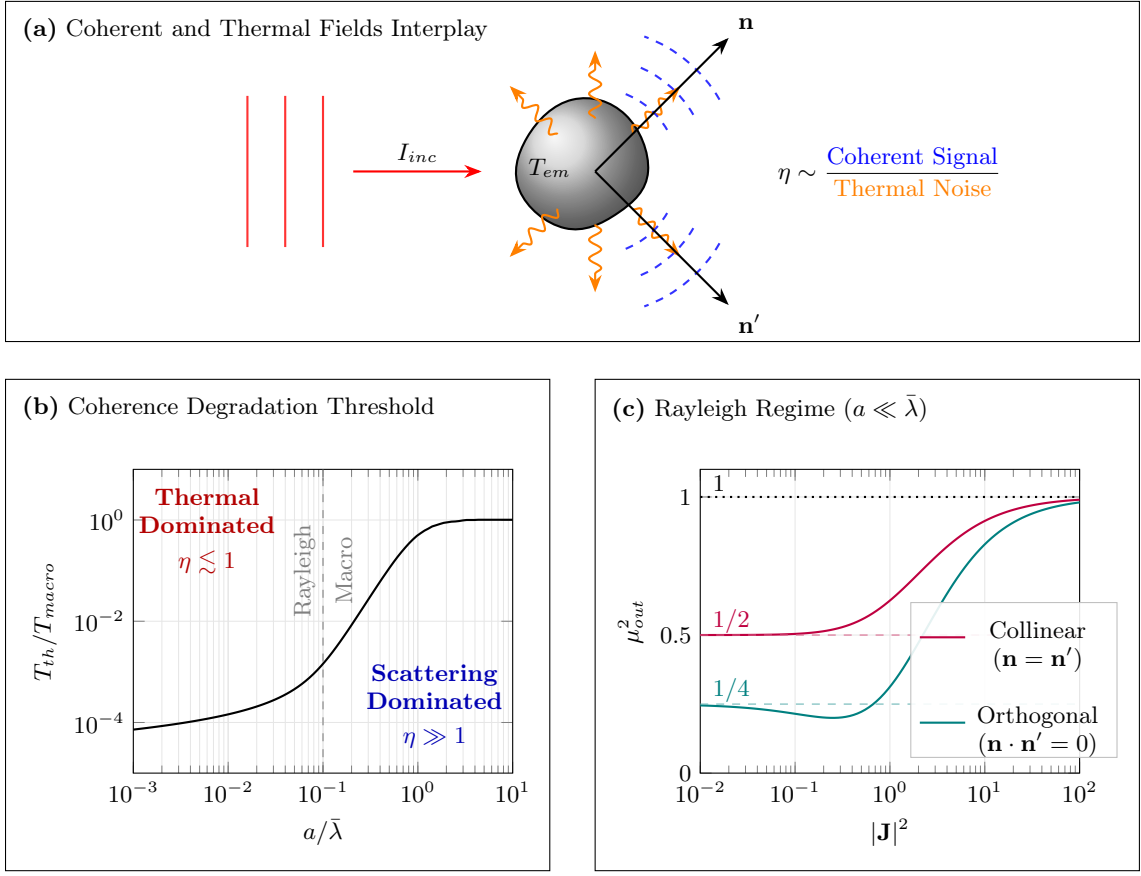

By substituting the factorized expectation value of Eq.~(\ref{COH_ing_spe_cor_dya}) directly into the general formula for the outgoing correlation dyadic in Eq.~(\ref{General_C_out}), the far-field spatial coherence simplifies to:
\begin{equation} \label{COH_Cout_Reduced}
{\cal C}_{out} \left( {{\bf{n}},{\bf{n}}';r,t} \right) = {\bf{A}}^* \left( {{\bf{n}},t - \frac{r}{c}} \right){\bf{A}}\left( {{\bf{n}}',t - \frac{r}{c}} \right) + \int d\omega \, \omega n_\omega  \left( {T_{em} } \right) {\cal V}_\omega  \left( {{\bf{n}},{\bf{n}}'} \right),
\end{equation}
where we have defined the vector field
\begin{equation} \label{A_field_definition}
{\bf A}({\bf n}, \tau) = \int_{\Delta \Omega_{\bf n}} do_{\bf m} \int d\omega \int do_{\bf p} \, \sqrt{\omega} e^{-i\omega \tau} {\cal T}_{\omega ss}({\bf m}|{\bf p}) \cdot \boldsymbol{\alpha}_\omega({\bf p}),
\end{equation}
and ${\cal V}_\omega \left( {{\bf{n}},{\bf{n}}'} \right)$ is the structured dyadic previously defined in Eq.~(\ref{DyadV}). Physically, ${\bf A}({\bf n}, \tau)$ represents the deterministic wavepacket collected over the detector aperture along the observation direction ${\bf n}$, arising solely from the elastic scattering of the incident laser light by the object's electromagnetic geometry. Equation~(\ref{COH_Cout_Reduced}) structurally separates the far-field spatial coherence into two distinct physical contributions: the purely deterministic elastic scattering of the incident laser field and the broadband stochastic thermal emission driven by the object's internal dissipation. This physical interplay is schematically illustrated in Fig.~\ref{fig:coherent_interplay}(a).

However, evaluating the generalized degree of spatial coherence $\mu_{out}^2$ directly from this full frequency-integrated dyadic requires careful operational consideration. In any practical realization, the incident coherent field is highly monochromatic (centered at a given frequency $\bar{\omega}$), whereas the intrinsic thermal emission of the object constitutes an ultra-broadband stochastic background. Consequently, a direct substitution of the general dyadic into the definition of $\mu^2$ would result in the broadband thermal noise dominating the deterministic scattered signal, thereby artificially degrading the observable spatial coherence. To obtain a physically meaningful measure of far-field spatial coherence, it is necessary to introduce a narrow-band spectral filter at the detection stage. By assuming a detector with a finite bandwidth $\Delta \omega$ centered at the incident frequency $\bar{\omega}$, the spectrally filtered outgoing correlation dyadic reduces to:
\begin{equation} \label{COH_Cout_Filtered}
{\cal C}_{out}^{(f)} \left( {{\bf{n}},{\bf{n}}';r,t} \right) = {\bf{A}}^* \left( {{\bf{n}},t - \frac{r}{c}} \right){\bf{A}}\left( {{\bf{n}}',t - \frac{r}{c}} \right) + \Delta \omega \, \bar{\omega} n_{\bar{\omega}}  \left( {T_{em} } \right) {\cal V}_{\bar{\omega}}  \left( {{\bf{n}},{\bf{n}}'} \right).
\end{equation}
The generalized degree of spatial coherence $\mu_{out}^2$, obtained by substituting this filtered correlation dyadic ${\cal C}_{out}^{(f)}$ into Eq.~(\ref{mu2_definition}), is intrinsically suppressed below unity by the thermal emission. Achieving perfect macroscopic spatial coherence ($\mu_{out}^2 = 1$) strictly requires the total correlation dyadic to factorize as a rank-one tensor. While the deterministic elastic scattering of the incident coherent state naturally satisfies this condition via the dyad ${\bf A}^* {\bf A}$, the intrinsic thermal contribution is inherently multi-rank ($\mu_{out}^2 < 1$). Consequently, the superposition of stochastic quantized material fluctuations algebraically breaks the rank-one factorization of the total field. Rather than merely inflating local uncorrelated intensities, thermal emission acts as a non-unitary decoherence channel. It fundamentally bounds the measurable macroscopic spatial coherence away from unity, formalizing the exact algebraic transition from a pure classical-like wavepacket to a partially coherent statistical mixture.

\paragraph{Operational thresholds for spatial coherence degradation.}
As established by Eq.~(\ref{COH_Cout_Filtered}), while thermal emission vanishes for cold ($T_{em} = 0$) or ideally transparent scatterers ($\text{Im}(\varepsilon_{\bar{\omega}}) = \text{Im}(\mu_{\bar{\omega}}) = 0$) to recover the purely classical limit $\mu_{out}^2 = 1$, macroscopic spatial coherence for finite-temperature dissipative bodies degrades due to the superposition of stochastic thermal noise. To estimate the physical threshold at which this thermal emission can no longer be neglected, we rely on the filtered correlation dyadic ${\cal C}_{out}^{(f)}$ to compare the deterministic signal against the thermal background. Within the spectral window $\Delta \omega$, the local coherent signal intensity and the filtered thermal intensity are proportional to $|{\bf{A}}({\bf{n}})|^2$ and $\Delta \omega \, \bar{\omega} n_{\bar{\omega}}(T_{em}) {\rm Tr}[{\cal V}_{\bar{\omega}}({\bf{n}}, {\bf{n}})]$, respectively. Upon angular integration, the coherent signal relates to the incident macroscopic intensity $I_{inc}$ via the scattering cross-section $\sigma_{\bar{\omega}}^{(sca)}$, whereas the absorption cross-section $\sigma_{\bar{\omega}}^{(abs)}$ directly emerges from the emission term integration as a mathematical consequence of Kirchhoff's law formalized in Eq.~(\ref{Kirchhoff_Symmetry}). The spatial coherence breakdown---the regime where thermal fluctuations significantly compromise macroscopic phase correlations---occurs when the operational thermodynamic signal-to-noise ratio falls below unity $\eta \lesssim 1$, where $\eta$ is defined as:
\begin{equation} \label{eta_Thresh}
\eta \sim \left[ \frac{c^2 I_{inc}}{\Delta \omega \hbar \bar{\omega}^3  n_{\bar{\omega}}(T_{em})} \right] \frac{\sigma_{\bar{\omega}}^{(sca)}}{\sigma_{\bar{\omega}}^{(abs)}}.
\end{equation}
By imposing the exact breakdown condition $\eta = 1$ and inverting the Bose-Einstein distribution $n_{\bar{\omega}}(T_{em}) = [\exp(\hbar\bar{\omega}/k_B T_{em}) - 1]^{-1}$, the threshold temperature $T_{th}$ at which spatial coherence is degraded can be analytically derived:
\begin{equation} \label{T_th}
T_{th} = \frac{{\hbar \bar \omega }}{{k_B }}\left\{ {\log \left[ {1 + \frac{{\Delta \omega \hbar \bar \omega ^3 }}{{c^2 I_{inc} }}\frac{{\sigma _{\bar \omega }^{\left( {abs} \right)} }}{{\sigma _{\bar \omega }^{\left( {sca} \right)} }}} \right]} \right\}^{ - 1}.
\end{equation}

This analytical bound establishes a universal thermodynamic phase diagram, illustrated in Fig.~\ref{fig:coherent_interplay}(b). For macroscopic bodies ($a \gg \bar{\lambda}$), optical extinction dictates that both cross sections scale proportionally to the geometric area ($\propto a^2$), forcing the absorption-to-scattering ratio $\sigma_{\bar{\omega}}^{(abs)} / \sigma_{\bar{\omega}}^{(sca)}$ to converge to a size-independent bulk constant. Consequently, the threshold temperature $T_{th}$ plateaus to a constant defined by $T_{macro}$, demonstrating that macroscopic coherence breakdown is driven strictly by the competition between the incident photon flux and the local thermal energy scale. Within this macroscopic limit, three physically relevant examples effectively illustrate the parametric dependencies of Eq.~(\ref{T_th}). First, in the visible and near-infrared, the high frequency $\bar{\omega}$ yields a large prefactor $\hbar\bar{\omega}/k_B$ that drives $T_{th}$ well above standard ambient conditions, thereby preserving robust coherence. Second, in the microwave and terahertz range, the small frequency minimizes this prefactor, drastically suppressing $T_{th}$ so that intrinsic thermal noise inherently bounds phase correlations. Third, under weak coherent illumination, a diminished incident intensity $I_{inc}$ inflates the argument of the logarithm, systematically lowering $T_{th}$ and rendering macroscopic coherence fragile even at moderate temperatures.

Conversely, in the subwavelength Rayleigh regime ($a \ll \bar{\lambda}$), the fundamental cross-sectional scaling alters this dynamic completely. Because elastic scattering is intrinsically inefficient compared to absorption ($\sigma_{\bar{\omega}}^{(sca)} \propto a^6 / \bar{\lambda}^4$ versus $\sigma_{\bar{\omega}}^{(abs)} \propto a^3 / \bar{\lambda}$), the geometric ratio diverges proportionally to $(\bar{\lambda}/a)^3$. As dictated by Eq.~(\ref{T_th}), this divergence systematically inflates the argument of the logarithm, thereby severely lowering the threshold temperature required to preserve macroscopic coherence. Specifically, when the second term in the logarithm's argument remains small, the logarithm linearizes and the threshold temperature follows the inverse of the geometric ratio, yielding a steep cubic suppression $T_{th} \propto (a/\bar{\lambda})^3$. However, as the size $a$ is further reduced, the geometric ratio diverges; the second term becomes dominant, and the denominator of Eq.~(\ref{T_th}) transitions the threshold suppression into a slow asymptotic logarithmic tail, $T_{th} \propto [\log(\bar{\lambda}/a)]^{-1}$. Ultimately, this explicit analytical scaling demonstrates that dissipative nanostructures act as highly efficient thermal antennas, imposing severe and fundamental operational constraints on the preservation of macroscopic spatial coherence in subwavelength optics.

\paragraph{Subwavelength spherical scatterer.}
To provide an analytical counterpart to the general bounds discussed above, we explicitly evaluate the spatial coherence for a dissipative spherical scatterer in the Rayleigh regime ($a \ll \bar{\lambda}$). The object is illuminated by a monochromatic coherent plane wave propagating along $\bar{\bf{p}}$ with transverse complex amplitude $\bar{\bf{A}}$. As detailed in Appendix C, the superposition of the deterministic scattered field and the intrinsic thermal emission within the detection bandwidth yields a closed-form expression for the generalized degree of spatial coherence:
\begin{align} \label{Coherent_Sphere_mu2}
\mu_{out}^2({\bf{n}}, {\bf{n}}') &= \frac{|{\bf{J}}({\bf{n}})|^2 |{\bf{J}}({\bf{n}}')|^2 + {\rm Re}[{\bf{J}}^*({\bf{n}}) \cdot {\bf{J}}({\bf{n}}')] + \frac{1}{4}[1 + ({\bf{n}} \cdot {\bf{n}}')^2]}{(|{\bf{J}}({\bf{n}})|^2 + 1)(|{\bf{J}}({\bf{n}}')|^2 + 1)}.
\end{align}
Here, ${\bf{J}}({\bf{n}})$ is a dimensionless vector encapsulating the angle-resolved coherent signal:
\begin{align} \label{Coherent_Sphere_J}
{\bf{J}}({\bf{n}}) &= \frac{\sqrt{\frac{\bar{\omega}}{2}} \left[ \delta_{{\bf{n}}, \bar{\bf{p}}} {\cal I} + \Delta \Omega \frac{i\bar{\omega}^3}{8\pi^2 c^3} \alpha_{\bar{\omega}} {\cal I}_{\bf{n}} \right] \cdot \bar{\bf{A}}}{\frac{\Delta \Omega}{2\pi c} \sqrt{\Delta \omega \, \bar{\omega}^3 n_{\bar{\omega}}(T_{em}) \sigma_{\bar{\omega}}^{(abs)}}}.
\end{align}
Crucially, the squared norm $|{\bf{J}}({\bf{n}})|^2$ constitutes the exact analytical realization of the operational thermodynamic signal-to-noise ratio $\eta$ introduced previously, quantifying the local competition between the elastically scattered coherent field and the filtered stochastic thermal intensity. 

The functional dependence of $\mu_{out}^2$ on the operational signal-to-noise ratio $|{\bf{J}}|^2$ is illustrated in Fig.~\ref{fig:coherent_interplay}(c), capturing the exact analytical transition between classical determinism and stochastic thermal emission. In the scattering-dominated limit ($|{\bf{J}}| \to \infty$), the spatial coherence saturates to unity ($\mu_{out}^2 \to 1$), recovering perfect classical phase preservation. Conversely, in the thermal-dominated limit ($|{\bf{J}}| \to 0$), the deterministic signal is overwhelmed by stochastic material fluctuations. In this regime, the spatial coherence does not merely vanish; it converges exactly to $\mu_{out}^2 = [1 + ({\bf{n}} \cdot {\bf{n}}')^2]/4$. This analytically recovers the pure thermal emission geometric signature previously established in Eq.~(\ref{mu2_Rayleigh}) for the cold vacuum scenario ($T_s = 0$). This rigorous convergence demonstrates that when macroscopic optical coherence is degraded, the residual far-field spatial correlations are entirely dictated by the intrinsic van Cittert-Zernike profile of the dissipative source. This explicit limit structurally unifies classical coherent scattering and thermal emission within a single, continuous quantum-optical formulation.

\section{Non-classical illumination under spatial entanglement}

\paragraph{Core setup.} To apply the proposed framework to a strictly non-classical illumination regime, we examine the far-field spatial coherence produced by illuminating the object with a continuous-variable entangled biphoton state, thereby providing a platform to systematically assess the interplay between non-local spatial correlations and local thermodynamic dissipation. We consider an incident optical field prepared in a pure two-photon state, typically generated via Spontaneous Parametric Down-Conversion (SPDC) pumped by a laser beam of finite spatial extent. To maintain consistency with the narrow-band detection framework and standard spatial entanglement models \cite{Law2004}, we restrict the analysis to the monochromatic regime at a degenerate frequency $\bar{\omega}$. Accordingly, the density operator in the scattering sector is defined as $\hat{\rho}_s = |\Psi_{2ph}\rangle \langle \Psi_{2ph}|$, where the biphoton state reads:
\begin{equation} \label{Biphoton_State}
\left| {\Psi _{2ph} } \right\rangle  = \frac{1}{{\sqrt 2 }}\int {d\omega } \phi \left( {\omega } \right)\int {d\omega' } \phi \left( {\omega' } \right)\int {do_{{\bf{n}}} } \int {do_{{\bf{n}}'} } {\bf{\hat g}}_{\omega s}^\dag  \left( {{\bf{n}}} \right) \cdot {\cal J}\left( {\left. {\bf{n}} \right|{\bf{n}}'} \right) \cdot {\bf{\hat g}}_{\omega' s}^\dag  \left( {{\bf{n}}'} \right)\left| 0 \right\rangle.
\end{equation}
Here, the complex function $\phi(\omega)$ defines a factorizable, narrow-band spectral envelope sharply peaked at the degenerate carrier frequency $\bar{\omega}$, satisfying $\int {d\omega } |\phi(\omega)|^2 = 1$ to ensure the normalizability of the continuous-variable state within the MLNF Fock space. The dyadic amplitude ${\cal J}\left( {\left. {\bf{n}} \right|{\bf{n}}'} \right)$ dictates the transverse spatial entanglement inherited from the pump \cite{Monken1998, Walborn2010} and satisfies three fundamental constraints: (i) transversality, ${\cal I}_{{\bf{n}} } \cdot {\cal J}\left( {\left. {\bf{n}} \right|{\bf{n}}'} \right) = {\cal J}\left( {\left. {\bf{n}} \right|{\bf{n}}'} \right) \cdot {\cal I}_{{\bf{n}}' } = {\cal J}\left( {\left. {\bf{n}} \right|{\bf{n}}'} \right)$; (ii) bosonic permutation symmetry, ${\cal J}\left( {\left. {\bf{n}} \right|{\bf{n}}'} \right) = {\cal J}^T\left( {\left. {\bf{n}}' \right|{\bf{n}}} \right)$; and (iii) normalization, $\int {do_{{\bf{n}}} } \int {do_{{\bf{n}}'} } {\rm tr}\left[ {{\cal J}^{T*}\left( {\left. {\bf{n}} \right|{\bf{n}}'} \right) \cdot {\cal J} \left( {\left. {\bf{n}} \right|{\bf{n}}'} \right)} \right] = 1$, derived from the state unit norm $\langle \Psi_{2ph} | \Psi_{2ph} \rangle = 1$. Accordingly, $\mathcal{J}$ constitutes the kernel of a Hilbert-Schmidt integral operator, which is consequently compact, acting upon the Hilbert space of vector fields ${\bf v}({\bf n})$ defined on the unit sphere that satisfy both the transversality condition ${\bf n} \cdot {\bf v}({\bf n}) = 0$ and the square-integrability requirement $\int do_{\bf n} \, |{\bf v}({\bf n})|^2 < +\infty$. This guarantees that ${\cal J}$ admits a rigorous continuous-variable Schmidt decomposition \cite{Law2004}: 
\begin{equation} \label{Schmidt_Decomp}
{\cal J}\left( {\left. {\bf n} \right| {\bf n}'} \right) = \sum_{k=1}^\infty \sqrt{\lambda_k} \, {\bf u}_k({\bf n}) {\bf v}_k({\bf n}'),
\end{equation}
where the transverse vector modes ${\bf u}_k$ and ${\bf v}_k$ constitute complete orthonormal bases satisfying $\int {do_{\bf p} } {\bf u}_k^*({\bf p}) \cdot {\bf u}_j({\bf p}) = \delta_{kj}$ (and analogously for ${\bf v}_k$). The real, non-negative Schmidt weights $\lambda_k$ fulfill the normalization constraint $\sum_{k=1}^\infty \lambda_k = 1$ imposed by the aforementioned normalization of the dyadic $\cal{J}$. The degree of spatial entanglement is formally quantified by the Schmidt rank $K$, defined as the total number of strictly positive Schmidt weights, where $K=1$ identifies a strictly separable state and $K>1$ designates an entangled bipartite state.

By utilizing the equivalence of the left and right Schmidt vector bases (${\bf u}_k = {\bf v}_k$), established in Appendix D as a consequence of bosonic permutation symmetry, a straightforward calculation shows that the ingoing spectral correlation dyadic takes the form:
\begin{equation} \label{Biphoton_Input_Dyad}
\langle \hat{\mathbf{g}}_{\omega s}^\dag(\mathbf{m}) \hat{\mathbf{g}}_{\omega' s}(\mathbf{m}') \rangle = 2 \phi^*(\omega) \phi(\omega') \sum_{k=1}^\infty \lambda_k \, \mathbf{u}_k^*(\mathbf{m}) \mathbf{u}_k(\mathbf{m}').
\end{equation}
Unlike the strictly factorized correlation dyadic of the macroscopic coherent state in Eq.~(\ref{COH_ing_spe_cor_dya}), Eq.~(\ref{Biphoton_Input_Dyad}) exhibits a non-factorizable structure for any entangled state ($K>1$), demonstrating that biphoton spatial entanglement physically manifests as a measurable lack of optical spatial coherence. Substituting Eq.~(\ref{Biphoton_Input_Dyad}) into the definition of the ingoing correlation dyadic ${\cal C}_{in}$ in Eqs.~(\ref{C_dyads}) and evaluating the required trace operations yields the explicit analytic expression for the local degree of spatial coherence:
\begin{equation} \label{Local_Mu_In}
\mu _{in}^2 \left( {{\bf{n}},{\bf{n}}'} \right) = \frac{{\sum\limits_{k,k' = 1}^\infty  {\lambda _k } \lambda _{k'} \left[ {{\bf{u}}_k \left( { - {\bf{n}}} \right) \cdot {\bf{u}}_{k'}^* \left( { - {\bf{n}}} \right)} \right]\left[ {{\bf{u}}_k^* \left( { - {\bf{n}}'} \right) \cdot {\bf{u}}_{k'} \left( { - {\bf{n}}'} \right)} \right]}}{{\sum\limits_{k,k' = 1}^\infty  {\lambda _k \lambda _{k'} } \left| {{\bf{u}}_k \left( { - {\bf{n}}} \right) \cdot {\bf{u}}_k^* \left( { - {\bf{n}}} \right)} \right| \left| {{\bf{u}}_{k'} \left( { - {\bf{n}}'} \right) \cdot {\bf{u}}_{k'}^* \left( { - {\bf{n}}'} \right)} \right| }}.
\end{equation}
Equation~(\ref{Local_Mu_In}) explicitly links the macroscopic degree of spatial coherence to the underlying non-local quantum statistics of the incident field. The spatial coherence is governed by the statistical mixture and interference of the transverse Schmidt modes ${\bf{u}}_k$, weighted by $\lambda_k \lambda_{k'}$. In the collinear limit (${\bf{n}} = {\bf{n}}'$), this expression evaluates the local degree of polarization, which is strictly bounded below unity by the inherent mixing of the polarization states. For non-collinear directions (${\bf{n}} \neq {\bf{n}}'$), a strictly separable biphoton state ($K=1$) trivially factorizes, yielding perfect macroscopic optical coherence ($\mu_{in}^2 = 1$). Conversely, strong continuous-variable spatial entanglement ($K \gg 1$) broadly populates multiple spatial modes characterized by rapidly oscillating, uncoordinated phases. The resulting destructive interference mathematically suppresses the spatial cross-correlations ($\mu_{in}^2 \ll 1$). This analytical formulation strictly formalizes the duality between quantum non-locality and classical-like phase correlation, establishing that the macroscopic manifestation of bipartite continuous-variable spatial entanglement is intrinsically equivalent to a severe degradation of the reduced transverse optical coherence.

To evaluate how the macroscopic dissipative object shapes the far-field spatial coherence under this non-classical illumination, we substitute the ingoing spectral correlation dyadic of Eq.~(\ref{Biphoton_Input_Dyad}) into the general expression for the outgoing correlation dyadic in Eq.~(\ref{General_C_out}). We implement the narrow-band detection framework introduced in Sec.~V with an observation bandwidth $\Delta \omega$ centered at the carrier frequency $\bar{\omega}$. Under the assumption that the detection bandwidth is sufficiently broad to fully encompass the narrow spectral envelope $|\phi(\omega)|^2$, the frequency integral $\int_{\Delta \omega} {d\omega } |\phi(\omega)|^2 \approx 1$ isolates the central frequency components of the scattering dyadics, yielding the filtered outgoing correlation dyadic:
\begin{equation} \label{Biphoton_Cout}
{\cal C}_{out}^{(f)} \left( {\mathbf{n}},{\mathbf{n}}' \right) = \sum_{k=1}^\infty \lambda_k \, \mathbf{W}_k^* \left( {\mathbf{n}} \right) \mathbf{W}_k \left( {\mathbf{n}}' \right) + \Delta \omega \, \bar{\omega} n_{\bar{\omega}}  \left( {T_{em} } \right) {\cal V}_{\bar{\omega}}  \left( {\mathbf{n}},{\mathbf{n}}' \right),
\end{equation}
where 
\begin{equation} \label{W_modes}
\mathbf{W}_k(\mathbf{n}) = \sqrt{\bar{\omega}} \Delta \Omega \int {do_{\mathbf{p}} } {\cal T}_{\bar{\omega} ss}(\mathbf{n}|\mathbf{p}) \cdot \mathbf{u}_k(\mathbf{p})
\end{equation}
are the monochromatic spatial Schmidt modes undergoing elastic scattering, obtained by exploiting the smallness of the detection solid angles and the equal-time phase cancellation that eliminates explicit space-time dependence. Due to the arbitrary scattering geometry and the non-unitary nature of the transmission operator caused by material absorption/emission (see Eq.~(\ref{Unitarity_Relation_Main})), the outgoing modes $\mathbf{W}_k$ generally lose the strict orthonormality characterizing the incident modes $\mathbf{u}_k$.

Equation~(\ref{Biphoton_Cout}) demonstrates that the outgoing far-field radiation is inherently partially coherent. For entangled incident states ($K > 1$), the outgoing correlation dyadic takes the form of a statistical mixture of scattered spatial modes ${\bf W}_k({\bf n})$, structurally bounding the macroscopic spatial coherence below unity ($\mu_{out}^2 < 1$) even at absolute zero ($T_{em} = 0$). Consequently, the far-field correlation profile is here governed by a dual statistical mixing mechanism: an intrinsic optical correlation deficit stemming from the multi-mode nature of the incident quantum state, and a local thermodynamic decoherence channel driven by the object's thermal emission.

\paragraph{Geometric and thermodynamic scaling of the quantum-thermal interplay.} 
Unlike classical coherent illumination—where the macroscopic incident intensity can be arbitrarily increased to suppress the relative impact of thermal noise—non-classical continuous-variable states possess a vanishing mean field ($\langle \hat{\mathbf{g}} \rangle = 0$). Crucially, they cannot be amplified to overcome thermal degradation, as increasing the generation rate inherently induces multiphoton emission, strictly destroying the requisite two-photon statistics. Consequently, Eq.~(\ref{Biphoton_Cout}) describes a strict, unscalable competition between the partial spatial coherence of the scattered Schmidt modes and the stochastic material thermal fluctuations. Observing the interplay of these two mechanisms requires the elastically scattered quantum signal to be of the same order as the intrinsic thermal emission, an operational balance quantified by the ratio $\sigma^{(sca)} / [\sigma^{(abs)} n_{\bar{\omega}}(T_{em})]$.

In the subwavelength Rayleigh limit ($a \ll \bar{\lambda}$), observing this interplay is constrained by the intrinsic scaling of the optical cross-sections. Because scattering efficiency scales as $a^6$ while material absorption scales as $a^3$, the ratio $\sigma^{(sca)}/\sigma^{(abs)}$ vanishes proportionally to $(a/\bar{\lambda})^3$. The subwavelength object effectively acts as a thermal emitter rather than a scatterer, and the thermal radiation masks the quantum spatial signature. Recovering any measurable non-local correlation requires operating in a deep-cryogenic regime ($\hbar \bar{\omega} \gg k_B T_{em}$) to exponentially suppress the local thermal population $n_{\bar{\omega}}(T_{em})$.

In the mesoscopic regime ($a \sim \bar{\lambda}$) or through engineered nanophotonics, this cross-sectional disparity can be circumvented. For instance, while highly dissipative plasmonic nanoparticles act primarily as thermal emitters, high-index dielectric nanostructures (e.g., silicon or gallium phosphide) support strong geometrical Mie resonances that break the strict volumetric Rayleigh scaling. By tailoring this electromagnetic geometry, the scattering efficiency can be enhanced to match or exceed material absorption ($\sigma^{(sca)} \gtrsim \sigma^{(abs)}$), facilitating the observation of the interplay between quantum interference and local dissipation even under moderate thermal noise.

In the macroscopic bulk limit ($a \gg \bar{\lambda}$), geometrical optics dictates that both cross-sections scale proportionally to the projected area ($\propto a^2$), fixing their ratio to a size-independent constant. Here, the competition between quantum correlations and thermal decoherence is entirely governed by the thermal energy scale relative to the incident photon energy. If $\hbar \bar{\omega} \gg k_B T_{em}$ (e.g., visible light at room temperature), the thermal population vanishes ($n_{\bar{\omega}} \to 0$), and the scattered non-classical state is preserved. Conversely, if $\hbar \bar{\omega} \ll k_B T_{em}$ (e.g., microwave frequencies), the thermal noise ($n_{\bar{\omega}} \gg 1$) acts as a dominant decoherence channel, degrading the spatial correlations. The interplay mathematically framed by our formalism explicitly emerges when $\hbar \bar{\omega} \approx k_B T_{em}$ (e.g., mid-infrared photons at room temperature). In this specific regime, the thermal bath populates the far-field modes at the single-photon level, causing the non-local partial coherence of the entangled modes and the geometry-driven thermal emission to mix with comparable statistical weights.

\begin{figure}[!t]
\centering
\begin{tikzpicture}[
    % Stili generali
    box/.style={draw=black!50, rounded corners, fill=white, thick, inner sep=10pt},
    mode1/.style={ultra thick, blue!80!black},
    mode2/.style={ultra thick, red!80!black, dashed},
    mode3/.style={ultra thick, green!60!black, densely dotted},
    incident/.style={decorate, decoration={snake, post length=1.5mm, amplitude=0.8mm, segment length=2.5mm}, ->, >=Stealth, thick, cyan!70!black},
    scattered/.style={decorate, decoration={snake, post length=1.5mm, amplitude=0.8mm, segment length=2.5mm}, ->, >=Stealth, thick, blue!70!black},
    labeltext/.style={font=\normalsize\sffamily, align=center},
    math/.style={font=\normalsize}
]

% =======================================================
% SFONDO PER DIVIDERE LE ZONE (Accorciato verticalmente)
% =======================================================
\fill[blue!2] (-7.5, -5.7) rectangle (-2.5, 4.3); % Zona Input
\fill[gray!5] (-2.5, -5.7) rectangle (2.5, 4.3);  % Zona Scatterer
\fill[blue!2] (2.5, -5.7) rectangle (7.5, 4.3);   % Zona Output

% Linee divisorie sfumate
\draw[dashed, gray!50, thick] (-2.5, -5.7) -- (-2.5, 4.3);
\draw[dashed, gray!50, thick] (2.5, -5.7) -- (2.5, 4.3);

% =======================================================
% TITOLI DELLE SEZIONI (Allineamento perfetto in alto usando anchor=north)
% =======================================================
\node[labeltext, font=\large\bfseries\sffamily, anchor=north] at (-5.0, 4.1) {INPUT:\\Ingoing\\Far-Field Coherence};
\node[labeltext, font=\large\bfseries\sffamily, anchor=north] at (0, 4.1) {PROCESS:\\Dissipative Scattering\\ with negligible \\ Thermal Emission};
\node[labeltext, font=\large\bfseries\sffamily, anchor=north] at (5.0, 4.1) {OUTPUT:\\Geometrically Purified\\Far-Field Coherence};

% =======================================================
% ZONA CENTRALE: LO SCATTERER
% =======================================================
% Sfera 3D 
\shade[ball color=blue!10!gray!40] (0, 0.8) circle (1.1);
\draw[thick, black!60] (0, 0.8) circle (1.1);

% ONDE INCIDENTI E SCATTERATE 
% Incidenti (da sinistra)
\draw[incident] (0, 0.8) ++(140:2.0) -- ++(-40:0.8);
\draw[incident] (0, 0.8) ++(180:2.0) -- ++(0:0.8);
\draw[incident] (0, 0.8) ++(220:2.0) -- ++(40:0.8);

% Scatterate (verso destra)
\draw[scattered] (0, 0.8) ++(40:1.2) -- ++(40:0.8);
\draw[scattered] (0, 0.8) ++(0:1.2) -- ++(0:0.8);
\draw[scattered] (0, 0.8) ++(320:1.2) -- ++(-40:0.8);

% Testi esplicativi centrali 
\node[labeltext, text=black, text width=4.6cm] at (0, -0.9) {\textbf{Mode-Selective Filter}};
\node[labeltext, text=black!80, text width=4.8cm] at (0, -1.9) {Resonant scattering of specific modes\\Irreversible absorption of others};

% Freccia logica centrale (Alzata a Y = -3.2)
\draw[->, >=Stealth, line width=2pt, draw=blue!60!black] (-1.8, -3.2) -- (1.8, -3.2);

% =======================================================
% ZONA SINISTRA: INPUT 
% =======================================================
\begin{scope}[shift={(-5.0, 0.8)}] 
    % Cono del fascio
    \fill[blue!10, opacity=0.6] (-2.0, 1.2) -- (2.0, 0.6) -- (2.0, -0.6) -- (-2.0, -1.2) -- cycle;
    
    % Modi spaziali
    \draw[mode1] plot[domain=-1.8:1.8, samples=50] (\x, {0.7*exp(-3*\x*\x)});
    \draw[mode2] plot[domain=-1.8:1.8, samples=50] (\x, {1.4*\x*exp(-3*\x*\x)});
    \draw[mode3] plot[domain=-1.8:1.8, samples=50] (\x, {0.5*(1-6*\x*\x)*exp(-3*\x*\x)});
    
    % Freccia di propagazione
    \draw[->, >=Stealth, ultra thick, black!60] (-1.8, -1.4) -- (1.8, -1.4) node[midway, below] {Propagation};
\end{scope}

% =======================================================
% ZONA DESTRA: OUTPUT 
% =======================================================
\begin{scope}[shift={(5.0, 0.8)}] 
    % Cono del fascio in uscita
    \fill[blue!6, opacity=0.8] (-2.0, 0.5) -- (2.0, 0.9) -- (2.0, -0.9) -- (-2.0, -0.5) -- cycle;
    
    % Modi spaziali
    \draw[mode1] plot[domain=-1.8:1.8, samples=50] (\x, {0.45*exp(-3*\x*\x)});
    \draw[mode2, opacity=0.25] plot[domain=-1.8:1.8, samples=50] (\x, {0.15*\x*exp(-3*\x*\x)});
    
    % Freccia di propagazione
    \draw[->, >=Stealth, ultra thick, black!60] (-1.8, -1.4) -- (1.8, -1.4);
\end{scope}

% =======================================================
% GRAFICI IN BASSO: SPETTRO DI SCHMIDT E COERENZA
% =======================================================
\def\ygraph{-3.7} % Grafici alzati significativamente (da -4.9 a -3.7)

% Grafico Input (Sinistra)
\begin{scope}[shift={(-7.2, \ygraph)}]
    \draw[->, thick] (0,0) -- (3.2,0) node[right, math] {$k$};
    \draw[->, thick] (0,0) -- (0,1.4) node[above, math] {$\lambda_k$};
    
    % Barre spettro Schmidt 
    \filldraw[fill=blue!80!black] (0.3, 0) rectangle (0.7, 1.0);
    \filldraw[fill=red!80!black]  (0.9, 0) rectangle (1.3, 0.75);
    \filldraw[fill=green!60!black](1.5, 0) rectangle (1.9, 0.5);
    \filldraw[fill=gray!50]       (2.1, 0) rectangle (2.5, 0.3);
    
    % Testi
    \node[math, right, align=left, text width=4.4cm] at (0, -0.5) {$K \gg 1$ (Multimode)};
    \node[math, right, red!80!black, align=left, text width=4.4cm] at (0, -1.0) {$\mu^2_{in} \ll 1$ (Low Coherence)};
    \node[math, right, red!80!black, align=left, text width=4.4cm] at (0, -1.5) {\small driven by non-local entanglement};
\end{scope}

% Grafico Output (Destra)
\begin{scope}[shift={(3.0, \ygraph)}]
    \draw[->, thick] (0,0) -- (3.2,0) node[right, math] {$k$};
    \draw[->, thick] (0,0) -- (0,1.4); 
    
    % Etichetta Filtered Weights
    \node[labeltext, align=left, anchor=south west] at (0.05, 1.45) {Filtered\\Weights};
    
    % Barre spettro Output
    \filldraw[fill=blue!80!black] (0.3, 0) rectangle (0.7, 1.1);
    \filldraw[fill=red!80!black, opacity=0.2]  (0.9, 0) rectangle (1.3, 0.1);
    \filldraw[fill=green!60!black, opacity=0.2](1.5, 0) rectangle (1.9, 0.05);
    
    % Testi Output 
    \node[math, right, align=left, text width=4.4cm] at (0, -0.6) {Dyadic Rank $\to 1$ (Single-mode)};
    \node[math, right, blue!80!black, align=left, text width=4.4cm] at (0, -1.4) {$\mu^2_{out} > \mu^2_{in}$ (High Coherence)};
\end{scope}

\end{tikzpicture}
\caption{\textbf{Geometric purification of spatially entangled light by a dissipative macroscopic scatterer.} (Left) The incident continuous-variable state is highly multimode (Schmidt rank $K \gg 1$). The severely reduced macroscopic spatial coherence ($\mu^2_{in} \ll 1$) physically manifests as the exact local counterpart of the non-local spatial entanglement. (Center) Operating in a regime where the thermal population at the carrier frequency is sufficiently small to neglect thermal emission, the dissipative object introduces no relevant thermal noise, instead acting as a passive, non-unitary spatial filter (cyan wavy arrows represent the incoming quantum field, while dark blue wavy arrows denote the elastically scattered modes). Crucially, this filtering is not universal but geometry-dependent: when the object's geometry features resonances that selectively scatter a specific mode while irreversibly absorbing the others, it enacts a differential mode attenuation. (Right) Under such selective filtering conditions, the irreversible absorption systematically reduces the algebraic rank of the outgoing correlation dyadic towards unity. Consequently, the scatterer geometrically purifies the reduced quantum field, enhancing the far-field spatial coherence ($\mu^2_{out} > \mu^2_{in}$) at the strict operational expense of the total transmitted biphoton flux.}
\label{fig:geometric_purification}
\end{figure}
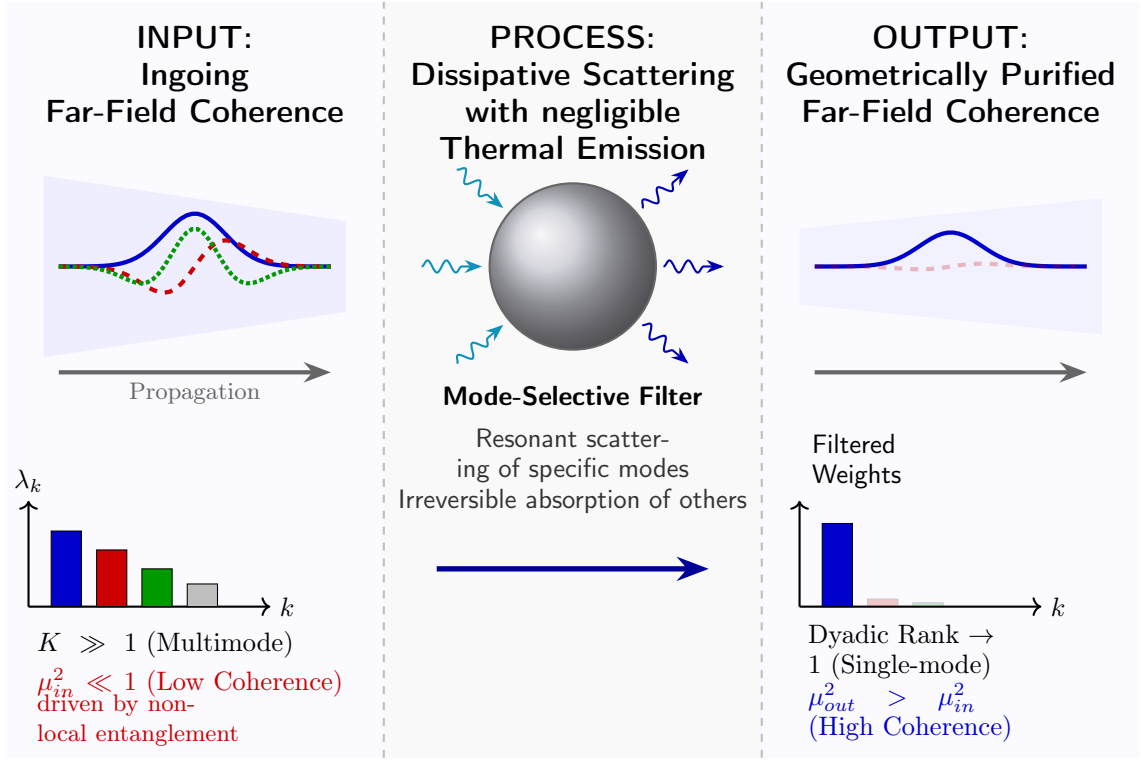

\paragraph{Geometric purification by a cold dissipative scatterer.} 
When the thermal population at the carrier frequency is sufficiently small to neglect intrinsic thermal emission, the outgoing correlation dyadic in Eq.~(\ref{Biphoton_Cout}) reduces exclusively to the pure elastic scattering of the incident biphoton field, given by the superposition of the outgoing modes ${\bf W}_k$. In this regime, the object introduces no additional stochastic noise; rather, it acts as a deterministic, macroscopic spatial filter on the incident continuous-variable state. The transformation from the incident coherence $\mu_{in}^2$ to the outgoing coherence $\mu_{out}^2$ is mathematically dictated by how the object's mode-dependent absorption and non-unitary transmission modify the algebraic rank of the ingoing correlation dyadic—which, for the incident biphoton field, coincides exactly with the spatial Schmidt rank $K$.

If the incident biphoton state is strictly separable ($K=1$), the field is fully coherent ($\mu_{in}^2 = 1$). The scattering process deterministically maps the single incident mode $\mathbf{u}_1$ into a single outgoing spatial mode $\mathbf{W}_1$. Consequently, the outgoing correlation dyadic retains its strict rank-one factorization, ensuring that macroscopic spatial coherence remains bounded to unity ($\mu_{out}^2 = 1$) regardless of the object's geometric complexity. 

Conversely, for highly spatially entangled states ($K > 1$), the incident correlation dyadic is intrinsically multimode, yielding partial spatial coherence ($\mu_{in}^2 < 1$), as illustrated in Fig.~\ref{fig:geometric_purification} (Left). Under this illumination, a dissipative scatterer functions as a mode-selective, non-unitary spatial filter [Fig.~\ref{fig:geometric_purification} (Center)]. Its effect depends entirely on the overlap between the incident Schmidt basis $\mathbf{u}_k$ and the classical transmission dyadic $\mathcal{T}_{\bar{\omega}ss}$. If the object's geometry selectively scatters a single dominant Schmidt mode while irreversibly absorbing all other modes, the associated outgoing modes $\mathbf{W}_{k>1}$ are heavily attenuated. This differential absorption drops the algebraic rank of the outgoing correlation dyadic towards unity [Fig.~\ref{fig:geometric_purification} (Right)]. By projecting the multimode statistical mixture into a lower-dimensional scattered subspace, the macroscopic scatterer geometrically purifies the reduced optical field. This leads to a net enhancement of the far-field spatial coherence ($\mu_{out}^2 > \mu_{in}^2$) at the strict expense of the overall transmitted biphoton flux, representing a fundamental operational trade-off. On the contrary, if the object scatters multiple Schmidt modes without a dominant filtering resonance, it geometrically superimposes these statistically independent channels. This routing sustains the multi-rank nature of the correlation dyadic, strictly preserving the macroscopic spatial incoherence inherent to the non-local entanglement.

\section{Conclusions}

In summary, we have formulated a comprehensive, first-principles framework that provides an exactly unified description of arbitrarily structured quantum light scattering and the intrinsic thermal emission of finite dissipative objects, thereby rigorously capturing their previously unexplored interplay. By explicitly anchoring our approach to the canonically founded modified Langevin noise formalism, we circumvented the fundamental limitations of both idealized lossless quantum models and semiclassical fluctuational electrodynamics. We demonstrated that the outgoing far-field spatial coherence strictly separates into an exact algebraic superposition of two geometry-driven contributions. The first contribution, arising from elastic scattering, acts as a deterministic, non-unitary spatial filter that mode-selectively attenuates incident quantum fluctuations; concurrently, the second contribution, originating from intrinsic thermal emission, projects the object's localized dissipation into the far field, yielding an exact, phenomenological-free quantum-vectorial generalization of the macroscopic van Cittert-Zernike theorem. Crucially, these two physical mechanisms are intimately connected by the global unitarity of the coupled radiation-matter quantum dynamics.

Applying this theoretical framework across fundamental optical regimes establishes rigorous operational bounds for dissipative quantum photonics. Under isotropic thermal illumination, our framework formally demonstrates that a cold passive scatterer ($T_{em} = 0$) breaks the spatial symmetry of the background bath through irreversible photon absorption. This localized energy sink casts a structured far-field radiation deficit---a ``thermal shadow''---whose transverse spatial coherence profile is structurally identical to that of an active primary emitter in vacuum. Under deterministic coherent illumination, we derive the exact thermodynamic thresholds governing the degradation of macroscopic phase correlations by stochastic thermal emission, proving that subwavelength nanostructures suffer a severe suppression of the coherence survival threshold compared to bulk objects due to their divergent absorption-to-scattering cross-sectional ratio. Finally, under spatially entangled continuous-variable illumination, we evaluate the geometric and thermodynamic scaling of the quantum-thermal interplay to establish the operational regimes governing the competition between non-local correlations and local dissipation. In the subwavelength Rayleigh regime, the vanishing scattering-to-absorption ratio masks the quantum signature under thermal noise unless the thermal population is heavily suppressed. Conversely, in the macroscopic bulk limit or via engineered mesoscopic Mie resonances, this continuous interplay explicitly emerges when the incident photon energy matches the thermal energy scale ($\hbar \bar{\omega} \approx k_B T_{em}$). Furthermore, we demonstrate that under operational conditions where intrinsic thermal emission is negligible, the dissipative scatterer can function as a selective non-unitary spatial filter. Provided the object features a specific geometric resonance that selectively scatters a single dominant Schmidt mode while irreversibly absorbing the others, this differential attenuation systematically reduces the algebraic rank of the outgoing correlation dyadic. Consequently, the scatterer geometrically purifies the far-field spatial coherence at the strict operational expense of the total transmitted biphoton flux.

Ultimately, this work conceptually and mathematically bridges the gap between quantum optics, classical coherence theory, and macroscopic fluctuational electrodynamics. It provides the rigorous analytical tools necessary to evaluate quantum spatial correlations in the presence of realistic, open dissipative boundaries. The ability to exactly quantify the continuous interplay between non-local quantum statistics, deterministic geometric routing, and stochastic thermal decoherence provides a rigorous foundation for the ab initio design of robust quantum optical components. Future extensions of this framework will naturally target near-field spatial coherence, where evanescent quantized material fluctuations dominate the local density of states, opening entirely new pathways for the rigorous engineering of loss-resilient structured-light networks, quantum imaging devices, and realistic thermal metamaterials. From an experimental perspective, recent advances in mid-infrared single-photon detection and intensity interferometry provide the immediate technological platforms to test these theoretical predictions, offering a direct route to observe both the predicted thermal shadows and the geometrically purified entanglement in laboratory settings.

\appendix

\section{Theoretical groundwork}
To render this article self-contained, this appendix briefly outlines the MLNF and the associated quantum scattering approach. The underlying definitions of the corresponding operators and classical dyadics introduced here are essential for evaluating the far-field spatial coherence in the main text. We consider an arbitrarily shaped, finite, and dissipative magnetodielectric object embedded in vacuum. Assuming an $e^{-i \omega t}$ time dependence, its causal linear optical response is governed by a globally defined complex permittivity $\varepsilon _\omega \left( {\bf{r}} \right)$ and permeability $\mu _\omega \left( {\bf{r}} \right)$, which match the material properties inside the object volume $V$ and reduce to unity outside.

\paragraph{Modified Langevin noise formalism.} Within the MLNF \cite{Ciattoni2024,Ciattoni2026}, the coupled radiation-matter system is described by bosonic operators: the scattering ($s$) polariton operators ${\bf{\hat g}}_{\omega s} \left( {\bf{n}} \right)$, defined for any spatial direction $\bf{n}$ and transverse to it (${{\bf{n}} \cdot {\bf{\hat g}}_{\omega s} \left( {\bf{n}} \right) = 0}$), alongside the electric ($e$) and magnetic ($m$) polariton operators, ${\bf{\hat f}}_{\omega e } \left( {\bf{r}} \right)$ and ${\bf{\hat f}}_{\omega m } \left( {\bf{r}} \right)$, which are confined to the material interior (${{\bf{r}} \in V}$) and vanish in vacuum. Adopting the convention of positive frequencies $\omega >0$, these polariton operators obey the canonical bosonic commutation rules:
\begin{align} \label{MLNF_Com_Rel}
\left[ {{\bf{\hat g}}_{\omega s} \left( {\bf{n}} \right),{\bf{\hat g}}_{\omega 's}^\dag  \left( {{\bf{n}}'} \right)} \right] &= \delta \left( {\omega  - \omega '} \right)\delta \left( {o_{\bf{n}}  - o_{{\bf{n}}'} } \right){\cal I}_{\bf{n}}, \nonumber \\
 \left[ {{\bf{\hat f}}_{\omega \nu } \left( {\bf{r}} \right),{\bf{\hat f}}_{\omega '\nu '}^\dag  \left( {{\bf{r}}'} \right)} \right] &= \delta \left( {\omega  - \omega '} \right)\delta _{\nu \nu '} \delta \left( {{\bf{r}} - {\bf{r}}'} \right){\cal I},
\end{align}
with all other commutators vanishing. In these expressions,
 $[ {{\bf{\hat A}},{\bf{\hat B}}} ] = [ {\hat A_i ,\hat B_j } ]{\bf{u}}_i {\bf{u}}_j$ is the dyadic commutator of the vector operators ${\bf{\hat A}} = \hat A_i {\bf{u}}_i$ and ${\bf{\hat B}} = \hat B_i {\bf{u}}_i$,  ${\cal I}$ denotes the identity dyadic,  ${\cal I}_{\bf{n}}  = {\cal I} - {\bf{nn}}$ is the dyadic projector onto the plane perpendicular to the propagation direction ${\bf{n}} = \sin \theta _{\bf{n}} \left( {\cos \varphi _{\bf{n}} {\bf{u}}_x  + \sin \varphi _{\bf{n}} {\bf{u}}_y } \right) + \cos \theta _{\bf{n}} {\bf{u}}_z$ and $\delta \left( {o_{\bf{n}}  - o_{{\bf{n}}'} } \right) = \delta \left( {\theta _{\bf{n}}  - \theta '_{\bf{n}} } \right)\delta \left( {\varphi _{\bf{n}}  - \varphi '_{\bf{n}} } \right)/\sin \theta _{\bf{n}}$ defines the standard Dirac delta function for solid angles. Working hereafter in the Heisenberg picture and adopting the shorthand notation $\int d\omega \equiv \int_0^\infty d\omega$, the quantized electric field vector is ${\bf{\hat E}}\left( {{\bf{r}},t} \right) = \int {d\omega } [ {e^{ - i\omega t} {\bf{\hat E}}_\omega  \left( {\bf{r}} \right) + e^{i\omega t} {\bf{\hat E}}_\omega ^\dag  \left( {\bf{r}} \right)} ]$, with its positive-frequency component given by: 

\begin{equation} \label{E_om}
{\bf{\hat E}}_\omega  \left( {\bf{r}} \right) = \int {do_{\bf{n}} } {\cal F}_{\omega s} \left( {\left. {\bf{r}} \right|{\bf{n}}} \right) \cdot {\bf{\hat g}}_{\omega s} \left( {\bf{n}} \right) + \sum\limits_\nu  {\int {d^3 {\bf{r}}'\,} {\cal G}_{\omega \nu } \left( {\left. {\bf{r}} \right|{\bf{r}}'} \right) \cdot {\bf{\hat f}}_{\omega \nu } \left( {{\bf{r}}'} \right)} 
\end{equation}
where $do_{\bf{n}}  = \sin \theta _{\bf{n}} d\theta _{\bf{n}} d\varphi _{\bf{n}}$ denotes the differential solid angle around the direction vector $\bf n$. The associated dyadic weighting functions, namely the scattering kernel ($s$) along with the electric and magnetic kernels ($\nu = e,m$), are defined as:
\begin{equation} \label{Kernels}
{\cal F}_{\omega s} \left( {\left. {\bf{r}} \right|{\bf{n}}} \right) = {\cal F}_\omega  \left( {\left. {\bf{r}} \right|{\bf{n}}} \right)\sqrt {\frac{{\hbar k_\omega ^3 }}{{16\pi ^3 \varepsilon _0 }}} ,\quad \quad \begin{aligned}
  {\cal G}_{\omega e} \left( {\left. {\bf{r}} \right|{\bf{r}}'} \right) &= {\cal G}_\omega  \left( {\left. {\bf{r}} \right|{\bf{r}}'} \right) i\sqrt { \displaystyle \frac{{\hbar k_\omega ^4 }}{{\pi \varepsilon _0 }}{\mathop{\rm Im}\nolimits} \left[ {\varepsilon _\omega  \left( {{\bf{r}}'} \right)} \right]},  \\
  {\cal G}_{\omega m} \left( {\left. {\bf{r}} \right|{\bf{r}}'} \right) &= {\cal G}_\omega  \left( {\left. {\bf{r}} \right|{\bf{r}}'} \right) \displaystyle \frac{{ \times \mathord{\buildrel{\lower3pt\hbox{$\scriptscriptstyle\leftarrow$}} 
\over \nabla } _{{\bf{r}}'} }}{{ik_\omega  }}\sqrt {\frac{{\hbar k_\omega ^4 }}{{\pi \varepsilon _0 }}{\mathop{\rm Im}\nolimits} \left[ {\frac{{ - 1}}{{\mu _\omega  \left( {{\bf{r}}'} \right)}}} \right]}.
\end{aligned}
\end{equation}
where $k_\omega = \omega /c$ is the free-space wavenumber, while the modal dyadic ${\cal F}_\omega \left( {\left. {\bf{r}} \right|{\bf{n}}} \right)$ and the dyadic Green's function ${\cal G}_\omega \left( {\left. {\bf{r}} \right|{\bf{r}}'} \right)$ are defined as the solutions to the boundary-value problems:
\begin{align} \label{GS}
\left[ {\left( {\nabla _{\bf{r}}  \times \frac{1}{{\mu _\omega  \left( {\bf{r}} \right)}}\nabla _{\bf{r}}  \times } \right) - k_\omega ^2 \varepsilon _\omega  \left( {\bf{r}} \right)} \right]{\cal F}_\omega  \left( {\left. {\bf{r}} \right|{\bf{n}}} \right) &=  0,   & 
{\cal F}_\omega  \left( {\left. {r{\bf{m}}} \right|{\bf{n}}} \right)  \mathop  \approx \limits_{r \to + \infty } & e^{i\left( {k_\omega  {\bf{n}}} \right) \cdot \left( {r{\bf{m}}} \right)} {\cal I}_{\bf{n}}   \nonumber \\
& &  +& \frac{{e^{ik_\omega  r} }}{r}{\cal S}_\omega  \left( {{\bf{m}}\left| {\bf{n}} \right.} \right) , \nonumber \\
\left[ {\left( {\nabla _{\bf{r}}  \times \frac{1}{{\mu _\omega  \left( {\bf{r}} \right)}}\nabla _{\bf{r}}  \times } \right) - k_\omega ^2 \varepsilon _\omega  \left( {\bf{r}} \right)} \right]{\cal G}_\omega  \left( {\left. {\bf{r}} \right|{\bf{r}}'} \right) & = \delta \left( {{\bf{r}} - {\bf{r}}'} \right){\cal I}, &
{\cal G}_\omega  \left( {\left. {r{\bf{m}}} \right|{\bf{r}}'} \right)  \mathop  \approx \limits_{r \to + \infty } &\frac{{e^{ik_\omega  r} }}{r}{\cal W}_\omega  \left( {\left. {\bf{m}} \right|{\bf{r}}'} \right).
\end{align}
Here, $\bf m$ denotes an arbitrary observation unit vector, and the notation $\mathop \approx \limits_{r \to + \infty }$ isolates the dominant asymptotic contribution in the far-zone. Furthermore, ${\cal S}_\omega \left( {{\bf{m}}\left| {\bf{n}} \right.} \right)$ identifies the classical scattering dyadic \cite{Kristensson2016}, while ${\cal W}_\omega \left( {{\bf{m}}\left| {\bf{r}} \right.} \right)$ characterizes the asymptotic amplitude distribution of the Green's tensor and is linked to the modal dyadic through the relation ${\cal F}_\omega  \left( {\left. {\bf{r}} \right|{\bf{n}}} \right) = 4\pi {\cal W}_\omega ^T \left( { - {\bf{n}}\left| {\bf{r}} \right.} \right)$. From a classical perspective, the modal dyadic ${\cal F}_\omega \left( {\left. {\bf{r}} \right|{\bf{n}}} \right)$ physically describes the scattering by the object of an incident plane wave propagating along the direction ${\bf n}$, whereas the dyadic Green's function ${\cal G}_\omega \left( {\left. {\bf{r}} \right|{\bf{r}}'} \right)$ governs the radiation emitted by a point dipole situated at ${\bf r}'$. This classical picture elucidates the bipartite structure of the quantized electric field ${\bf{\hat E}}_\omega \left( {\bf{r}} \right)$ in Eq.(\ref{E_om}), clarifying both the physical interpretation and the nomenclature of the associated polaritonic operators. Specifically, the contribution driven by the modal dyadic represents the background \textit{scattering field}, where the operators ${\bf{\hat g}}_{\omega s}$ describe the quantized excitations of the external incoming vacuum modes. Concurrently, the term governed by the dyadic Green's function constitutes the \textit{medium-assisted field}, in which the operators ${\bf{\hat f}}_{\omega e}$ and ${\bf{\hat f}}_{\omega m}$ embody the quantized electric and magnetic dipolar sources whose quantum existence is dictated by the internal losses of the object. Indeed, as shown in Eqs.(\ref{Kernels}), the corresponding dyadic kernels are proportional to the imaginary parts of the object permittivity and permeability, ensuring that these material excitations vanish in the lossless limit. The independent physical reality of these polaritonic excitations is reflected by the diagonal form of the Hamiltonian operator, which decompose into the sum of distinct contributions corresponding to the scattering and material polaritons, i.e. $\hat{H} = \hat{H}_s + \hat{H}_{em}$, given by:
\begin{align} \label{Hs_Hem}
\hat{H}_s &= \int {d\omega } \;\hbar \omega \int {do_{\bf{n}} } {\bf{\hat g}}_{\omega s}^\dag  \left( {\bf{n}} \right) \cdot {\bf{\hat g}}_{\omega s} \left( {\bf{n}} \right), \nonumber \\
\hat{H}_{em} &= \int {d\omega } \;\hbar \omega \sum\limits_\nu  \int {d^3 {\bf{r}} \,}  {{\bf{\hat f}}_{\omega \nu }^\dag  \left( {\bf{r}} \right) \cdot {\bf{\hat f}}_{\omega \nu } \left( {\bf{r}} \right)}.
\end{align}
We highlight a fundamental and classical integral relation:
\begin{equation}
\int {do_{\bf{n}} } {\cal F}_{\omega s} \left( {\left. {\bf{r}} \right|{\bf{n}}} \right) \cdot {\cal F}_{\omega s}^{T*} \left( {\left. {{\bf{r}}'} \right|{\bf{n}}} \right) + \sum\limits_\nu \int {d^3 {\bf{s}} \,}   {{\cal G}_{\omega \nu } \left( {\left. {\bf{r}} \right|{\bf{s}}} \right) \cdot {\cal G}_{\omega \nu }^{T*} \left( {\left. {{\bf{r}}'} \right|{\bf{s}}} \right)}  = \frac{{\hbar k_\omega ^2 }}{{\pi \varepsilon _0 }}{\mathop{\rm Im}\nolimits} \left[ {{\cal G}_\omega  \left( {\left. {\bf{r}} \right|{\bf{r}}'} \right)} \right],
\end{equation}
which, within the quantum framework of the MLNF, provides the mathematical foundation to prove the electric field commutation relation $[ {{\bf{\hat E}}_\omega  \left( {\bf{r}} \right),{\bf{\hat E}}_{\omega '}^\dag  \left( {{\bf{r}}'} \right)} ] = \delta \left( {\omega  - \omega '} \right)\frac{{\hbar k_\omega ^2 }}{{\pi \varepsilon _0 }}{\mathop{\rm Im}\nolimits} [ {{\cal G}_\omega  \left( {\left. {\bf{r}} \right|{\bf{r}}'} \right)} ]$. This result is essential for establishing a direct connection with the fluctuation-dissipation theorem and for ensuring the overall structural consistency of the entire formalism.

\paragraph{Quantum scattering approach.} The MLNF enables the development of a general approach to quantum scattering, as extensively detailed in Ref.\cite{Ciattoni2025}. For our present purposes, we briefly outline the essential features required to analyze spatial coherence in the far field. Crucially, the positive frequency part of the electric field operator, defined as ${\bf{\hat E}}^{(+)}({\bf{r}},t) = \int d\omega \, e^{-i\omega t} {\bf{\hat E}}_\omega({\bf{r}})$, admits distinct far-field ($r \to +\infty$) asymptotic behaviors in the far-past ($t \to -\infty$) and far-future ($t \to +\infty$) limits, namely:
\begin{equation} \label{E_asymptotic}
{\bf{\hat E}}^{(+)}(r{\bf{n}},t) \mathop \approx \limits_{r \to +\infty} \frac{1}{ir} \int d\omega \sqrt{\frac{\hbar k_\omega}{4\pi \varepsilon_0}} 
\begin{cases} 
- e^{-ik_\omega(r + ct)} {\bf{\hat g}}_{\omega s}(-{\bf{n}}), & t \to -\infty, \\
\phantom{-} e^{ik_\omega(r - ct)} {\bf{\hat G}}_{\omega s}({\bf{n}}), & t \to +\infty,
\end{cases}
\end{equation}
where the outgoing scattering polariton operator ${\bf{\hat G}}_{\omega s} \left( {\bf{n}} \right)$ is defined as:
\begin{equation} \label{G_operator}
{\bf{\hat G}}_{\omega s} \left( {\bf{n}} \right) = \int do_{\bf{n}'} \, {\cal T}_{\omega ss} \left( {\left. {\bf{n}} \right|{\bf{n}}'} \right) \cdot {\bf{\hat g}}_{\omega s} \left( {{\bf{n}}'} \right) + \sum\limits_\nu  \int d^3 {\bf{r}}\, {{\cal E}_{\omega s\nu } \left( {\left. {\bf{n}} \right|{\bf{r}}} \right) \cdot {\bf{\hat f}}_{\omega \nu } \left( {\bf{r}} \right)},
\end{equation}
in terms of the transmission dyadic ${\cal T}_{\omega ss}$ and the electric and magnetic emission dyadics ${\cal E}_{\omega s\nu }$ ($\nu = e,m$), which are given by:
\begin{equation} \label{Scattering_Dyadics}
{\cal T}_{\omega ss} \left( {\left. {\bf{n}} \right|{\bf{n}}'} \right) = \delta \left( {o_{\bf{n}}  - o_{{\bf{n}}'} } \right){\cal I}_{\bf{n}}  + \frac{{ik_\omega  }}{{2\pi }}{\cal S}_\omega  \left( {\left. {\bf{n}} \right|{\bf{n}}'} \right), \quad \quad \begin{aligned}
  {\cal E}_{\omega se} \left( {\left. {\bf{n}} \right|{\bf{r}}} \right) &=  - {\cal W}_\omega  \left( {\left. {\bf{n}} \right|{\bf{r}}} \right)\sqrt {4k_\omega ^3 {\mathop{\rm Im}\nolimits} \left[ {\varepsilon _\omega  \left( {\bf{r}} \right)} \right]},  \\
  {\cal E}_{\omega sm} \left( {\left. {\bf{n}} \right|{\bf{r}}} \right) &= {\cal W}_\omega  \left( {\left. {\bf{n}} \right|{\bf{r}}} \right) \times \mathord{\buildrel{\lower3pt\hbox{$\scriptscriptstyle\leftarrow$}} \over \nabla } _{\bf{r}} \sqrt {4k_\omega  {\mathop{\rm Im}\nolimits} \left[ {\frac{{ - 1}}{{\mu _\omega  \left( {\bf{r}} \right)}}} \right]}.
\end{aligned}
\end{equation}
Note that the two temporal limits of the electric field align with the physical fact that converging and diverging spherical waves can only contribute in the far-past and far-future, respectively, an observation that justifies the algebraic form of these two limits. Indeed, the far-past asymptotic behavior is governed by the original ingoing polariton operators ${\bf{\hat g}}_{\omega s}$ because the only converging spherical-wave contribution originates from the plane-wave term within the asymptotic expansion of the modal dyadic. Furthermore, the far-future asymptotic behavior is governed by the outgoing polariton operators ${\bf{\hat G}}_{\omega s}$ since the explicit terms account for all diverging spherical-wave contributions contained within the asymptotic limits of both the modal dyadic (which generates the transmission dyadic) and the dyadic Green's function (which generates the emission dyadics).

The main result of the approach developed in Ref.\cite{Ciattoni2025} lies in supplementing the outgoing scattering polariton operators ${\bf{\hat G}}_{\omega s}$ with the corresponding outgoing electric and magnetic polariton operators ${\bf{\hat F}}_{\omega \nu}$ ($\nu = e,m$), defined explicitly as:
\begin{equation} \label{F_operator}
{\bf{\hat F}}_{\omega \nu } \left( {\bf{r}} \right) = \int do_{\bf{n}} \, {\cal A}_{\omega \nu s} \left( {\left. {\bf{r}} \right|{\bf{n}}} \right) \cdot {\bf{\hat g}}_{\omega s} \left( {\bf{n}} \right) + \sum\limits_{\nu '} \int d^3 {\bf{r}}' \, {\cal Q}_{\omega \nu \nu '} \left( {\left. {\bf{r}} \right|{\bf{r}}'} \right) \cdot {\bf{\hat f}}_{\omega \nu '} \left( {{\bf{r}}'} \right),
\end{equation}
where the absorption dyadics ${\cal A}_{\omega \nu s}$ are defined as:
\begin{equation} \label{Absorption_Dyadics}
\begin{aligned}
 {\cal A}_{\omega es} \left( {\left. {\bf{r}} \right|{\bf{n}}} \right) &=  - \sqrt {4k_\omega ^3 {\mathop{\rm Im}\nolimits} \left[ {\varepsilon _\omega  \left( {\bf{r}} \right)} \right]} {\cal W}_\omega ^T \left( { - {\bf{n}}\left| {\bf{r}} \right.} \right), \\ 
 {\cal A}_{\omega ms} \left( {\left. {\bf{r}} \right|{\bf{n}}} \right) &=  - \sqrt {4k_\omega  {\mathop{\rm Im}\nolimits} \left[ {\frac{{ - 1}}{{\mu _\omega  \left( {\bf{r}} \right)}}} \right]} \nabla _{\bf{r}}  \times {\cal W}_\omega ^T \left( {\left. { - {\bf{n}}} \right|{\bf{r}}} \right),
\end{aligned}
\end{equation}
and the dyadics ${\cal Q}_{\omega \nu \nu'}$ governing the redistribution of the object's internal energy take the form:
\begin{equation} \label{Q_Dyadics}
\begin{aligned}
 {\cal Q}_{\omega ee} \left( {\left. {\bf{r}} \right|{\bf{r}}'} \right) &=  - \delta \left( {{\bf{r}} - {\bf{r}}'} \right){\cal I} - 2ik_\omega ^2 \sqrt {{\mathop{\rm Im}\nolimits} \left[ {\varepsilon _\omega  \left( {\bf{r}} \right)} \right]} {\cal G}_\omega  \left( {\left. {\bf{r}} \right|{\bf{r}}'} \right)\sqrt {{\mathop{\rm Im}\nolimits} \left[ {\varepsilon _\omega  \left( {{\bf{r}}'} \right)} \right]},  \\ 
 {\cal Q}_{\omega em} \left( {\left. {\bf{r}} \right|{\bf{r}}'} \right) &= 2ik_\omega  \sqrt {{\mathop{\rm Im}\nolimits} \left[ {\varepsilon _\omega  \left( {\bf{r}} \right)} \right]} {\cal G}_\omega  \left( {\left. {\bf{r}} \right|{\bf{r}}'} \right) \times \mathord{\buildrel{\lower3pt\hbox{$\scriptscriptstyle\leftarrow$}} \over \nabla } _{{\bf{r}}'} \sqrt {{\mathop{\rm Im}\nolimits} \left[ {\frac{{ - 1}}{{\mu _\omega  \left( {{\bf{r}}'} \right)}}} \right]},  \\ 
 {\cal Q}_{\omega me} \left( {\left. {\bf{r}} \right|{\bf{r}}'} \right) &=  - 2ik_\omega  \sqrt {{\mathop{\rm Im}\nolimits} \left[ {\frac{{ - 1}}{{\mu _\omega  \left( {\bf{r}} \right)}}} \right]} \nabla _{\bf{r}}  \times {\cal G}_\omega  \left( {\left. {\bf{r}} \right|{\bf{r}}'} \right)\sqrt {{\mathop{\rm Im}\nolimits} \left[ {\varepsilon _\omega  \left( {{\bf{r}}'} \right)} \right]},  \\ 
 {\cal Q}_{\omega mm} \left( {\left. {\bf{r}} \right|{\bf{r}}'} \right) &=  - \delta \left( {{\bf{r}} - {\bf{r}}'} \right){\cal I} + 2i\sqrt {{\mathop{\rm Im}\nolimits} \left[ {\frac{{ - 1}}{{\mu _\omega  \left( {\bf{r}} \right)}}} \right]} \nabla _{\bf{r}}  \times {\cal G}_\omega  \left( {\left. {\bf{r}} \right|{\bf{r}}'} \right) \times \mathord{\buildrel{\lower3pt\hbox{$\scriptscriptstyle\leftarrow$}} \over \nabla } _{{\bf{r}}'} \sqrt {{\mathop{\rm Im}\nolimits} \left[ {\frac{{ - 1}}{{\mu _\omega  \left( {{\bf{r}}'} \right)}}} \right]}. 
\end{aligned}
\end{equation}
It is worth emphasizing that, while the transmission dyadic ${\cal T}_{\omega ss}$ and the emission dyadics ${\cal E}_{\omega s\nu}$ arise directly from the far-future asymptotic behavior of the electric field in Eq.(\ref{E_asymptotic}), the absorption dyadics ${\cal A}_{\omega \nu s}$ and the energy redistribution dyadics ${\cal Q}_{\omega \nu \nu'}$ are obtained by explicitly enforcing the unitarity of the input-output mapping connecting the ingoing operators $({\bf{\hat g}}_{\omega s}, {\bf{\hat f}}_{\omega e}, {\bf{\hat f}}_{\omega m})$ to the outgoing operators $({\bf{\hat G}}_{\omega s}, {\bf{\hat F}}_{\omega e}, {\bf{\hat F}}_{\omega m})$. Explicitly, this mapping can be expressed in matrix form as:
\begin{equation}
\begin{pmatrix}
 {\bf{\hat G}}_{\omega s} \\
 {\bf{\hat F}}_{\omega e} \\
 {\bf{\hat F}}_{\omega m}
\end{pmatrix}
= 
\begin{pmatrix}
 \mathsf{T}_{\omega ss} & \mathsf{E}_{\omega se} & \mathsf{E}_{\omega sm} \\
 \mathsf{A}_{\omega es} & \mathsf{Q}_{\omega ee} & \mathsf{Q}_{\omega em} \\
 \mathsf{A}_{\omega ms} & \mathsf{Q}_{\omega me} & \mathsf{Q}_{\omega mm}
\end{pmatrix}
\begin{pmatrix}
 {\bf{\hat g}}_{\omega s} \\
 {\bf{\hat f}}_{\omega e} \\
 {\bf{\hat f}}_{\omega m}
\end{pmatrix},
\end{equation}
where the capital letters denote integral operators whose kernels are the corresponding classical dyadics. For instance, the transmission operator $\mathsf{T}_{\omega ss}$ acts on an arbitrary vector field ${\bf{v}}({\bf{n}})$, defined on the unit sphere and everywhere tangent to it, according to
\begin{equation} \label{Toss}
\left[ {\mathsf{T}_{\omega ss} {\bf{v}}} \right]\left( {\bf{n}} \right) = \int do_{\bf{n}'} {\cal T}_{\omega ss} \left( {\left. {\bf{n}} \right|{\bf{n}}'} \right) \cdot {\bf{v}}\left( {{\bf{n}}'} \right).
\end{equation}
The enforced unitarity of the input-output transformation requires that:
\begin{equation} \label{in-out_unitarity}
\begin{aligned}
\begin{pmatrix}
 \mathsf{T}_{\omega ss} & \mathsf{E}_{\omega se} & \mathsf{E}_{\omega sm} \\
 \mathsf{A}_{\omega es} & \mathsf{Q}_{\omega ee} & \mathsf{Q}_{\omega em} \\
 \mathsf{A}_{\omega ms} & \mathsf{Q}_{\omega me} & \mathsf{Q}_{\omega mm}
\end{pmatrix}
\begin{pmatrix}
 \mathsf{T}_{\omega ss}^+ & \mathsf{A}_{\omega es}^+ & \mathsf{A}_{\omega ms}^+ \\
 \mathsf{E}_{\omega se}^+ & \mathsf{Q}_{\omega ee}^+ & \mathsf{Q}_{\omega me}^+ \\
 \mathsf{E}_{\omega sm}^+ & \mathsf{Q}_{\omega em}^+ & \mathsf{Q}_{\omega mm}^+
\end{pmatrix}
&= 
\begin{pmatrix}
 \mathsf{I}_{ss} & 0 & 0 \\
 0 & \mathsf{I}_{ee} & 0 \\
 0 & 0 & \mathsf{I}_{mm}
\end{pmatrix}, \\
\begin{pmatrix}
 \mathsf{T}_{\omega ss}^+ & \mathsf{A}_{\omega es}^+ & \mathsf{A}_{\omega ms}^+ \\
 \mathsf{E}_{\omega se}^+ & \mathsf{Q}_{\omega ee}^+ & \mathsf{Q}_{\omega me}^+ \\
 \mathsf{E}_{\omega sm}^+ & \mathsf{Q}_{\omega em}^+ & \mathsf{Q}_{\omega mm}^+
\end{pmatrix}
\begin{pmatrix}
 \mathsf{T}_{\omega ss} & \mathsf{E}_{\omega se} & \mathsf{E}_{\omega sm} \\
 \mathsf{A}_{\omega es} & \mathsf{Q}_{\omega ee} & \mathsf{Q}_{\omega em} \\
 \mathsf{A}_{\omega ms} & \mathsf{Q}_{\omega me} & \mathsf{Q}_{\omega mm}
\end{pmatrix}
&= 
\begin{pmatrix}
 \mathsf{I}_{ss} & 0 & 0 \\
 0 & \mathsf{I}_{ee} & 0 \\
 0 & 0 & \mathsf{I}_{mm}
\end{pmatrix},
\end{aligned}
\end{equation}
where $\mathsf{M}^+$ indicates the Hermitian adjoint of the integral operator $\mathsf{M}$, defined in terms of the underlying spatial dyadic kernels as ${\cal M}^+ \left( \left. {\bf{x}} \right|{\bf{x}}' \right) = {\cal M}^{T*} \left( \left. {\bf{x}}' \right|{\bf{x}} \right)$ with the superscript $T*$ denoting the transposed complex conjugate, and the identity operators appearing on the right-hand side are defined by their respective kernels as:
\begin{equation} \label{Identity_Kernels}
\begin{aligned}
 {\cal I}_{ss} \left( {\left. {\bf{n}} \right|{\bf{n}}'} \right) &= \delta \left( {o_{\bf{n}}  - o_{{\bf{n}}'} } \right){\cal I}_{\bf{n}},  \\ 
 {\cal I}_{ee} \left( {\left. {\bf{r}} \right|{\bf{r}}'} \right) &= {\cal I}_{mm} \left( {\left. {\bf{r}} \right|{\bf{r}}'} \right) = \delta \left( {{\bf{r}} - {\bf{r}}'} \right){\cal I}.
\end{aligned}
\end{equation}
This unitarity condition is mathematically equivalent to the requirement that the outgoing polariton operators remain bosonic, thereby satisfying the canonical commutation relations:
\begin{equation} \label{Outgoing_Commutators}
\begin{aligned}
 \left[ {{\bf{\hat G}}_{\omega s} \left( {\bf{n}} \right),{\bf{\hat G}}_{\omega 's}^\dag  \left( {{\bf{n}}'} \right)} \right] &= \delta \left( {\omega  - \omega '} \right)\delta \left( {o_{\bf{n}}  - o_{{\bf{n}}'} } \right){\cal I}_{\bf{n}},  \\ 
 \left[ {{\bf{\hat F}}_{\omega \nu } \left( {\bf{r}} \right),{\bf{\hat F}}_{\omega '\nu '}^\dag  \left( {{\bf{r}}'} \right)} \right] &= \delta \left( {\omega  - \omega '} \right)\delta _{\nu \nu '} \delta \left( {{\bf{r}} - {\bf{r}}'} \right){\cal I}.
\end{aligned}
\end{equation}
Furthermore, this unitarity ensures the exact invertibility of the input-output mapping, allowing the ingoing operators to be expressed in terms of the outgoing ones as:
\begin{equation} \label{Inverse_Mapping}
\begin{pmatrix}
 {\bf{\hat g}}_{\omega s} \\
 {\bf{\hat f}}_{\omega e} \\
 {\bf{\hat f}}_{\omega m}
\end{pmatrix}
= 
\begin{pmatrix}
 \mathsf{T}_{\omega ss}^+ & \mathsf{A}_{\omega es}^+ & \mathsf{A}_{\omega ms}^+ \\
 \mathsf{E}_{\omega se}^+ & \mathsf{Q}_{\omega ee}^+ & \mathsf{Q}_{\omega me}^+ \\
 \mathsf{E}_{\omega sm}^+ & \mathsf{Q}_{\omega em}^+ & \mathsf{Q}_{\omega mm}^+
\end{pmatrix}
\begin{pmatrix}
 {\bf{\hat G}}_{\omega s} \\
 {\bf{\hat F}}_{\omega e} \\
 {\bf{\hat F}}_{\omega m}
\end{pmatrix}.
\end{equation}
Accordingly, these properties establish that the ingoing and outgoing representations provide two distinct but quantum-mechanically equivalent descriptions of the global radiation-matter system.

Physically, this unitary input-output relation arises from the conservation of the global energy of the radiation-matter system within the quantum scattering dynamics. Owing to the macroscopic nature of the MLNF, this quantum conservation is inherently tied to its classical counterpart, such that the integral relations in Eqs.(\ref{in-out_unitarity}) enforcing unitarity exactly coincide with the classical identities governing the physical balance between scattered, emitted, and absorbed energy. In particular, the analysis developed in the main text specifically relies on the relation corresponding to the $(1,1)$ element of the first matrix identity in Eqs.(\ref{in-out_unitarity}), which connects the transmission and emission operators as:
\begin{equation} \label{Operator_Unitarity}
 \mathsf{T}_{\omega ss} \mathsf{T}_{\omega ss}^+ + \mathsf{E}_{\omega se} \mathsf{E}_{\omega se}^+ + \mathsf{E}_{\omega sm} \mathsf{E}_{\omega sm}^+ = \mathsf{I}_{ss},
\end{equation}
which in dyadic notation is equivalent to the integral relation:
\begin{equation} \label{Unitarity_Relation}
\int do_{\bf{m}} \, {\cal T}_{\omega ss} \left( {\left. {\bf{n}} \right|{\bf{m}}} \right) \cdot {\cal T}_{\omega ss}^{T*} \left( {\left. {{\bf{n}}'} \right|{\bf{m}}} \right) + \sum\limits_\nu  \int d^3 {\bf{r}} \, {{\cal E}_{\omega s\nu } \left( {\left. {\bf{n}} \right|{\bf{r}}} \right) \cdot {\cal E}_{\omega s\nu }^{T*} \left( {\left. {{\bf{n}}'} \right|{\bf{r}}} \right)}  = \delta \left( {o_{\bf{n}}  - o_{{\bf{n}}'} } \right){\cal I}_{\bf{n}}.
\end{equation}
This relation demonstrates that the thermal emission term mathematically governs the deviation of the transmission operator from unitarity; for a dissipative object, this unitarity defect is compensated by the material's emission channels.

Finally, comparing the explicit definitions of the emission dyadics in Eqs.(\ref{Scattering_Dyadics}) and the absorption dyadics in Eqs.(\ref{Absorption_Dyadics}), one can directly verify the symmetry relation:
\begin{equation} \label{Kirchhoff_Symmetry_App}
 {\cal A}_{\omega \nu s} \left( {\left. {\bf{r}} \right|{\bf{n}}} \right) = {\cal E}_{\omega s\nu }^T \left( { - {\bf{n}}\left| {\bf{r}} \right.} \right).
\end{equation}
This identity formally encapsulates Kirchhoff's law, equating the object's capacity to absorb radiation from a given direction with its capacity to emit radiation in the opposite direction. Unlike standard semiclassical approaches where this equivalence is typically introduced as a phenomenological thermodynamic postulate, here it is established as an exact algebraic consequence of the unitary input-output relations governing the macroscopic quantum dynamics. Furthermore, as dictated by their explicit expressions, all these emission and absorption dyadics are inherently proportional to the square root of the imaginary parts of the object's complex permittivity and permeability. Physically, this scaling rigidly anchors the macroscopic thermal radiation exchange to the localized internal electromagnetic dissipation. It ensures that in the ideal lossless limit (${\mathop{\rm Im}\nolimits} [\varepsilon_\omega] \to 0$ and ${\mathop{\rm Im}\nolimits} [\mu_\omega] \to 0$), both the absorption and emission channels exactly vanish, reflecting the fundamental fact that the internal material polaritons lose their physical consistency and entirely disappear from the theoretical framework.

\section{Mathematical Bound on the Generalized Degree of Spatial Coherence in the Pure Thermal Emission Regime} \label{App:CoherenceBound}

In this Appendix, we provide the rigorous mathematical proof that the generalized degree of spatial coherence in the pure thermal emission regime is strictly bounded below unity ($\mu^2_{out} < 1$).

Substituting the definition of the spatial structured dyadic ${\cal V}_\omega$ [Eq.~(\ref{DyadV})] directly into the expression for the outgoing correlation dyadic [Eq.~(\ref{Emission_C_out})], we get
\begin{equation} \label{EqApp:C_out_combined}
{\cal C}_{out}({\bf n}, {\bf n}') = \int d\omega \, \omega n_\omega(T_{em}) \int_{\Delta \Omega_{\bf n}} do_{\bf m} \int_{\Delta \Omega_{{\bf n}'}} do_{{\bf m}'} \int_V d^3{\bf r} \sum\limits_{\nu} {\cal E}_{\omega s\nu}^*({\bf m}|{\bf r}) \cdot {\cal E}_{\omega s\nu}^T({\bf m}'|{\bf r}).
\end{equation}
It is worth emphasizing parenthetically that the structural form of Eq.~(\ref{EqApp:C_out_combined})---featuring a single volume integral over ${\bf r}$ and a single summation over the polaritonic index $\nu$ rather than double integrations---is a direct algebraic consequence of the strict spatial uncorrelation of the underlying internal thermal sources, as formally established by the Dirac deltas in Eq.~(\ref{Thermal_Averages}).

To evaluate the numerator of the generalized degree of coherence $\mu^2_{out}$, we must compute the squared Frobenius norm of this dyadic. For this purpose, we consider the Cartesian components of Eq.~(\ref{EqApp:C_out_combined}), explicitly writing out the summations over the discrete indices:
\begin{equation} \label{EqApp:C_components}
{\cal C}_{out, ij}({\bf n}, {\bf n}') = \sum\limits_{\nu } \sum_{k} \int d\omega \, \omega n_\omega(T_{em}) \int_V d^3{\bf r} \left[ \int_{\Delta \Omega_{\bf n}} do_{\bf m} {\cal E}_{\omega s\nu, ik}^*({\bf m}|{\bf r}) \right] \left[ \int_{\Delta \Omega_{{\bf n}'}} do_{{\bf m}'} {\cal E}_{\omega s\nu, jk}({\bf m}'|{\bf r}) \right],
\end{equation}
where $i, j, k \in \{1, 2, 3\}$ denote the Cartesian coordinate indices. Because the thermal weight function $\omega n_\omega(T_{em})$ is strictly positive, Eq.~(\ref{EqApp:C_components}) behaves mathematically as an inner product. Let us formally define the Hilbert space $\mathscr{H}$ of square-integrable complex functions $f(\nu, k,\omega, {\bf r})$ defined over the discrete indices $\nu$ and $k$ and the continuous variables $(\omega, {\bf r})$. The inner product in this space is explicitly given by:
\begin{equation} \label{EqApp:InnerProductDef}
(f, g)_{\mathscr{H}} = \sum\limits_{\nu} \sum_{k} \int d\omega \int_V d^3{\bf r} \, f^*(\nu, k, \omega, {\bf r}) g(\nu, k, \omega, {\bf r}).
\end{equation}
Within this functional space, we can define an effective complex field function $U_{{\bf n}, i}$ associated with the fixed observation direction ${\bf n}$ and the Cartesian component $i$, which naturally incorporates the thermal weight and the angular integration over the detector's aperture:
\begin{equation} \label{EqApp:U_function}
U_{{\bf n}, i}(\nu, k, \omega, {\bf r}) = \sqrt{\omega n_\omega(T_{em})} \int_{\Delta \Omega_{\bf n}} do_{\bf m} \, {\cal E}_{\omega s\nu, ik}({\bf m}|{\bf r}).
\end{equation}
This definition allows us to recast the Cartesian matrix elements of the macroscopic outgoing dyadic from Eq.~(\ref{EqApp:C_components}) directly as functional inner products in $\mathscr{H}$:
\begin{equation} \label{EqApp:C_InnerProduct}
{\cal C}_{out, ij}({\bf n}, {\bf n}') = (U_{{\bf n}, i}, U_{{\bf n}', j})_{\mathscr{H}}.
\end{equation}
At this juncture, we evaluate the squared Frobenius norm of the dyadic, defined in the matrix space as 
\begin{equation}
{\rm tr}\left[ {\cal C}_{out}({\bf n}, {\bf n}') \cdot {\cal C}_{out}^{T*}({\bf n}, {\bf n}') \right] = \sum_{i,j} |{\cal C}_{out, ij}({\bf n}, {\bf n}')|^2.
\end{equation}
Substituting Eq.~(\ref{EqApp:C_InnerProduct}), we apply the standard Cauchy-Schwarz inequality in $\mathscr{H}$, namely $|(U_{{\bf n}, i}, U_{{\bf n}', j})_{\mathscr{H}}|^2 \le (U_{{\bf n}, i}, U_{{\bf n}, i})_{\mathscr{H}} (U_{{\bf n}', j}, U_{{\bf n}', j})_{\mathscr{H}}$, to each Cartesian pair. Summing over both free indices $i$ and $j$, the squared matrix norm strictly obeys the bound:
\begin{equation} \label{EqApp:CauchySchwarz}
\sum_{i,j=1}^3 \left| {\cal C}_{out, ij}({\bf n}, {\bf n}') \right|^2 \le \left[ \sum_{i=1}^3 (U_{{\bf n}, i}, U_{{\bf n}, i})_{\mathscr{H}} \right] \left[ \sum_{j=1}^3 (U_{{\bf n}', j}, U_{{\bf n}', j})_{\mathscr{H}} \right].
\end{equation}
The summed indices on the right-hand side of Eq.~(\ref{EqApp:CauchySchwarz}) mathematically reconstruct the local traces evaluated at the individual detectors. Consequently, the inequality is exactly equivalent to the coordinate-free dyadic expression:
\begin{equation} \label{EqApp:TraceFinal}
{\rm tr}\left[ {\cal C}_{out}({\bf n}, {\bf n}') \cdot {\cal C}_{out}^{T*}({\bf n}, {\bf n}') \right] \le {\rm tr}\left[ {\cal C}_{out}({\bf n}, {\bf n}) \right] {\rm tr}\left[ {\cal C}_{out}({\bf n}', {\bf n}') \right].
\end{equation}
While the general validity of this dyadic inequality was established by Tervo \textit{et al.} \cite{Tervo2003} to ensure that the degree of coherence is universally well-defined ($\mu^2 \le 1$), explicitly re-deriving it for the specific algebraic structure of the 
outgoing correlation dyadic of Eq.(\ref{EqApp:C_out_combined}) is mathematically essential here. Indeed, mapping the macroscopic correlation traces directly onto the underlying functional inner products of $\mathscr{H}$ provides the exact mathematical machinery required to unambiguously evaluate the limiting condition $\mu^2_{out} = 1$. 

Specifically, the equality in the Cauchy-Schwarz bound (and thus $\mu^2_{out} = 1$) is satisfied if and only if the effective field functions are linearly dependent almost everywhere in $\mathscr{H}$, dictating a strict global proportionality $U_{{\bf n}, i} \propto U_{{\bf n}', i}$ across all discrete indices and continuous spatial coordinates. Physically and geometrically, since the integration in Eq.~(\ref{EqApp:U_function}) is performed over strictly disjoint solid angles ($\Delta \Omega_{\bf n} \cap \Delta \Omega_{{\bf n}'} = \emptyset$ for distinct observation directions), and due to the complex directional phase accumulation and vectorial amplitude variations during wave propagation, such strict global proportionality is rigorously precluded. This intrinsic macroscopic incoherence naturally imposes the strict condition $\mu^2_{out} < 1$.

\section{Optical and thermodynamic properties of a Rayleigh scatterer}

Throughout this appendix, we consider a homogeneous, isotropic, and nonmagnetic spherical scatterer of radius $a$ and dielectric permittivity $\varepsilon_\omega^s$  in the subwavelength regime $k_\omega a \ll 1$.

\paragraph{Validity of the Rayleigh approximation.} It is worth beginning with a brief discussion of the regime of validity of the Rayleigh approximation concerning the object's intrinsic thermal emission. The subwavelength condition $2\pi a \ll \lambda$ must hold for the dominant spectral components of the emitted thermal field. According to Wien's displacement law, the peak emission wavelength of a thermal bath at temperature $T_{em}$ is given by $\lambda_{\max} = b/T_{em}$, where $b \simeq 2.89 \times 10^{-3}\text{ m K}$ is Wien's displacement constant. Consequently, the dipole approximation remains physically justified and accurate provided the sphere radius satisfies the bound
\begin{equation} \label{Rayleigh_Validity_Tem}
a \ll \frac{b}{{2\pi T_{em}}}.
\end{equation}
To provide a practical perspective on this geometric constraint, we evaluate the upper bound $b/(2\pi T_{em})$ across three physical regimes. At cryogenic liquid nitrogen temperatures ($77\text{ K}$), the bound evaluates to approximately $6.0\ \mu\text{m}$, allowing for the treatment of relatively large microscale scatterers. For macroscopic setups at room temperature ($300\text{ K}$), the limit drops to about $1.5\ \mu\text{m}$, restricting the validity to nanoparticles with radii up to a few hundred nanometers. Finally, for an incandescent source in the high-temperature regime ($3000\text{ K}$), the peak emission shifts toward the near-infrared and visible spectrum, rendering the bound restrictive at roughly $150\text{ nm}$ and requiring the scatterer radius to be merely a few tens of nanometers. These examples illustrate how the maximum permissible size of the object for the validity of the quantum scattering dipole model scales inversely with its internal temperature.

\paragraph{Scattering dyadics and cross-sections.} The classical scattering dyadic is given by
\begin{equation} \label{Dip_Scat_Dyad}
{\cal S}_\omega \left( {\left. {\bf{n}} \right|{\bf{n}}'} \right) = \frac{k_\omega^2}{4\pi} \alpha_\omega {\cal I}_{\bf{n}} \cdot {\cal I}_{{\bf{n}}'},
\end{equation}
where $\alpha_\omega$ is the scalar polarizability \cite{Ciattoni2025b}. To guarantee energy conservation within the scattering process, the polarizability employed here takes the form
\begin{align}
\alpha_\omega &= \frac{\alpha_\omega^{(0)}}{ 1 - \alpha_\omega^{(0)} \left( \frac{k_\omega^2}{4\pi a} + \frac{i k_\omega^3}{6\pi} \right) }, & {\rm with} \; \alpha_\omega^{(0)} &= 4\pi a^3 \frac{\varepsilon_\omega^s - 1}{\varepsilon_\omega^s + 2},
\end{align}
where the static Clausius-Mossotti term $\alpha_\omega^{(0)} $ is corrected by $k_\omega^2/(4\pi a)$ and $i k_\omega^3/(6\pi)$ to enforce the dynamic depolarization shift and the radiative damping, respectively \cite{Albaladejo2010}. The polarizability determines the optical cross-sections of the scatterer. According to classical electrodynamics and the optical theorem for a subwavelength dipole, the scattering and extinction cross-sections are respectively given by $\sigma _\omega ^{\left( {sca} \right)}  = \frac{k_\omega ^4}{{6\pi }} \left| {\alpha _\omega  } \right|^2$ and $
\sigma _\omega ^{\left( {ext} \right)}  = k_\omega  {\rm Im} \left( {\alpha _\omega  } \right)$. Energy conservation dictates that the net energy dissipated as heat within the material must equal the difference between the total energy extracted from the incident field and the energy elastically re-radiated. Therefore, the absorption cross-section is given by
\begin{equation} \label{sigmabs}
\sigma _\omega ^{\left( {abs} \right)}  = \sigma _\omega ^{\left( {ext} \right)}- \sigma _\omega ^{\left( {sca} \right)} = k_\omega  {\rm Im} \left( {\alpha _\omega  } \right) - \frac{k_\omega ^4 }{{6\pi }}\left| { \alpha _\omega  } \right|^2 .
\end{equation}
The inclusion of the radiative damping term in the dressed polarizability $\alpha_\omega$ is required to ensure $\sigma _\omega ^{\left( {abs} \right)} \ge 0$, thereby maintaining thermodynamic validity. Specifically, in the ideally transparent limit ($\text{Im}(\varepsilon_\omega) = \text{Im}(\mu_\omega) = 0$), this  formulation guarantees that total extinction matches elastic scattering as dictated by the optical theorem, so that the absorption cross-section vanishes ($\sigma _\omega ^{\left( {abs} \right)} = 0$).

\paragraph{Evaluation of the dyadic ${\cal V}_\omega({\bf n}, {\bf n}')$.} To evaluate the structured dyadic ${\cal V}_\omega({\bf n}, {\bf n}')$ for the subwavelength sphere, we first note that, from an operational perspective, the thermal emission contribution can be simplified by invoking Eq.~(\ref{Unitarity_Relation_Main}), which, when combined with relation~(\ref{T_dyadic}), yields
\begin{align} \label{Emission_Unitarity}
\sum\limits_\nu \int_V d^3 {\bf{r}} \, {\cal E}_{\omega s\nu }^* \left( {\left. {\bf{m}} \right|{\bf{r}}} \right) \cdot {\cal E}_{\omega s\nu }^T \left( {\left. {{\bf{m}}'} \right|{\bf{r}}} \right) &= \frac{{ik_\omega  }}{{2\pi }}\left[ {{\cal S}_\omega ^* \left( {\left. {\bf{m}} \right|{\bf{m}}'} \right) - {\cal S}_\omega ^T \left( {\left. {{\bf{m}}'} \right|{\bf{m}}} \right)} \right] \nonumber \\
&\quad - \left( {\frac{{k_\omega  }}{{2\pi }}} \right)^2 \int do_{\bf{p}} \, {\cal S}_\omega ^* \left( {\left. {\bf{m}} \right|{\bf{p}}} \right) \cdot {\cal S}_\omega ^T \left( {\left. {{\bf{m}}'} \right|{\bf{p}}} \right).
\end{align}
By substituting this operational relation into the definition of Eq.~(\ref{DyadV}), we can express the volume integral of the emission dyadics in terms of the classical scattering dyadic, yielding an interference cross-term and a quadratic scattering term:
\begin{align}
{\cal V}_\omega \left( {{\bf{n}},{\bf{n}}'} \right) &= \int\limits_{\Delta \Omega _{\bf{n}} } do_{\bf{m}} \int\limits_{\Delta \Omega _{{\bf{n}}'} } do_{{\bf{m}}'} \left\{ \frac{{ik_\omega  }}{{2\pi }}\left[ {{\cal S}_\omega ^* \left( {\left. {\bf{m}} \right|{\bf{m}}'} \right) - {\cal S}_\omega ^T \left( {\left. {{\bf{m}}'} \right|{\bf{m}}} \right)} \right] \right. \nonumber \\
&\quad \left. - \left( {\frac{{k_\omega  }}{{2\pi }}} \right)^2 \int do_{\bf{p}} \, {\cal S}_\omega ^* \left( {\left. {\bf{m}} \right|{\bf{p}}} \right) \cdot {\cal S}_\omega ^T \left( {\left. {{\bf{m}}'} \right|{\bf{p}}} \right) \right\}.
\end{align}
By introducing the classical dipole scattering dyadic of Eq.~(\ref{Dip_Scat_Dyad}), the structured dyadic becomes:
\begin{align}
{\cal V}_\omega \left( {{\bf{n}},{\bf{n}}'} \right) &= 2\pi {\rm Im} \left( \frac{{k_\omega ^3 \alpha _\omega  }}{{8\pi ^3 }} \right) \left( \int\limits_{\Delta \Omega _{\bf{n}} } do_{\bf{m}} {\cal I}_{\bf{m}} \right) \cdot \left( \int\limits_{\Delta \Omega _{{\bf{n}}'} } do_{{\bf{m}}'} {\cal I}_{{\bf{m}}'} \right) \nonumber \\
&\quad - \pi ^2 \left| \frac{{k_\omega ^3 \alpha _\omega  }}{{8\pi ^3 }} \right|^2 \int\limits_{\Delta \Omega _{\bf{n}} } do_{\bf{m}} \int\limits_{\Delta \Omega _{{\bf{n}}'} } do_{{\bf{m}}'} {\cal I}_{\bf{m}} \cdot \left( \int do_{\bf p} {\cal I}_{\bf p} \right) \cdot {\cal I}_{{\bf{m}}'},
\end{align}
where the nested angular integrations become tractable. Specifically, leveraging the assumption of small detection solid angles to evaluate the local integrals as $\int_{\Delta \Omega_{\bf n}} do_{\bf m} {\cal I}_{\bf m} \simeq \Delta \Omega {\cal I}_{\bf n}$, alongside the full solid-angle identity $\int do_{\bf p} {\cal I}_{\bf p} = \frac{8\pi}{3}{\cal I}$, the algebraic reductions yield:
\begin{align}
{\cal V}_\omega \left( {{\bf{n}},{\bf{n}}'} \right) &= \Delta \Omega ^2 \frac{{k_\omega ^2 }}{{4\pi ^2 }}\left[ {k_\omega  {\rm Im} \left( {\alpha _\omega  } \right) - \frac{{k_\omega ^4 }}{{6\pi }}\left| {\alpha _\omega  } \right|^2 } \right]{\cal I}_{\bf{n}}  \cdot {\cal I}_{{\bf{n}}'}.
\end{align}
The terms inside the brackets recombine to reconstruct the absorption cross-section derived in Eq.~(\ref{sigmabs}). Consequently, the structured dyadic evaluates to the analytical form:
\begin{equation} \label{Dip_V_Dyad}
{\cal V}_\omega \left( {{\bf{n}},{\bf{n}}'} \right) = \Delta \Omega ^2 \frac{{k_\omega ^2 }}{{4\pi ^2 }}\sigma _\omega ^{\left( {abs} \right)} {\cal I}_{\bf{n}}  \cdot {\cal I}_{{\bf{n}}'}.
\end{equation}

\paragraph{Evaluation of the spatial coherence degree under thermal illumination.}
To evaluate the far-field spatial coherence under thermal illumination, we substitute the analytical expression of the structured dyadic ${\cal V}_\omega \left( {{\bf{n}},{\bf{n}}'} \right)$ of Eq.~(\ref{Dip_V_Dyad}) into the general expression for the outgoing correlation dyadic of Eq.~(\ref{mu2_C_out_therm}). By substituting $k_\omega = \omega/c$ to group the frequency dependence, the outgoing correlation dyadic evaluates to:
\begin{equation}
{\cal C}_{out} \left( {{\bf{n}},{\bf{n}}'} \right) = \left[ \int d\omega \, \omega n_\omega \left( {T_s } \right) \right] \Delta \Omega \delta _{{\bf{n}},{\bf{n}}'} {\cal I}_{\bf{n}}  + \left( \frac{{\Delta \Omega }}{{2\pi c}} \right)^2 \int d\omega \, \omega ^3 \left[ {n_\omega  \left( {T_{em} } \right) - n_\omega  \left( {T_s } \right)} \right]\sigma _\omega ^{\left( {abs} \right)} {\cal I}_{\bf{n}}  \cdot {\cal I}_{{\bf{n}}'}.
\end{equation}
By taking the definition of the generalized degree of spatial coherence in Eq.~(\ref{mu2_definition}), we evaluate the trace of the outgoing correlation dyadic ${\cal C}_{out}$ and its transposed conjugate over the distinct polarization states. For distinct observation directions (${\bf n} \neq {\bf n}'$), the incoherent delta-correlated background vanishes, and the numerator is dictated by the cross-correlation term ${\rm tr} \left[ \left( {\cal I}_{\bf n} \cdot {\cal I}_{{\bf n}'} \right) \cdot \left( {\cal I}_{{\bf n}'} \cdot {\cal I}_{\bf n} \right) \right] = 1 + \left( {\bf n} \cdot {\bf n}' \right)^2$. Conversely, the local intensities normalizing the expression at the denominator are computed by setting ${\bf n} = {\bf n}'$, where the trace of the transverse projector yields ${\rm tr} \left[ {\cal I}_{\bf n} \right] = 2$, representing the two independent orthogonal polarization degrees of freedom of the far-field radiation. Carrying out these algebraic contractions, the generalized degree of spatial coherence evaluates to:
\begin{equation}
\mu ^2_{out} \left( {{\bf{n}},{\bf{n}}'} \right) = \frac{1}{2}\delta _{{\bf{n}},{\bf{n}}'}  + \left( {\frac{\Gamma }{{1 + \Gamma }}} \right)^2 \frac{{1 + \left( {{\bf{n}} \cdot {\bf{n}}'} \right)^2 }}{4}\left( {1 - \delta _{{\bf{n}},{\bf{n}}'} } \right),
\end{equation}
where the dimensionless parameter $\Gamma$ dictating the magnitude of the far-field spatial correlations is defined as:
\begin{equation}
\Gamma  = \Delta \Omega \frac{{\int d\omega \, \omega ^3 \left[ {n_\omega  \left( {T_{em} } \right) - n_\omega  \left( {T_s } \right)} \right]\sigma _\omega ^{\left( {abs} \right)} }}{{\left( {2\pi c} \right)^2 \int d\omega \, \omega n_\omega  \left( {T_s } \right)}}.
\end{equation}

\paragraph{Evaluation of the spatial coherence degree under coherent illumination.}
To evaluate the far-field spatial coherence under coherent illumination, we consider an incident monochromatic plane wave propagating along the direction $\bar{\bf p}$, characterized by the spectral amplitude
\begin{align}
{\bf{\alpha}}_\omega({\bf{p}}) &= \delta(\omega - \bar{\omega})\delta(o_{\bf{p}} - o_{\bar{\bf{p}}}){\bf{\bar A}}.
\end{align}
Here, the complex vector amplitude ${\bf{\bar A}}$ satisfies the transversality condition ${\bf{\bar A}} \cdot \bar{\bf p} = 0$, a geometric constraint enforcing the transverse polarization of propagating electromagnetic vacuum modes. By substituting this profile into the definition of the coherent scattered field amplitude given in Eq.~(\ref{A_field_definition}), and evaluating the integral over a small detection solid angle $\Delta \Omega$, we obtain the angularly resolved vector field
\begin{align}
{\bf{A}}({\bf{n}}, \tau) &= \sqrt{\bar{\omega}} e^{-i\bar{\omega}\tau} \left[ \delta_{{\bf{n}}, \bar{\bf{p}}} {\cal I} + \Delta \Omega \frac{ik_{\bar{\omega}}^3}{8\pi^2} \alpha_{\bar{\omega}} {\cal I}_{\bf{n}} \right] \cdot {\bf{\bar A}}.
\end{align}
The purely coherent contribution to the outgoing correlation dyadic evaluates to the time-independent rank-one tensor ${\bf A}^*({\bf n}, \tau) {\bf A}({\bf n}', \tau)$. Following the narrow-band detection framework introduced in the main text, we evaluate the correlation dyadic over a finite bandwidth $\Delta \omega$ centered at the incident frequency $\bar{\omega}$. By superimposing the deterministic coherent contribution with the filtered thermal emission dyadic, the operational far-field correlation dyadic of Eq.(\ref{COH_Cout_Filtered}) yields
\begin{align}
{\cal C}_{out}^{(f)}({\bf{n}}, {\bf{n}}'; r, t) &= \bar{\omega} \left( \left[ \delta_{{\bf{n}}, \bar{\bf{p}}} {\cal I} - \Delta \Omega \frac{ik_{\bar{\omega}}^3}{8\pi^2} \alpha_{\bar{\omega}}^* {\cal I}_{\bf{n}} \right] \cdot {\bf{\bar A}}^* \right) \left( \left[ \delta_{{\bf{n}}', \bar{\bf{p}}} {\cal I} + \Delta \Omega \frac{ik_{\bar{\omega}}^3}{8\pi^2} \alpha_{\bar{\omega}} {\cal I}_{{\bf{n}}'} \right] \cdot {\bf{\bar A}} \right) \nonumber \\
&\quad + \left[ \Delta \omega \, \bar{\omega} n_{\bar{\omega}}(T_{em}) \Delta \Omega^2 \frac{k_{\bar{\omega}}^2}{4\pi^2} \sigma_{\bar{\omega}}^{(abs)} \right] {\cal I}_{\bf{n}} \cdot {\cal I}_{{\bf{n}}'}.
\end{align}
To systematically extract the generalized degree of spatial coherence via Eq.~(\ref{mu2_definition}), it is convenient to define the dimensionless coherent vector ${\bf{J}}({\bf{n}})$, which embeds the thermodynamic signal-to-noise ratio by normalizing the local coherent field against the filtered isotropic thermal noise:
\begin{align}
{\bf{J}}({\bf{n}}) &= \frac{\sqrt{\frac{\bar{\omega}}{2}} \left[ \delta_{{\bf{n}}, \bar{\bf{p}}} {\cal I} + \Delta \Omega \frac{ik_{\bar{\omega}}^3}{8\pi^2} \alpha_{\bar{\omega}} {\cal I}_{\bf{n}} \right] \cdot {\bf{\bar A}}}{\frac{\Delta \Omega}{2\pi c} \sqrt{\Delta \omega \, \bar{\omega}^3 n_{\bar{\omega}}(T_{em}) \sigma_{\bar{\omega}}^{(abs)}}}.
\end{align}
By substituting these defined quantities and evaluating the dyadic traces, the analytical expression for the generalized degree of spatial coherence simplifies to the closed form:
\begin{align}
\mu^2({\bf{n}}, {\bf{n}}'; r, t) &= \frac{|{\bf{J}}({\bf{n}})|^2 |{\bf{J}}({\bf{n}}')|^2 + {\rm Re}[{\bf{J}}^*({\bf{n}}) \cdot {\bf{J}}({\bf{n}}')] + \frac{1}{4}[1 + ({\bf{n}} \cdot {\bf{n}}')^2]}{(|{\bf{J}}({\bf{n}})|^2 + 1)(|{\bf{J}}({\bf{n}}')|^2 + 1)}.
\end{align}
This formulation captures the macroscopic spatial correlations across the entire far field, incorporating the forward-scattering interference and the transverse polarization overlaps. Crucially, the local parameter ${\bf{J}}({\bf{n}})$ governs the transition between a fully coherent elastic scattering regime ($|{\bf{J}}| \to \infty$) and a geometry-driven thermal emission background ($|{\bf{J}}| \to 0$).

\section{Bosonic symmetry and Schmidt mode reduction}
This appendix demonstrates that the exchange symmetry ${\cal J}({\bf{n}}, {\bf{n}}') = {\cal J}^T({\bf{n}}', {\bf{n}})$ analytically enforces the equality of the left and right bases, ${\bf{u}}_k({\bf{n}}) = {\bf{v}}_k({\bf{n}})$, in the Schmidt decomposition of Eq.~(\ref{Schmidt_Decomp}).

The Schmidt decomposition implies that the vector modes ${\bf{u}}_k$ and ${\bf{v}}_k^*$ are, respectively, the orthonormal eigenfunctions of the self-adjoint integral operators characterized by the dyadic kernels:
\begin{align} \label{Kernel_1_2}
{\cal K}_1({\bf{n}}, {\bf{n}}') &= \int {do_{\bf{p}} } {\cal J}({\bf{n}}, {\bf{p}}) \cdot {\cal J}^{T*}({\bf{n}}', {\bf{p}}), \nonumber \\
{\cal K}_2({\bf{n}}, {\bf{n}}') &= \int {do_{\bf{p}} } {\cal J}^{T*}({\bf{p}}, {\bf{n}}) \cdot {\cal J}({\bf{p}}, {\bf{n}}').
\end{align}
Enforcing the exchange symmetry of the dyadic $\cal{J}$ yields:
\begin{align}
{\cal K}_1({\bf{n}}, {\bf{n}}') &= \int {do_{\bf{p}} } {\cal J}({\bf{n}}, {\bf{p}}) \cdot {\cal J}^*({\bf{p}}, {\bf{n}}'), \nonumber \\
{\cal K}_2({\bf{n}}, {\bf{n}}') &= \int {do_{\bf{p}} } {\cal J}^*({\bf{n}}, {\bf{p}}) \cdot {\cal J}({\bf{p}}, {\bf{n}}'),
\end{align}
whose comparison establishes that the two kernels are exact complex conjugates: ${\cal K}_2({\bf{n}}, {\bf{n}}') = {\cal K}_1^*({\bf{n}}, {\bf{n}}')$. The eigenvalue equations for the Schmidt modes are given by:
\begin{align}
\int {do_{{\bf{n}}'} } {\cal K}_1({\bf{n}}, {\bf{n}}') \cdot {\bf{u}}_k({\bf{n}}') &= \lambda_k {\bf{u}}_k({\bf{n}}), \nonumber \\
\int {do_{{\bf{n}}'} } {\cal K}_2({\bf{n}}, {\bf{n}}') \cdot {\bf{v}}_k^*({\bf{n}}') &= \lambda_k {\bf{v}}_k^*({\bf{n}}).
\end{align}
By substituting ${\cal K}_2 = {\cal K}_1^*$ into the second equation and taking the complex conjugate of both sides, one obtains:
\begin{equation} \label{Eigenvalue_v}
\int {do_{{\bf{n}}'} } {\cal K}_1({\bf{n}}, {\bf{n}}') \cdot {\bf{v}}_k({\bf{n}}') = \lambda_k {\bf{v}}_k({\bf{n}}).
\end{equation}
Equation~(\ref{Eigenvalue_v}) demonstrates that the vector field ${\bf{v}}_k({\bf{n}})$ satisfies the identical eigenvalue equation as ${\bf{u}}_k({\bf{n}})$, proving that both bases span the same eigenspaces of the operator ${\cal K}_1$.

\textit {Non-degenerate eigenvalues.} If the eigenvalue $\lambda_k$ is strictly non-degenerate, its single contribution to the Schmidt decomposition in Eq.~(\ref{Schmidt_Decomp}) is isolated as:
\begin{equation}
{\cal J}_k \left( {{\bf{n}},{\bf{n}}'} \right) = \sqrt {\lambda _k } {\bf{u}}_k \left( {\bf{n}} \right){\bf{v}}_k \left( {{\bf{n}}'} \right).
\end{equation}
The corresponding eigenspace is one-dimensional. Consequently, the normalized eigenvectors ${\bf{u}}_k$ and ${\bf{v}}_k$ can differ at most by a constant global phase factor, ${\bf{v}}_k({\bf{n}}) = e^{i\theta_k} {\bf{u}}_k({\bf{n}})$. Substituting this relation into the mode contribution yields:
\begin{equation}
{\cal J}_k \left( {{\bf{n}},{\bf{n}}'} \right) = \sqrt{\lambda_k} e^{i\theta_k} {\bf{u}}_k({\bf{n}}) {\bf{u}}_k({\bf{n}}').
\end{equation}
Because the phase $\theta_k$ is arbitrary in the definition of the single basis vectors, one defines a gauge-transformed orthonormal mode $\tilde{{\bf{u}}}_k({\bf{n}}) = e^{i\theta_k/2} {\bf{u}}_k({\bf{n}})$. Orthonormality is strictly preserved, and the single-mode contribution to the Schmidt decomposition in Eq.~(\ref{Schmidt_Decomp}) assumes the symmetric form:
\begin{equation}
{\cal J}_k \left( {{\bf{n}},{\bf{n}}'} \right) = \sqrt{\lambda_k} \tilde{{\bf{u}}}_k({\bf{n}}) \tilde{{\bf{u}}}_k({\bf{n}}').
\end{equation}
Thus, in the appropriate gauge, ${\bf{u}}_k({\bf{n}}) = {\bf{v}}_k({\bf{n}})$ identically.

\textit{Degenerate eigenspaces.} If an eigenvalue $\lambda_k$ exhibits a degeneracy of multiplicity $M > 1$, the corresponding $M$-dimensional eigenspace is spanned by the subsets $\{{\bf{u}}_1, \dots, {\bf{u}}_M\}$ and $\{{\bf{v}}_1, \dots, {\bf{v}}_M\}$. Since both sets constitute orthonormal bases for the identical vector subspace, they are rigorously connected by a unitary transformation:
\begin{equation}
{\bf{v}}_a({\bf{n}}) = \sum_{b=1}^M U_{ab} {\bf{u}}_b({\bf{n}}),
\end{equation}
where $U$ is a complex unitary matrix of dimension $M \times M$. The contribution of this degenerate eigenvalue to the Schmidt decomposition in Eq.~(\ref{Schmidt_Decomp}) is isolated as:
\begin{equation}
{\cal J}_k \left( {{\bf{n}},{\bf{n}}'} \right) = \sqrt{\lambda_k} \sum_{a=1}^M {\bf{u}}_a({\bf{n}}) {\bf{v}}_a({\bf{n}}') = \sqrt{\lambda_k} \sum_{a=1}^M \sum_{b=1}^M U_{ab} {\bf{u}}_a({\bf{n}}) {\bf{u}}_b({\bf{n}}').
\end{equation}
The bosonic symmetry constraint requires this specific block amplitude to be symmetric, ${\cal J}_k \left( {{\bf{n}},{\bf{n}}'} \right) = {\cal J}_k^T \left( {{\bf{n}}',{\bf{n}}} \right)$. Comparing the expansion coefficients enforces $U_{ab} = U_{ba}$, proving that $U$ is both unitary and symmetric ($U = U^T$).

According to the Autonne-Takagi factorization theorem, any complex symmetric unitary matrix admits the exact factorization $U = W^T W$, where $W$ is a unitary matrix. Inserting this factorization into the degenerate dyadic block reorganizes the sums as follows:
\begin{align}
{\cal J}_k \left( {{\bf{n}},{\bf{n}}'} \right) &= \sqrt{\lambda_k} \sum_{a,b=1}^M \left( \sum_{c=1}^M W_{ca} W_{cb} \right) {\bf{u}}_a({\bf{n}}) {\bf{u}}_b({\bf{n}}') \nonumber \\
&= \sqrt{\lambda_k} \sum_{c=1}^M \left( \sum_{a=1}^M W_{ca} {\bf{u}}_a({\bf{n}}) \right) \left( \sum_{b=1}^M W_{cb} {\bf{u}}_b({\bf{n}}') \right).
\end{align}
The terms inside the parentheses establish a new set of spatial modes defined by the unitary change of basis $\tilde{{\bf{u}}}_c({\bf{n}}) = \sum_{a=1}^M W_{ca} {\bf{u}}_a({\bf{n}})$. Expressing the degenerate contribution to the Schmidt decomposition in Eq.~(\ref{Schmidt_Decomp}) in terms of this new orthonormal basis removes the cross terms, yielding a strictly diagonal structure:
\begin{equation}
{\cal J}_k \left( {{\bf{n}},{\bf{n}}'} \right) = \sqrt{\lambda_k} \sum_{c=1}^M \tilde{{\bf{u}}}_c({\bf{n}}) \tilde{{\bf{u}}}_c({\bf{n}}').
\end{equation}
This confirms that the degeneracy space possesses sufficient degrees of freedom to analytically rotate the orthonormal basis until the left and right modes coincide identically (${\bf{u}}_a = {\bf{v}}_a$) across the corresponding subset of the Schmidt decomposition.


\begin{thebibliography}{10}

\bibitem{Mandel1995} 
L. Mandel and E. Wolf, 
Optical Coherence and Quantum Optics 
(Cambridge University Press, Cambridge, 1995). 

\bibitem{Glauber1963} 
R. J. Glauber, 
The Quantum Theory of Optical Coherence, 
Phys. Rev. \textbf{130}, 2529 (1963). 

\bibitem{Greffet2002} 
J.-J. Greffet, R. Carminati, K. Joulain, J.-P. Mulet, S. Mainguy, and Y. Chen, 
Coherent emission of light by thermal sources, 
Nature \textbf{416}, 61-64 (2002). 

\bibitem{Rotter2017} 
S. Rotter and S. Gigan, 
Light fields in complex media: Mesoscopic scattering meets wave control, 
Rev. Mod. Phys. \textbf{89}, 015005 (2017). 

\bibitem{Michelson1921}
A. A. Michelson and F. G. Pease, 
Measurement of the diameter of $\alpha$ Orionis with the interferometer, 
Astrophys. J. \textbf{53}, 249-259 (1921). 

\bibitem{Monnier2003}
J. D. Monnier, 
Optical interferometry in astronomy, 
Rep. Prog. Phys. \textbf{66}, 789 (2003). 

\bibitem{Zheng2013}
G. Zheng, R. Horstmeyer, and C. Yang, 
Wide-field, high-resolution Fourier ptychographic microscopy, 
Nat. Photon. \textbf{7}, 739-745 (2013). 

\bibitem{Schnars1994}
U. Schnars and W. P. O. Jüptner, 
Direct recording of holograms by a CCD target and numerical reconstruction, 
Appl. Opt. \textbf{33}, 179-181 (1994). 

\bibitem{Gisin2002}
N. Gisin, G. Ribordy, W. Tittel, and H. Zbinden, 
Quantum cryptography, 
Rev. Mod. Phys. \textbf{74}, 145 (2002). 

\bibitem{Walborn2006}
S. P. Walborn, D. S. Lemelle, M. P. Almeida, and P. H. Souto Ribeiro, 
Quantum Key Distribution with Higher-Order Spatial Modes of a Photon, 
Phys. Rev. Lett. \textbf{96}, 090501 (2006). 

\bibitem{Braunstein2005}
S. L. Braunstein and P. van Loock, 
Quantum information with continuous variables, 
Rev. Mod. Phys. \textbf{77}, 513 (2005). 

\bibitem{Carney1998}
P. S. Carney and E. Wolf, 
An energy theorem for scattering of partially coherent beams, 
Opt. Commun. \textbf{155}, 1-6 (1998). 

\bibitem{Fischer2012}
D. G. Fischer, T. van Dijk, T. D. Visser, and E. Wolf, 
Coherence effects in Mie scattering, 
J. Opt. Soc. Am. A \textbf{29}, 78-84 (2012). 

\bibitem{Schouten2024}
H. F. Schouten and T. D. Visser, 
Scattering of partially coherent electromagnetic beams by a sphere, 
Opt. Express \textbf{32}, 10690-10702 (2024). 

\bibitem{Premaratne2012}
M. Premaratne, 
Partially coherent light interaction with nano objects, 
14th International Conference on Transparent Optical Networks (ICTON), 1-4 (2012). 

\bibitem{Glauber1991}
R. J. Glauber and M. Lewenstein, 
Quantum optics of dielectric media, 
Phys. Rev. A \textbf{43}, 467-491 (1991). 

\bibitem{Fearn1987}
H. Fearn and R. Loudon, 
Quantum theory of the lossless beam splitter, 
Opt. Commun. \textbf{64}, 485-490 (1987). 

\bibitem{Blow1990}
K. J. Blow, R. Loudon, C. R. Phoenix, and T. J. Shepherd, 
Continuum quantum optics, 
Phys. Rev. A \textbf{42}, 4102-4114 (1990). 

\bibitem{Rytov1989}
S. M. Rytov, Y. A. Kravtsov, and V. I. Tatarskii, 
Principles of Statistical Radiophysics 3: Elements of Random Fields 
(Springer-Verlag, Berlin, 1989). 

\bibitem{Joulain2005}
K. Joulain, J.-P. Mulet, F. Marquier, R. Carminati, and J.-J. Greffet, 
Surface electromagnetic waves thermally excited: Radiative heat transfer, coherence properties and Casimir forces revisited in the light of near field optics, 
Surf. Sci. Rep. \textbf{57}, 59-112 (2005). 

\bibitem{Bertilone1997}
D. C. Bertilone, 
Radiometric coherence tensors for thermal radiation emission from an opaque specular surface, 
J. Opt. Soc. Am. A \textbf{14}, 693-702 (1997). 

\bibitem{Polder1971}
D. Polder and M. Van Hove, 
Theory of Radiative Heat Transfer between Closely Spaced Bodies, 
Phys. Rev. B \textbf{4}, 3303 (1971). 

\bibitem{Eckhardt1984}
W. Eckhardt, 
Macroscopic theory of electromagnetic fluctuations and stationary radiative heat transfer, 
Phys. Rev. A \textbf{29}, 1991 (1984). 

\bibitem{Gruner1996}
T. Gruner and D.-G. Welsch, 
Green-function approach to the radiation-field quantization for homogeneous and inhomogeneous Kramers-Kronig dielectrics, 
Phys. Rev. A \textbf{53}, 1818 (1996). 

\bibitem{Matloob1995}
R. Matloob, R. Loudon, S. M. Barnett, and J. Jeffers, 
Electromagnetic field quantization in absorbing dielectrics, 
Phys. Rev. A \textbf{52}, 4823 (1995). 

\bibitem{Gruner1996b}
T. Gruner and D.-G. Welsch, 
Quantum-optical input-output relations for dispersive and lossy multilayer dielectric plates, 
Phys. Rev. A \textbf{54}, 1661-1677 (1996). 

\bibitem{Knoll1999}
L. Knöll, S. Scheel, E. Schmidt, D.-G. Welsch, and A. V. Chizhov, 
Quantum-state transformation by dispersive and absorbing four-port devices, 
Phys. Rev. A \textbf{59}, 4716-4727 (1999). 

\bibitem{Khanbekyan2003}
M. Khanbekyan, L. Knöll, and D.-G. Welsch, 
Input-output relations at dispersing and absorbing planar multilayers for the quantized electromagnetic field containing evanescent components, 
Phys. Rev. A \textbf{67}, 063812 (2003). 

\bibitem{DiStefano2001}
O. Di Stefano, S. Savasta, and R. Girlanda, 
Mode expansion and photon operators in dispersive and absorbing dielectrics, 
J. Mod. Opt. \textbf{48}, 67-84 (2001). 

\bibitem{Drezet2017}
A. Drezet, 
Quantizing polaritons in inhomogeneous dissipative systems, 
Phys. Rev. A \textbf{95}, 023831 (2017). 

\bibitem{Na2023}
D. Y. Na, T. E. Roth, J. Zhu, W. C. Chew, and C. J. Ryu, 
Numerical framework for modeling quantum electromagnetic systems involving finite-sized lossy dielectric objects in free space, 
Phys. Rev. A \textbf{107}, 063702 (2023). 

\bibitem{Ciattoni2024}
A. Ciattoni, 
Quantum electrodynamics of lossy magnetodielectric samples in vacuum: Modified Langevin noise formalism, 
Phys. Rev. A \textbf{110}, 013707 (2024). 

\bibitem{Philbin2010}
T. G. Philbin, 
Canonical quantization of macroscopic electromagnetism, 
New J. Phys. \textbf{12}, 123008 (2010). 

\bibitem{Ciattoni2026}
A. Ciattoni, 
Direct derivation of the modified Langevin noise formalism from the canonical quantization of macroscopic electromagnetism, 
New J. Phys. \textbf{28}, 064513 (2026).

\bibitem{Miano2025a}
G. Miano, L. M. Cangemi, and C. Forestiere, 
Quantum emitter interacting with a dispersive dielectric object: a model based on the modified Langevin noise formalism, 
Nanophotonics (2025). 

\bibitem{Miano2025b}
G. Miano, L. M. Cangemi, and C. Forestiere, 
Spectral densities of a dispersive dielectric sphere in the modified Langevin noise formalism, 
Phys. Rev. A \textbf{112}, 033712 (2025). 

\bibitem{Miano2026}
G. Miano, L. M. Cangemi, and C. Forestiere, 
Modified Langevin noise formalism for multiple quantum emitters in dispersive electromagnetic environments out of equilibrium, 
Phys. Rev. A \textbf{113}, 023720 (2026). 

\bibitem{Ciattoni2026b}
A. Ciattoni, 
Quantum optomechanics of lossy bodies: general approach and structured squeezed vacuum effects, 
Submitted for publication on Physical Review A.

\bibitem{Ciattoni2025}
A. Ciattoni, 
Quantum-optical scattering by macroscopic lossy objects: A general approach, 
Phys. Rev. A \textbf{112}, 013704 (2025). 

\bibitem{Ciattoni2025b}
A. Ciattoni, 
Classically tuning the quantum interference of two photons scattered by a macroscopic lossy sphere, Accepted for publication on Phys. Rev. A.

\bibitem{Kristensson2016}
G. Kristensson, 
Scattering of Electromagnetic Waves by Obstacles, 
Scitech Publishing, New York (2016).

\bibitem{Albaladejo2010}
S. Albaladejo, R. G\'{o}mez-Medina, L. S. Froufe-P\'{e}rez, H. Marinchio, R. Carminati, J. F. Torrado, G. Armelles, A. García-Martín, and J. J. S\'{a}enz,
Radiative corrections to the polarizability tensor of an electrically small anisotropic dielectric particle, 
Opt. Express \textbf{18}, 3556-3567 (2010). 

\bibitem{Tervo2003}
J. Tervo, T. Setälä, and A. T. Friberg, 
Degree of coherence for electromagnetic fields, 
Opt. Express \textbf{11}, 1137-1143 (2003). 

\bibitem{Planck1914} 
M. Planck, 
The Theory of Heat Radiation 
(P. Blakiston's Son \& Co., Philadelphia, 1914). 

\bibitem{Bykovskii1985}
Y. A. Bykovskiĭ, A. M. Zarubin, and A. I. Larkin,
Partially coherent holography: Its properties and applications
Sov. J. Quantum Electron. \textbf{16}, 1165-1173 (1986). 

\bibitem{Zarubin1993}
A. M. Zarubin,
Three-dimensional generalization of the Van Cittert-Zernike theorem to wave and particle scattering,
Opt. Commun. \textbf{100} 491-507 (1993). 

\bibitem{Carter1981}
W.~H. Carter and E.~Wolf, 
Correlation theory of wavefields generated by fluctuating, three-dimensional, primary, scalar sources,
Opt. Acta \textbf{28}, 227 (1981).

\bibitem{Law2004}
C. K. Law and J. H. Eberly, 
Analysis and Interpretation of High Transverse Entanglement in Optical Parametric Down Conversion, 
Phys. Rev. Lett. \textbf{92}, 127903 (2004). 

\bibitem{Walborn2010}
S. P. Walborn, C. H. Monken, S. Pádua, and P. H. Souto Ribeiro, 
Spatial correlations in parametric down-conversion, 
Phys. Rep. \textbf{495}, 87-139 (2010). 

\bibitem{Monken1998}
C. H. Monken, P. H. S. Ribeiro, and S. Pádua, 
Transfer of angular spectrum and image formation in spontaneous parametric down-conversion, 
Phys. Rev. A \textbf{57}, 3123-3126 (1998). 

\bibitem{Saleh2000}
B. E. A. Saleh, A. F. Abouraddy, A. V. Sergienko, and M. C. Teich, 
Duality between partial coherence and partial entanglement, 
Phys. Rev. A \textbf{62}, 043816 (2000). 

\bibitem{Joobeur1994}
A. Joobeur, B. E. A. Saleh, and M. C. Teich, 
Spatiotemporal coherence properties of entangled light beams generated by parametric down-conversion, 
Phys. Rev. A \textbf{50}, 3349-3361 (1994). 

\end{thebibliography}
\end{document}